\documentclass{article}
\usepackage[a4paper, margin=1.0in]{geometry}
\usepackage{graphicx} 
\usepackage{subfiles}
\usepackage{blindtext}
\usepackage{url}
\usepackage[hidelinks]{hyperref}
\usepackage{listings}
\usepackage{amsmath}
\usepackage{authblk}
\usepackage{float}
\usepackage{comment}
\usepackage{microtype}

\newenvironment{courier}{\fontfamily{pcr}\selectfont}{\par}

\title {Analysis Preservation and Reproducibility in Collider Physics: An ADL/CutLang Implementation of ATLAS Open Data Analyses}

\author[1,2]{A. Ad{\i}g\"uzel}
\author[1]{A. Sansar}
\author[1]{K. Sahan}
\author[1]{K. Karaca}
\affil[1]{Istanbul University, Department of Physics, Istanbul, Turkey}
\affil[2]{Boğaziçi University, Feza Gürsey Center for Physics and Mathematics, Istanbul, Turkey}

\date{\today}

\begin{document}

\maketitle

\begin{abstract}
   Analysis Description Language (ADL) and the CutLang runtime interpreter offer an innovative and sustainable solution for describing, executing, and preserving the physics content of collider data analyses. In this solution, all physics objects and event selection criteria are expressed in a human-readable form using ADL, a domain-specific language designed for collider physics. ADL descriptions are interpreted directly at runtime by CutLang without translation into a general-purpose programming language, producing event selections that can be readily fed into statistical analysis tools. This approach addresses the long-standing challenge of analysis preservation and analysis reproducibility in high energy physics by decoupling the physics logic from experiment-specific software infrastructures.
   
    In this study, several benchmark  analyses within ATLAS Open Data C++ Framework are reimplemented in ADL and executed with CutLang using  data provided by ATLAS Open Data at a center-of-mass energy of 13 TeV and corresponding to an integrated luminosity of 10 fb$^{-1}$. The reimplemented analyses yield results in good agreement with the original publications, validating the ADL/CutLang workflow as a reliable alternative to traditional analysis frameworks. The study also provides a detailed assessment of CutLang's current capabilities and identifies areas for further development, offering a roadmap toward broader adoption of ADL-based analysis preservation in the LHC community.
\end{abstract}

\section{Introduction}
Data analysis in collider physics relies heavily on statistics and event classification. Collider events are classified as signal-like or background-like based on the properties of physics objects such as electrons or jets, derived quantities such as the angular isolation of leptons from nearby jets, or global event variables such as the solid angle between the highest-momentum jet and the highest-momentum lepton. In analyses aiming for discovery or exclusion, events are sorted into signal-enriched and background-enriched regions, and various statistical tools are then applied to draw conclusions from the contents of these regions.
Such an analysis workflow demands an extensive software infrastructure. Corrections and calibrations must be applied to raw detector data, physics objects must be defined and selected according to quality criteria, and an event selection algorithm must be constructed to classify each event. Executing this workflow over tens of millions of simulated and recorded collider events requires both substantial computing resources and large-scale software frameworks such as ATHENA, used by the ATLAS experiment.

Beyond the computational infrastructure, each analysis group typically develops its own analysis-specific code for every study. Both the generic data handling frameworks and the analysis-specific codes are almost universally written in general-purpose programming languages such as C$^{++}$ or Python.
Since these languages are not specifically designed for particle physics analyses, several challenges arise over time, including difficulties in analysis preservation, communication between different groups, and adapting to new software releases. To address these challenges, the Analysis Description Language (ADL) was developed. As a domain-specific and declarative language, ADL aims to describe the physics content of a collider analysis in a standard and unambiguous way. 

Since its development, the Analysis Description Language and CutLang, the runtime interpreter for ADL \cite{Unel_2018,Unel_2021}, have been used in various ATLAS and CMS analyses, reinterpretation studies and validation efforts. By providing a common language for describing analyses, ADL/CutLang facilitates reproducibility and transparency, both of which are crucial for result validation and fostering collaboration within the scientific community.  ADL/CutLang has also been introduced to students at workshops and schools over the past several years \cite{Adiguzel_2021}, demonstrating its potential as a pedagogical tool.

In the meantime, other computational tools have been provided for educational use and to
enhance the accessibility of collider data analysis. These tools include the 13 TeV ATLAS Open Data analysis framework written in C++ and integrated with ROOT, as well as Jupyter notebooks utilizing RDataFrame, Uproot, and PyROOT libraries \cite{ATLASCppFramework13TeV}.
The Open Data analysis framework can be used to access and analyze the publicly available portion of the  data collected by the ATLAS Collaboration.
In 2020, the ATLAS Experiment Collaboration publicly released proton-proton collision data corresponding to an integrated luminosity of 10 fb$^{-1}$, collected by the ATLAS detector at the Large Hadron Collider (LHC) at a center-of-mass energy of 13 TeV during the year 2016. This release also included simulated events from the Standard Model and Beyond the Standard Model (BSM) \cite{ATL-OREACH-PUB-2020-001}. The CMS Experiment Collaboration has publicly released proton-proton and heavy-ion collision data collected by the CMS detector for the periods 2010-2012 and 2015-2018. In addition, the CMS Collaboration has actively promoted the use of these datasets for educational and research purposes by organizing dedicated workshops and providing documentation to help new users get started with CMS Open Data \cite{cms_opendata_guide}.

A closer examination of the 13 TeV ATLAS Open Data framework reveals a notable limitation: in educational settings, the focus tends to fall on technical programming skills rather than on the physics logic underlying the analyses. Moreover, the available descriptions of the analyses, including their selection procedures and methodological choices, are sometimes not fully documented, which makes it difficult for students and newcomers to develop a clear understanding of the physics being studied.

The objective of this paper is therefore twofold. First, we demonstrate the advantages of ADL in educational environments by re-implementing a set of analyses from the ATLAS Open Data C$^{++}$ framework in ADL and executing them with CutLang. The results are shown to be in good agreement with those obtained from the original framework, while the ADL-based descriptions provide a more transparent, standardized, and pedagogically accessible representation of the analysis logic. 
Second, we use this study to validate ADL and CutLang as practical tools for communicating physics analyses, ensuring reproducibility, and supporting long-term preservation, while highlighting their current strengths and areas where further development of CutLang could expand its applicability.

\section{Analysis description language and CutLang runtime interpreter}

The Analysis Description Language \cite{ADL_CERN_WEB} is a domain-specific language that enables the standardized and unambiguous description of the physics algorithm underlying a High Energy Physics (HEP) analysis. In ADL, the physics content of an analysis is defined in a human-readable plain text file, where objects, variables, and event selection criteria are clearly separated through a structured system of blocks and keywords. The blocks, keywords, and their respective functions in a standard ADL file are described in Tables \ref{tab:ADL_blocks} and \ref{tab:ADL_keywords}.

CutLang \cite{Unel_2021} is a runtime interpreter for ADL that executes analysis descriptions directly, without requiring prior compilation. Written in C++ and built on the ROOT data analysis framework \cite{BRUN199781}, CutLang parses ADL files and maps all analysis objects and variables onto predefined internal physics object representations. This design allows a single ADL file to be executed over different input data formats without any additional coding, making the ADL/CutLang workflow both flexible and portable. The source code of CutLang is publicly available in a dedicated GitHub repository~\cite{cutlang_source}.

\begin{table}[ht]
\caption{Blocks and their Purposes in ADL \cite{ADL_CERN_WEB}.} 
\label{tab:ADL_blocks}
\centering
\begin{tabular}{|l|c|}
\hline
Blocks & Purpose \\
\hline\hline
object & Object definitions \\
region & Event selections \\
info & Analysis or ADL information  \\
table & Parameter definitions \\
\hline
\end{tabular}
\end{table}

\begin{table}[!ht]
\caption{Main keywords and their functions in ADL. \cite{ADL_CERN_WEB}.} 
\label{tab:ADL_keywords}
\centering
\begin{tabular}{|l|c|}
\hline
Keyword & Purpose \\
\hline\hline
\multicolumn{2}{|c|}{Analysis description} \\
\hline
define & Event variable and constant definitions \\
select & Selecting objects or events \\
reject & Rejecting objects or events\\
take & Calling the baseline object\\
weight & Applying event weights \\
sort & Sorting analysis objects \\
bin & Defining bins boundaries for variables \\
\hline
\multicolumn{2}{|c|}{Auxiliary} \\
\hline
print & Printing event variables \\
histo & Defining histograms \\
save & Saving event variables \\
\hline
\end{tabular}
\end{table}

\section{Benchmark analyses}

Nine analyses from the ATLAS Open Data portal were selected for re-implementation in ADL \cite{ADL4LHC_ATLAS13TeV} and execution with CutLang. The selection covers a broad range of standard model (SM) processes as well as Beyond the standard model (BSM) searches, providing a comprehensive test of the ADL/CutLang workflow across diverse final state topologies. The chosen analyses are: 1) SM Z boson production in the two-lepton final state, 2) SM W boson production in the leptonic final state, 3) SM single top quark production in the single-lepton final state, 4) SM top quark pair production in the single-lepton final state, 5) SM WZ diboson production in the three-lepton final state, 6) SM Higgs boson production in the $H \rightarrow ZZ*$ decay channel, 7) SM ZZ diboson production, 8) a search for direct production of pairs of sleptons, and 9) a Beyond the SM $Z'$ search in the $Z' \rightarrow t\bar{t}$ final state. 
These analyses were selected to provide a balanced representation across different levels of complexity. The Z boson, W boson, and single-top analyses have relatively simple analysis workflows, while the top quark pair, SUSY and WZ analyses are considered to have intermediate complexity, as they require the definition of multiple derived event variables. Finally, the ZZ diboson, Higgs boson, and $Z'$ analyses are classified as more complex due to the need for mass reconstruction via minimization techniques or numerous operations on the angular properties of the analysis objects. Both the collider data and various SM simulations, tuned for the ATLAS detector, were obtained from the ATLAS Open Data portal \cite{Opendata_web}. 
Additionally, a list of the SM simulation samples used in this study is given in Appendix \ref{appendix_opendata_samples}.

\subsection{Object definitions and initialization} \label{section_objects}
The first stage of each analysis algorithm involves filtering, grouping, and combining physics objects such as electrons, muons, photons and jets according to specific criteria. Filtering refers to the process of selecting particles that satisfy certain physical requirements (such as transverse momentum thresholds) while excluding those that may be misidentified or poorly reconstructed.

Objects are filtered and grouped to meet the requirements of the analysis by applying a set of selection criteria within object selection blocks, which derive a target object collection from an input object collection. Each selection in an {\tt object} block is applied individually to each object in the input collection.

Reconstructed physics objects and their attributes in the 13 TeV ATLAS Open Data ROOT n-tuples \cite{ATL-OREACH-PUB-2020-001} are accessible via the main objects predefined in CutLang, along with the functions listed in the table provided in Appendix \ref{cutlang_funcs}. In addition, CutLang can directly access all variables in the ROOT n-tuples. This feature is especially useful when working with variables from the 13 TeV ATLAS Open Data format that are not predefined in CutLang.

The selections on reconstructed physics objects are based on standard requirements such as pseudorapidity and transverse momenta thresholds, as well as additional criteria such as tracking and energy isolation \cite{ATL-OREACH-PUB-2020-001}. The refined objects, common to all analyses considered in this study, are expressed in ADL syntax as follows: 

\begin{courier} \begin{lstlisting}
object goodEles # Target 
 take ELE       # Input
 select isTightID(ELE) == 1
 select Pt(ELE) > 25.0
 select AbsEta(ELE) < 2.47
 select AbsEta(ELE) ][ 1.37 1.52
 select Ptcone30(ELE) < 0.15
 select Etcone20(ELE) < 0.15

object goodMuos
 take MUO
 select lep_isTightID(MUO) == 1
 select Pt(MUO) > 25.0
 select AbsEta(MUO) < 2.5
 select Ptcone30(MUO) < 0.15
 select Etcone20(MUO) < 0.15

object goodJets
 take JET
 select Pt(JET) > 25.0 
 select jet_jvt(JET) > 0.59
 select AbsEta(JET) < 2.5

object goodPhos
 take PHO
 select photon_isTightID(PHO) == 1
 select Pt(PHO) > 25.0
 select AbsEta(JET) < 2.37
 select AbsEta(ELE) ][ 1.37 1.52
 select Ptcone30(PHO) < 0.065
 select Etcone20(PHO) < 0.065

object goodFJets
 select FatJet
 select Pt(FatJet) > 250.0
 select m(FatJet) > 50.0 
\end{lstlisting} \end{courier}

In the examples used in this study, refined objects are prefixed with {\tt good}, such as {\tt goodMuos} or {\tt goodJets}; however, this is not a requirement of the language. The {\tt goodEles} defined above, for instance, are the electron candidates that will subsequently be used to reconstruct more complex objects such as $Z$ bosons.

It is also worth noting a deliberate design choice in ADL: following extensive discussion on the distinction between functions and attributes, it was decided that all properties of physics objects should be accessible uniformly as functions. This simplifies the learning curve, particularly for newcomers. 
In the listing above, for example, {\tt Pt()} returns the transverse momentum of the object passed as its argument. Furthermore, the reader will notice that only object names appear as arguments to these functions, with no explicit index (or loop variable):  the iteration over the object collection is handled implicitly by CutLang, keeping the ADL description concise and free of procedural bookkeeping.

In many collider data analyses, electrons and muons are collectively treated as a single lepton object, combining {\tt goodEles} and {\tt goodMuos} into a unified {\tt goodLepts} collection that simplifies subsequent event selection criteria.
In ADL notation such a unified collection is written as follows:
\begin{courier} 
    \begin{lstlisting}
    object goodLepts : Union(goodEles, goodMuos)
    \end{lstlisting} 
\end{courier}

In many ADL implementations, after defining physics objects, event selection begins with a base region block, usually named {\tt preselections}. In ADL, calling a previously defined region within a new one allows the new block inherit its selections. This preselection block typically contains the baseline event cuts and sorting rules. For example, it checks whether the required objects are present and sorts a combined lepton collection by descending transverse momentum. By simply including {\tt preselections} in a specific analysis region, these baseline criteria are automatically applied. This block helps avoid code duplication and keeps the analysis description short and readable. For an analysis with leptonic final states, a typical initialization block can be written as follows: 
\begin{courier} 
    \begin{lstlisting}
region preselections
  select ALL
  select Size(goodEles) >= 0
  select Size(goodMuos) >= 0
  select Size(goodLepts) >= 0
  sort Pt(goodLepts) descend
    \end{lstlisting} 
\end{courier}

\subsection{Event weighting}

Simulated events require weighting based on the production cross section, integrated luminosity, and Monte Carlo (MC) weights in order to accurately represent the number of events expected in experimental data. Additionally, scale factors accounting for detector efficiencies and trigger performance must also be taken into account. In this study, a standard event weight definition has been used consistently across all analyses, ensuring a uniform treatment of simulated samples. An ADL code snippet illustrating this definition and its usage in a \texttt{region} block is shown below:

\begin{courier}
    \begin{lstlisting}
define Sfactor : scaleFactor_ELE*scaleFactor_MUON*scaleFactor_LepTRIGGER*
scaleFactor_PILEUP
define Lumi : 10064 
define totalWeight : XSection*mcWeight*Sfactor*Lumi/SumWeights

region Signal
  weight evtweight totalWeight
    \end{lstlisting}
\end{courier} 

The variables \texttt{XSection}, \texttt{mcWeight}, \texttt{SumWeights}, and all components used in the \texttt{Sfactor} definition correspond to branches stored directly in the ATLAS Open Data samples. Detailed descriptions of these variables can be found in Ref.~\cite{ATL-OREACH-PUB-2020-001}. CutLang  can read these branches and automatically apply the total event weight defined as {\tt totalWeight} to each simulated event, requiring no additional user intervention.


\section{Event selection in each benchmark analysis}

In this section, the event variables and event selection criteria implemented in each benchmark analysis are described in detail. The focus is on how analysis variables are defined, their roles in the event selection and classification, and the specific criteria applied at each stage of the analyses. The complete ADL file for each analysis is publicly available in a dedicated GitHub repository~\cite{ADL4LHC_ATLAS13TeV}.
The first example is presented in greater detail, including the explanation of various keywords and the command line for any Unix-like systems to assist the reader. The remaining examples are presented in a more physics-focused manner.

\subsection{Standard model Z boson production in the two lepton final state} \label{section_Zanalysis}
The analysis focuses on events in which the $Z$ boson decays into either an electron-positron or muon-antimuon pair. The selection criteria are based on early 13 TeV data from the ATLAS detector and involve standard object selection requirements together with tight lepton identification requirements.

The central  part of event selection consists of defining analysis variables and constructing event selection blocks. The event variable definitions for this analysis are given below:
\begin{courier} \begin{lstlisting}
define Lepton1  : goodLepts[0]
define Lepton2  : goodLepts[1]
define Lepton12 : Lepton1 Lepton2
define mLL : m(Lepton12)
define zMassWindow : abs(mLL - 91.18)
\end{lstlisting} \end{courier}

{\tt Lepton1} and {\tt Lepton2},  are defined as the first and second elements of the {\tt goodLepts} collection, respectively, sorted in descending order of transverse momentum ($p_T$) magnitude. These two leptons are then combined to form a $Z$ boson candidate, referred to as {\tt Lepton12}. This variable can also be defined directly as \texttt{Lepton12}: {\tt goodLepts[0] goodLepts[1]}; thus, intermediate aliases such as {\tt lepton1} and {\tt lepton2} are not required for computing event-level variables. The invariant mass of this lepton pair is calculated and stored in the variable {\tt mLL}, which represents reconstructed mass of the $Z$ boson candidate.

In all analyses considered in this study, the variables that could be used directly in the event selection block were not  declared separately, except when explicit declaration improves the readability.
For this reason, in the listing above the reconstructed $Z$ boson candidate mass and the width of the window around the known $Z$ boson  mass are each clearly defined.
Similarly, additional criteria for $Z$ boson event selection that can be specified and applied directly within the selection block were not defined externally, keeping the ADL descriptions concise and readable.
 
\begin{courier} \begin{lstlisting}
region ZBosonAnalysis_ll
  preselections
  select trigE == 1 OR trigM == 1
  select Size(goodLepts) == 2
  select Size(JET) == 0
  select q(Lepton1)*q(Lepton2) < 0
  select Abs(pdgID(Lepton1)) == Abs(pdgID(Lepton2))
  select zMassWindow < 25
\end{lstlisting} \end{courier}

In the given event selection block provided above, the selection starts with the previously discussed initialization section, as an example of a block being called by another block.
In the following lines, after making sure that the event was triggered by either electrons  {\tt trigE} or muons {\tt trigM},
the final criteria targeting Z boson decays into lepton pairs are applied. First,  the number of elements in the {\tt goodLepts} collection must be exactly 2, meaning the event must contain two well defined leptons. The {\tt JET} collection must not contain any jets, ensuring that hadronic activity is excluded. The expression containing the lepton charges ({\tt q(Lepton1)}) and {\tt q(Lepton2)}) must be negative, indicating that these two leptons have opposite charges. Additionally, {\tt Abs(pdgID(Lepton1))} and {\tt Abs(pdgID(Lepton2))} must be equal, meaning both leptons are of the same type (either electrons or muons). Finally, the invariant mass of the lepton pair, {\tt mLL}, must satisfy {\tt zMassWindow} $<$ 25 GeV—corresponding to the range 66 to 116 GeV, close to the mass of the Z boson. These criteria are designed to isolate events that exclusively involve Z boson decays.

\begin{courier} \begin{lstlisting}
region ZBosonAnalysis_ee
  ZBosonAnalysis_ll
  select Abs(pdgID(Lepton1)) == 11
  
region ZBosonAnalysis_mm
  ZBosonAnalysis_ll
  select Abs(pdgID(Lepton1)) == 13
\end{lstlisting} \end{courier}

There are also two additional decay channel–specific event selection blocks, in which the Z boson decays exclusively into lepton pairs. The {\tt ZBosonAnalysis\_ee} region is defined by the condition {\tt Abs(pdgID(leadLept)) == 11}, corresponding to Z boson decays into electron–positron pairs, while the {\tt ZBosonAnalysis\_mm} region is defined by {\tt Abs(pdgID(leadLept)) == 13}, corresponding to muon–antimuon pairs.
Such an analysis can be executed with CutLang run-time interpreter using a simple command-line expression in the following format:\\ 

\hspace{1cm} \texttt{CLA} \texttt{data\_A.2lep.root} \texttt{ATLASODR2} \texttt{-i} \texttt{ZBosonAnalysis.adl}\\ 

\noindent where \texttt{CLA} is a script that wraps the CutLang executable; the second and third arguments represent the input ROOT file name and the input file format, respectively, and \texttt{-i} \texttt{ZBosonAnalysis.adl} argument specifies the input ADL file for the analysis.

\begin{figure}[!ht]
\centering
\includegraphics[width=0.9\textwidth]{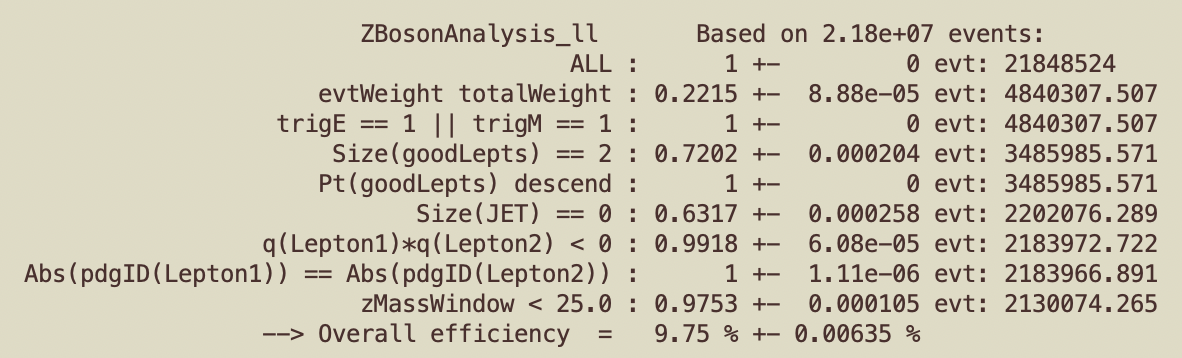}
\caption{Example of CutLang output showing the event-count cutflow, including statistical uncertainties, for the $Z \rightarrow \ell \ell$ region using a $Z+jets$ simulated sample.}
\label{fig:cutflowSS_Z}
\end{figure}

After the execution, CutLang produces the analysis output as a \texttt{.root} file containing cutflows and histograms for each region organized a \texttt{TDirectory} structure. In addition, CutLang prints the event counts cutflow for each region, along with their statistical uncertainties in the terminal. A screenshot of an example output is shown in Figure \ref{fig:cutflowSS_Z}.

\begin{figure}[!ht]
\centering
\includegraphics[width=0.55\textwidth]{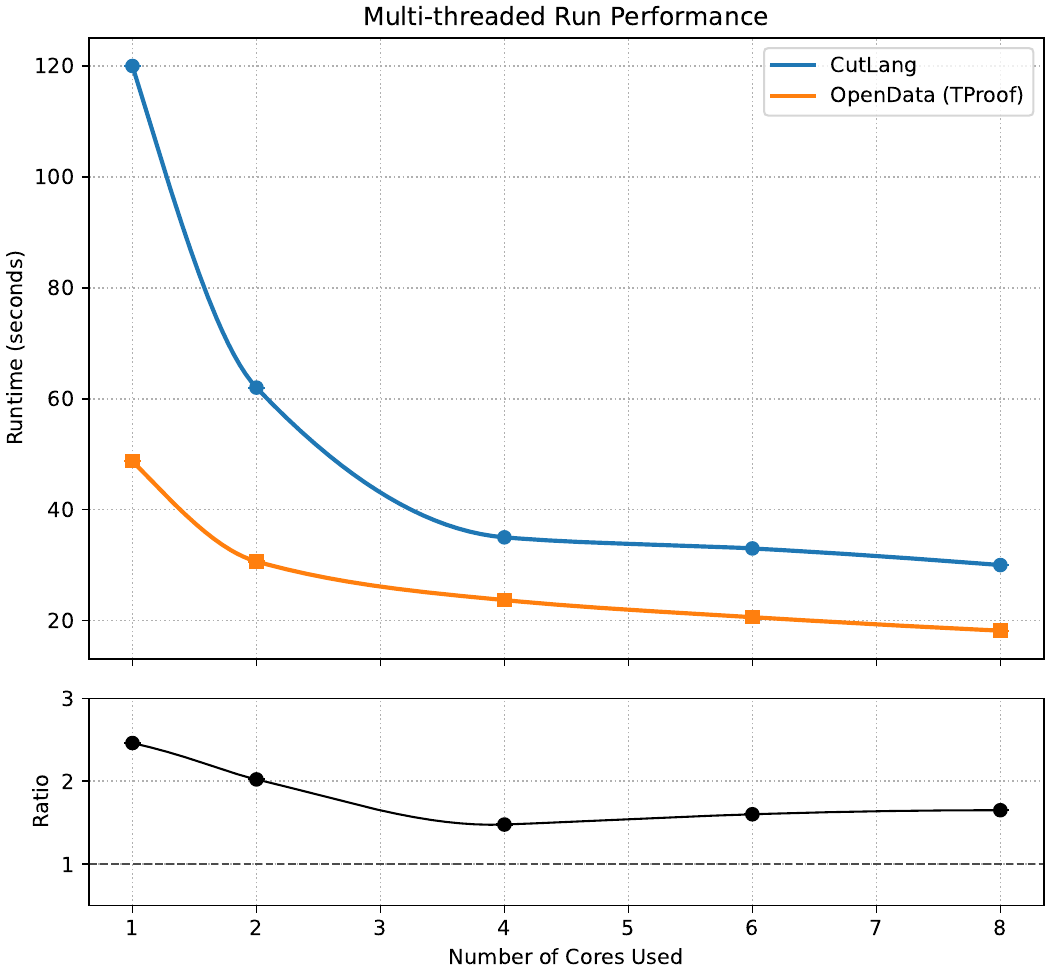}
\caption{Comparison of the run times obtained using the ADL/CutLang and Open Data frameworks across different numbers of threads.}
\label{fig:multithread}
\end{figure}

CutLang is also able to benefit from the multi-core architecture of modern computers and execute analyses in parallel. Since these analyses typically involve large amounts of collected data and simulation events, the total runtime is a critical factor. To evaluate CutLang's multi-threaded performance, a series of analysis runs were executed using different numbers of cores. For this test, 10 million events from  a $Z\rightarrow e^+e^-$ simulation sample were selected, and the run time from the Open Data framework, which uses ROOT's TPROOF module for parallelization, was used as a reference. The computer used for this test has an M1 silicon processor and runs macOS Sequoia 15.7.3. Figure \ref{fig:multithread} shows the multi-threaded runtimes obtained from both the ADL/CutLang and Open Data frameworks. 

\begin{figure}[!ht]
\centering
\includegraphics[width=0.40\textwidth]{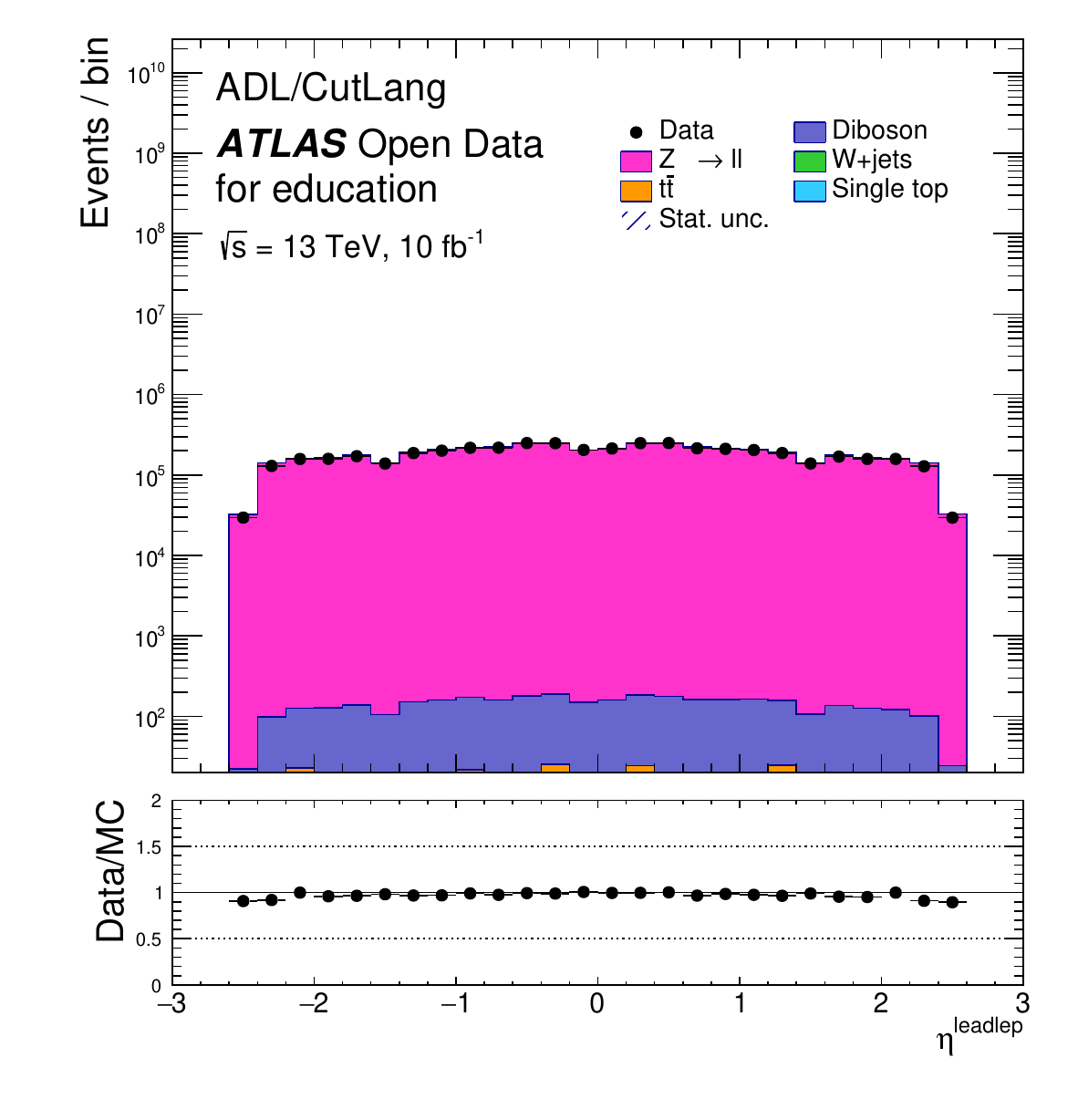}
\includegraphics[width=0.40\textwidth]{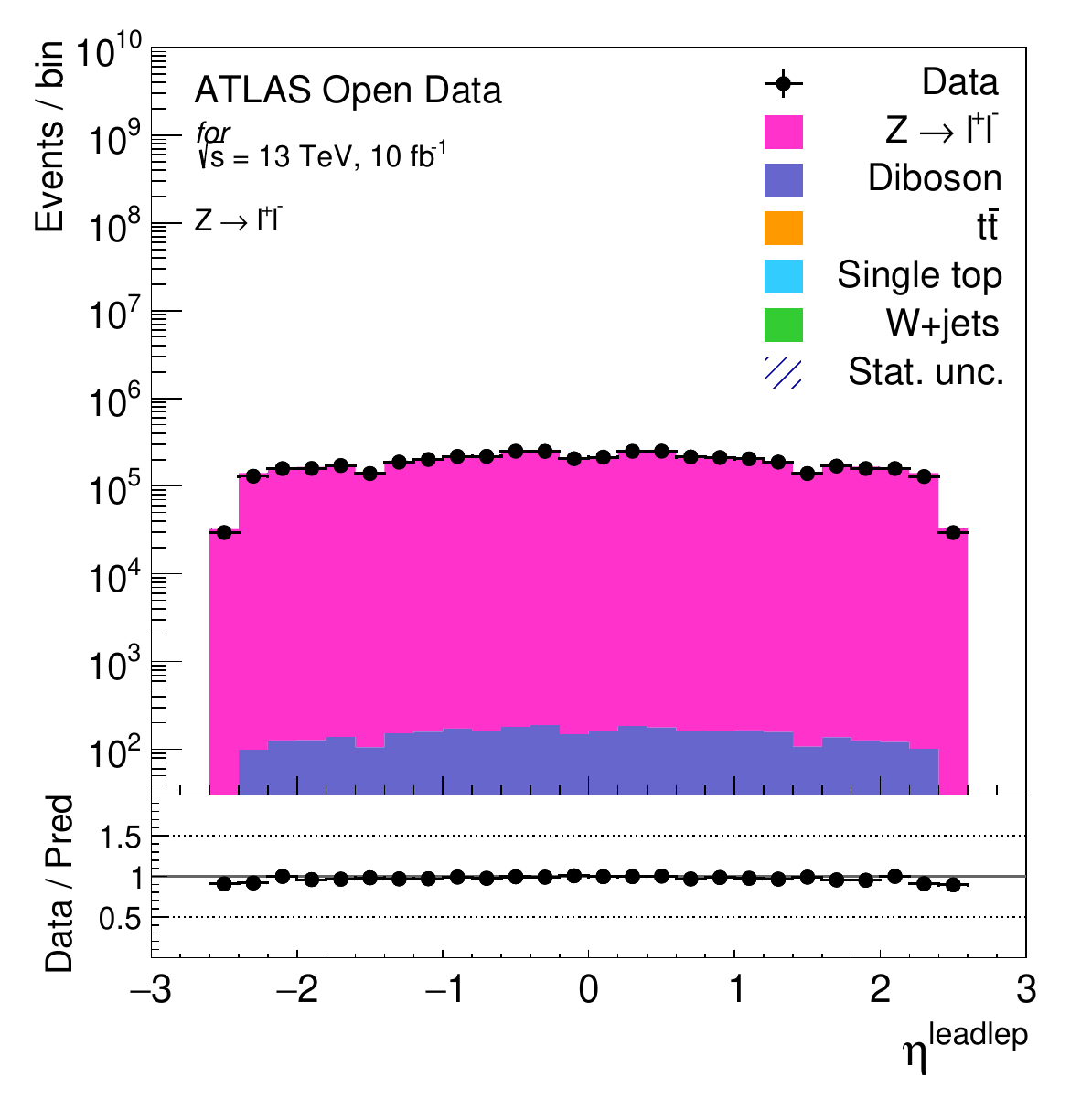}
\includegraphics[width=0.40\textwidth]{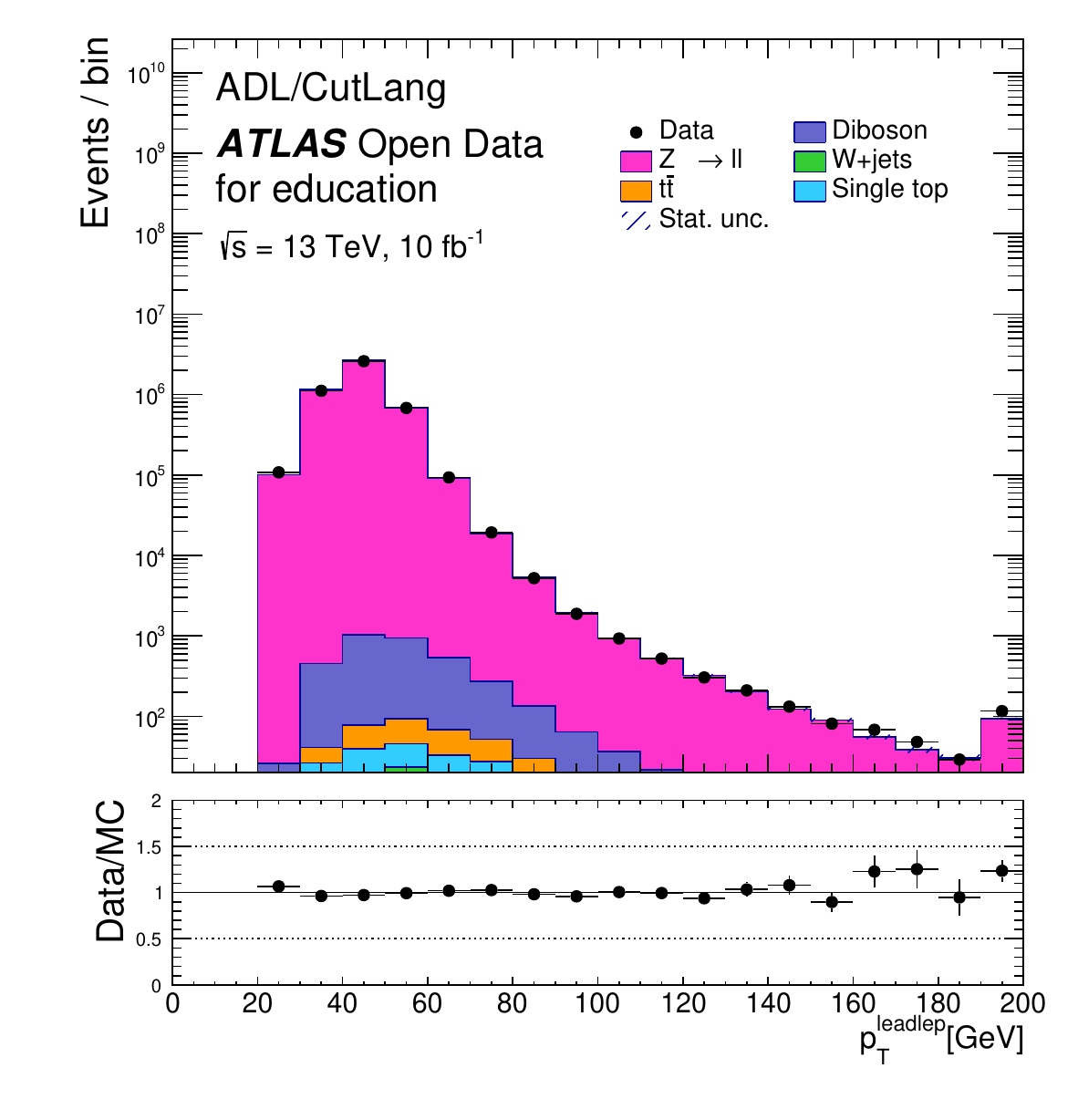}
\includegraphics[width=0.40\textwidth]{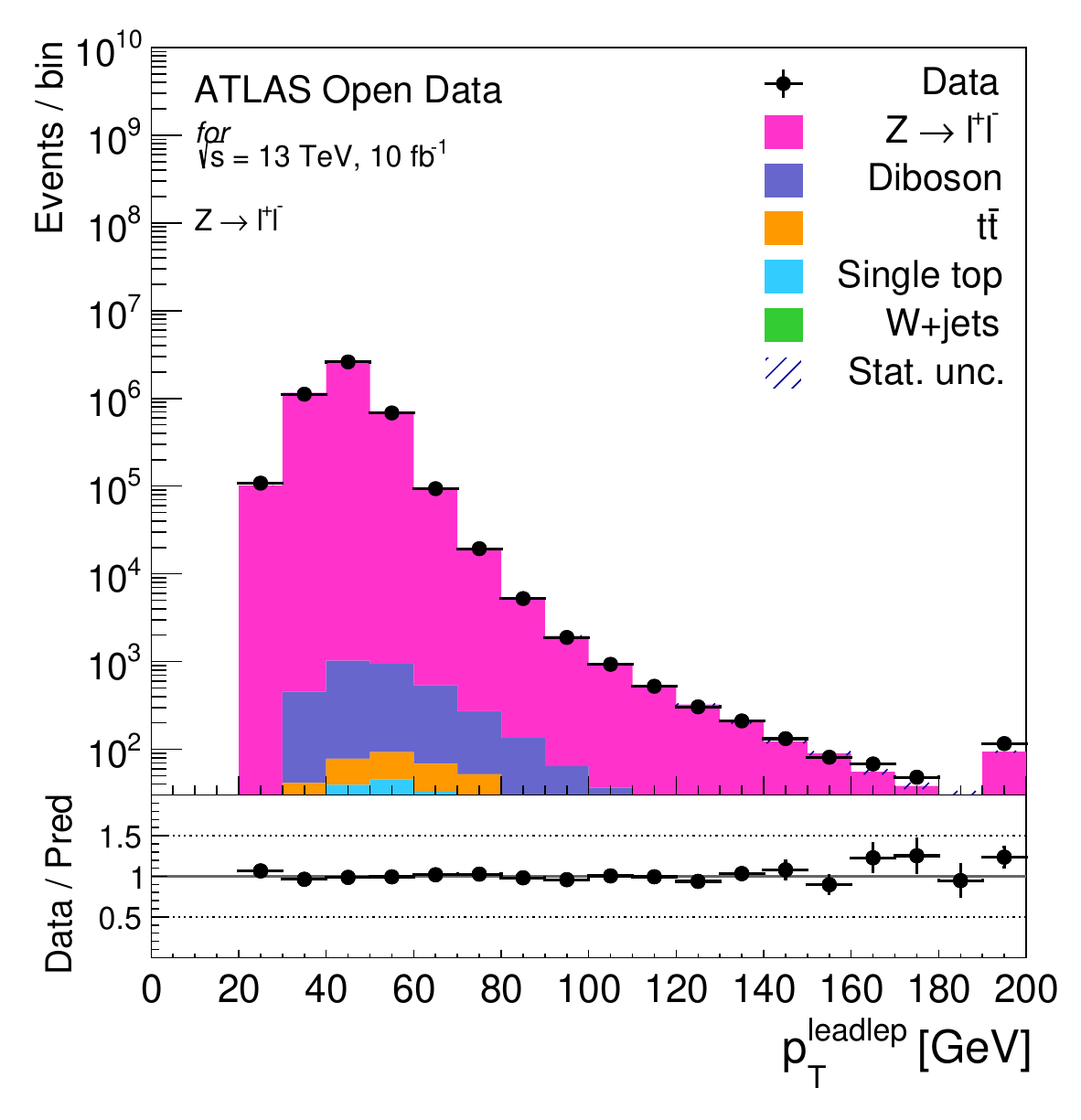}
\caption{Comparison of the leading lepton $\eta$ (top) and $p_T$ (bottom) distributions obtained with the ADL/CutLang (left) and Open Data (right) frameworks for $Z \rightarrow \ell\ell$ selections. The lower pad in each plot represents the Data/MC ratio. The last bin in each plot includes the overflow.}
\label{fig:CL_vs_OD_kinematics}
\end{figure}

\begin{figure}[!ht]
\centering
\includegraphics[width=0.40\textwidth]{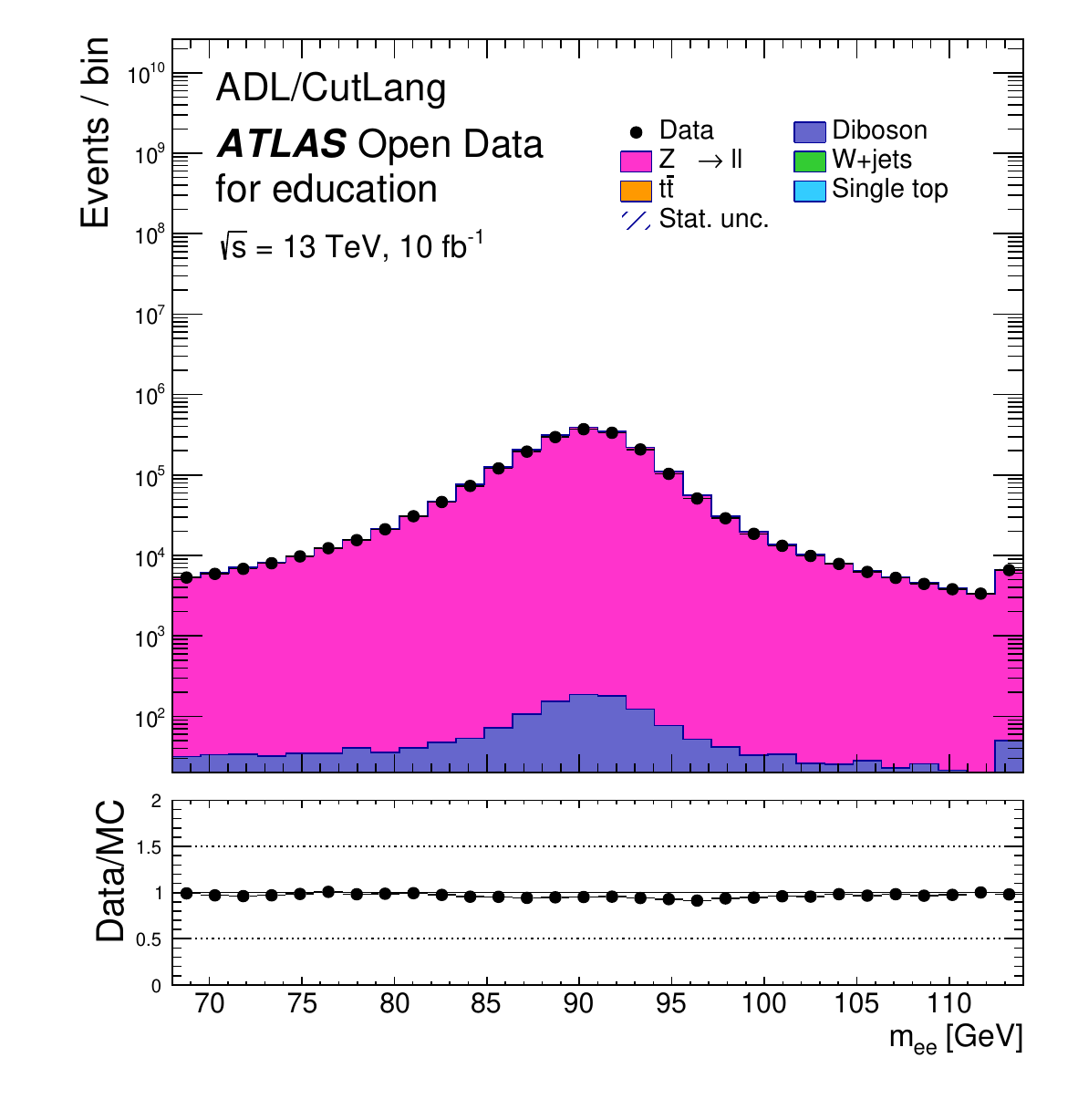}
\includegraphics[width=0.40\textwidth]{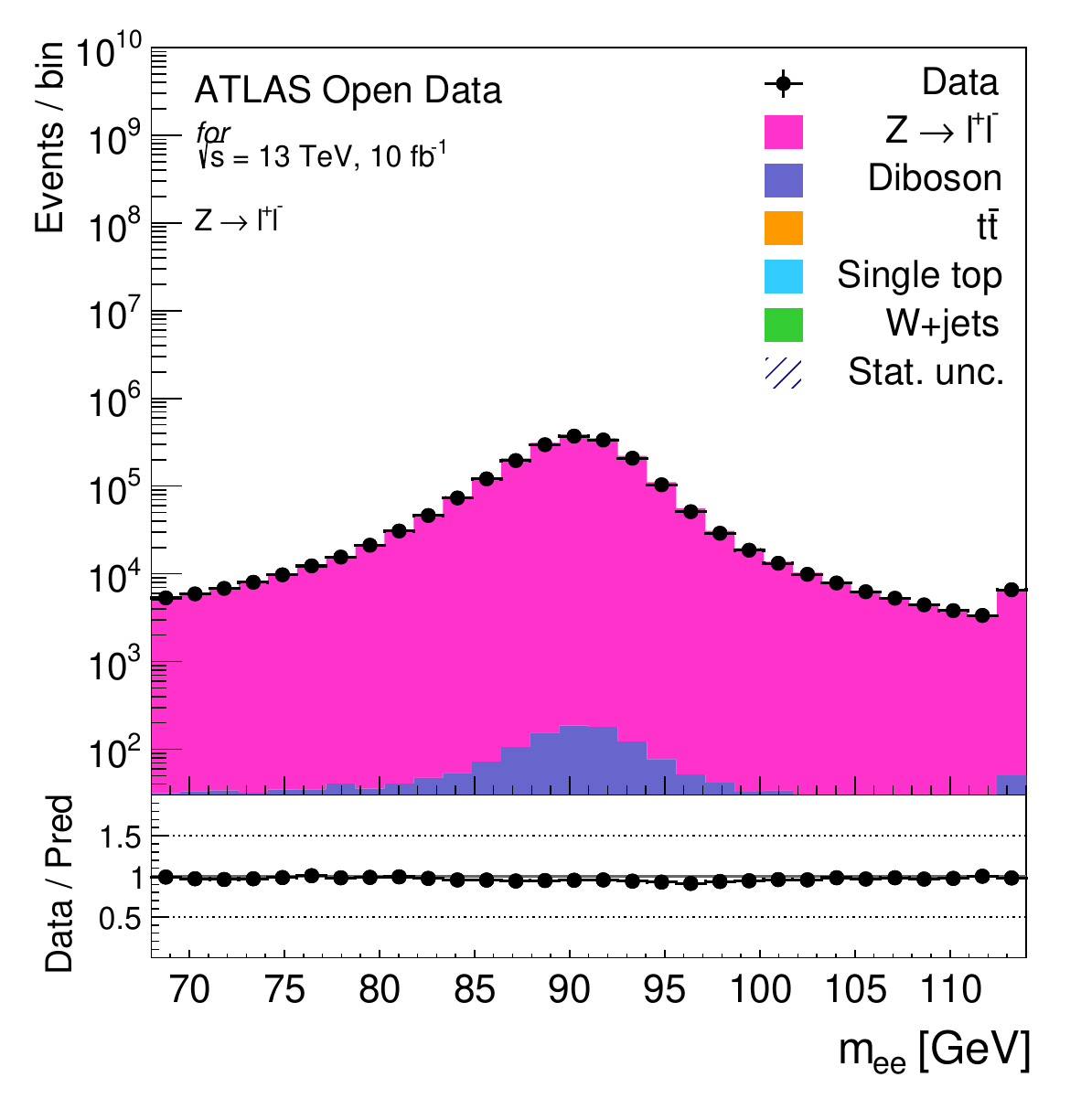}
\includegraphics[width=0.40\textwidth]{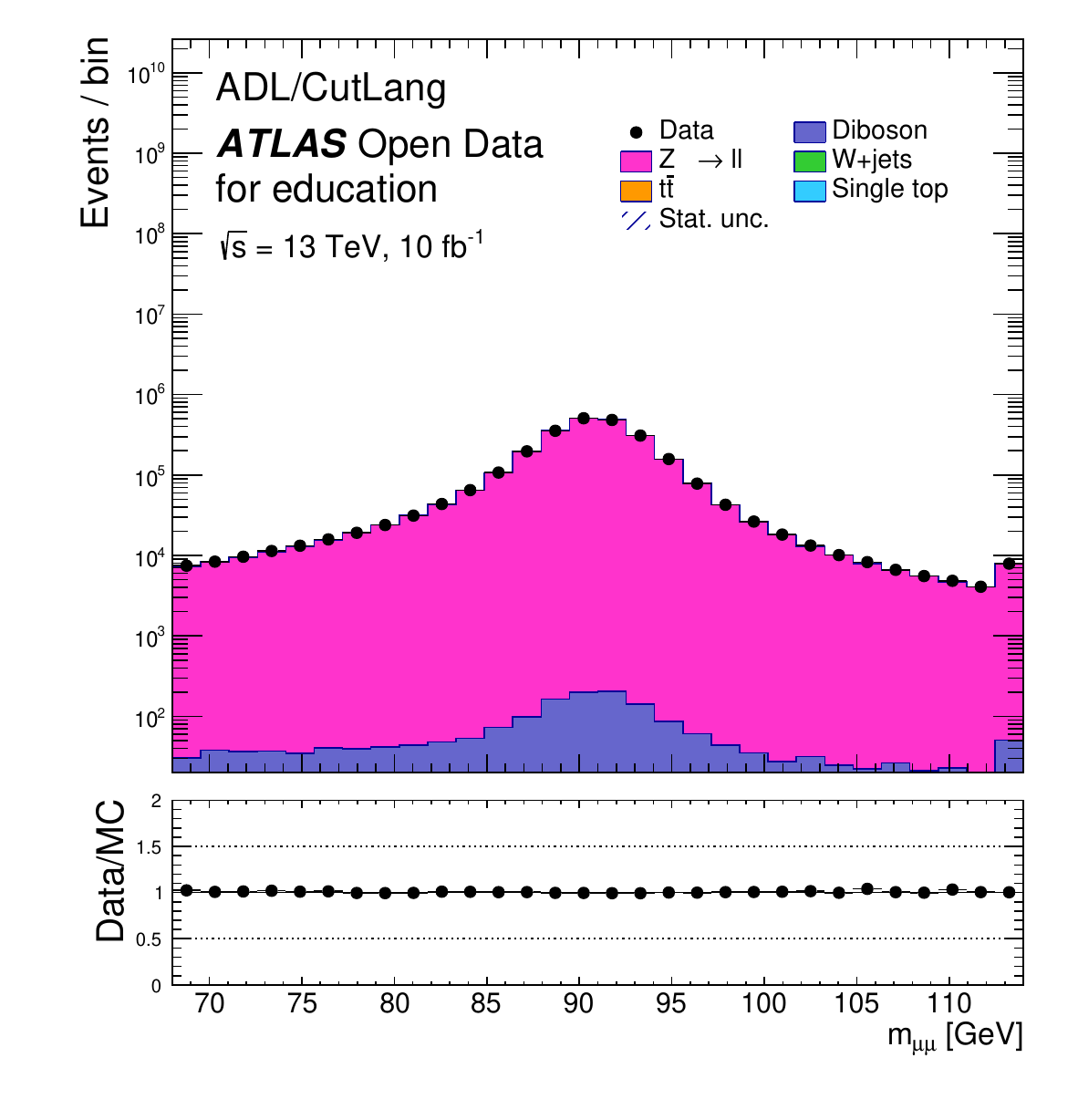}
\includegraphics[width=0.40\textwidth]{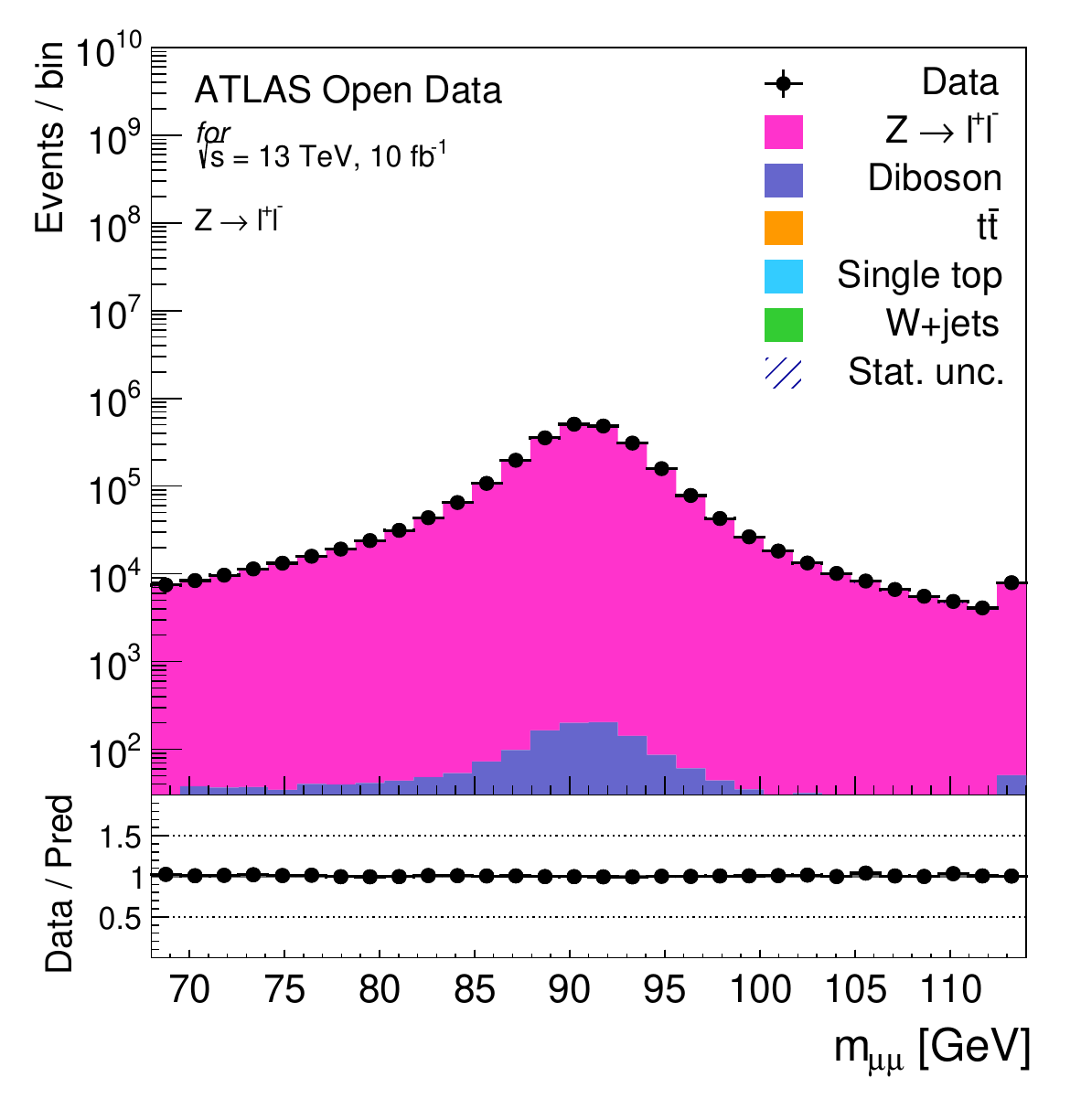}
\includegraphics[width=0.40\textwidth]{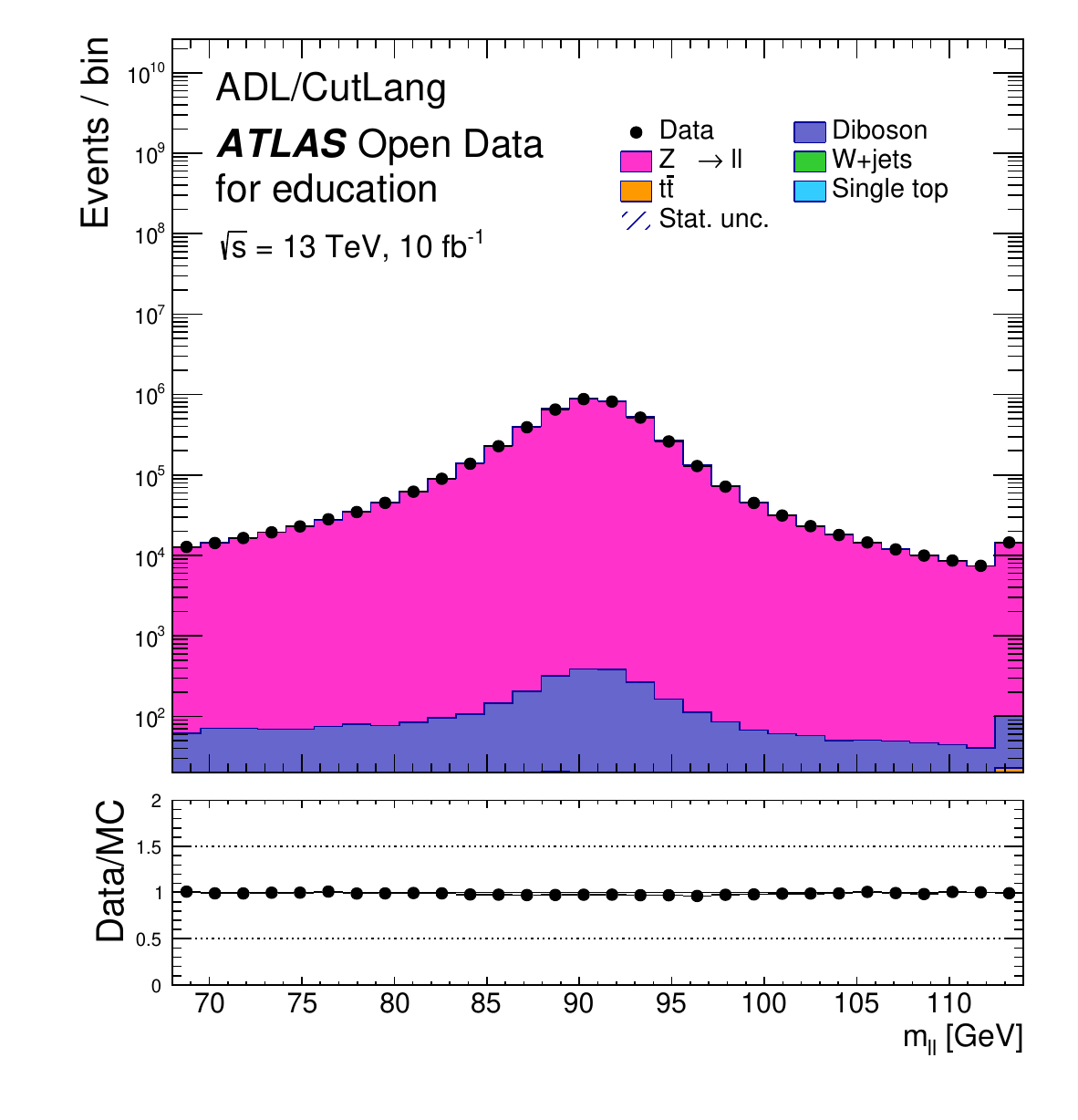}
\includegraphics[width=0.40\textwidth]{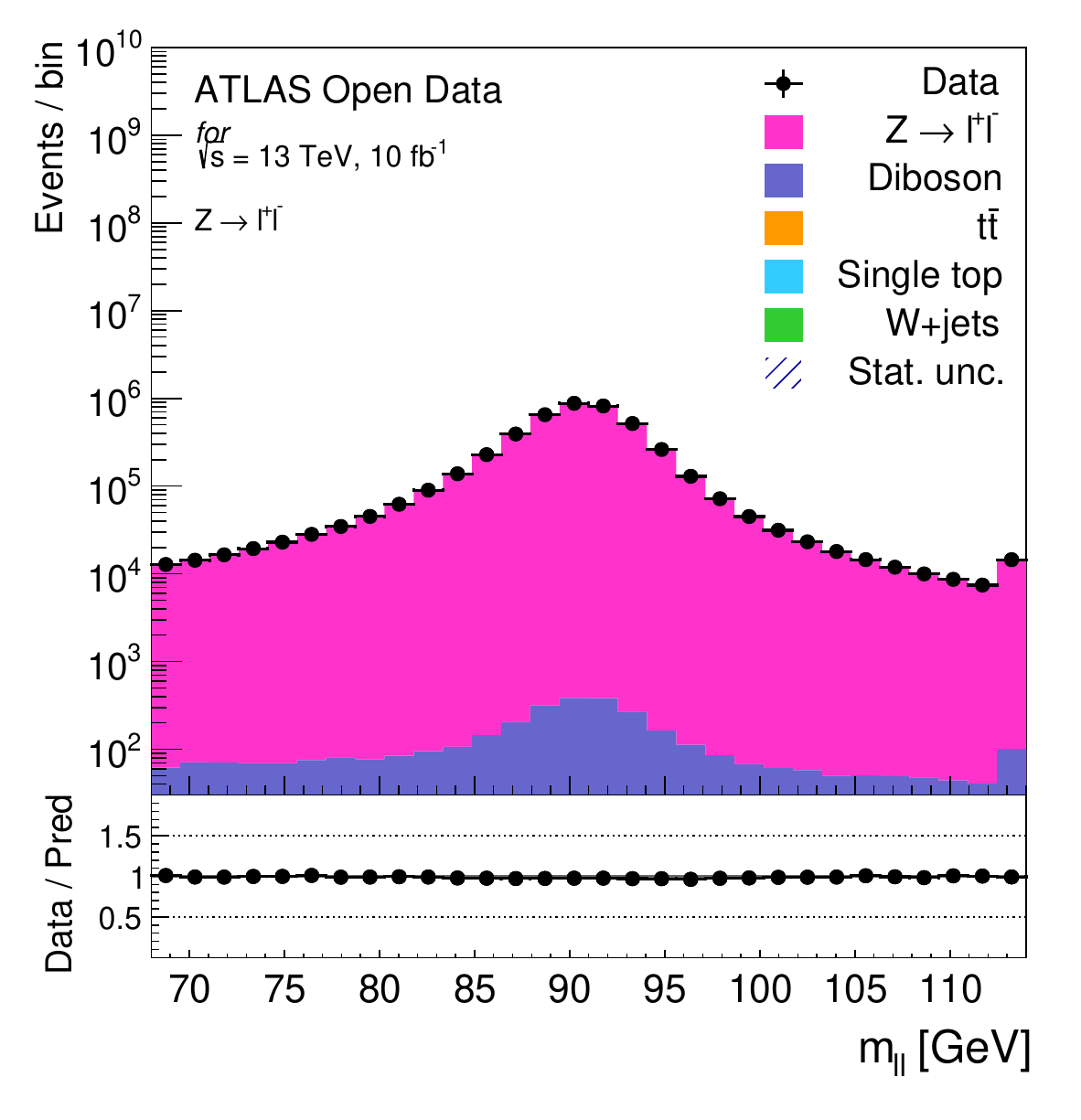}
\caption{Comparison of the Z candidate mass distributions in the $Z \rightarrow \ell \ell$ (top) and $Z \rightarrow \mu \mu$ (right) channels obtained with the ADL/CutLang (left) and Open Data (right) frameworks. The lower pad in each plot represents the Data/MC ratio. The last bin in each plot includes the overflow.}
\label{fig:CL_vs_OD_Z_masses}
\end{figure}

Figure \ref{fig:CL_vs_OD_kinematics} shows the pseudorapidity ($\eta$) and $p_T$ distributions for the leading lepton in the $Z \rightarrow \ell\ell$ channel obtained from ADL/CutLang and Open Data frameworks. 
The overall shape, normalization, and data-to-MC event ratio are consistent across the two frameworks.
Figure \ref{fig:CL_vs_OD_Z_masses} presents the invariant mass of the reconstructed Z candidates in the $Z \rightarrow \ell \ell$ and $Z \rightarrow ee$ and $Z \rightarrow \mu\mu$ channels, demonstrating once more consistent results between the two frameworks. Additionally, the cutflow of event counts obtained with CutLang for the $Z \rightarrow \ell \ell$ channel is presented in Table \ref{tab:cutflow_Z}. Table \ref{tab:yields_Zyields} shows the event yields for each sample in the $Z \rightarrow \ell \ell$ channel from both ADL/CutLang and Open Data frameworks. For a fair comparison of the event yields, the integral of the $m_{Z \rightarrow \ell \ell}$ distribution was used to determine the yields for each sample. This demonstrates that it is possible to reproduce the same analysis logic with ADL/CutLang and obtain consistent results.

\begin{table}[!ht]
\begin{center}
\caption{Cutflow of event counts for each sample in the $Z \rightarrow \ell\ell$ channel.}
\resizebox{\textwidth}{!}{
\begin{tabular}{|l|c|c|c|c|c|c|}
\hline
Selection & Data & Diboson & $Z+jets$ & $W+jets$ & singletop & $t\bar{t}$ \\ \hline 
ALL & $12205790.00 \pm 3493.68$ & $14254335.00 \pm 3775.49$ & $44153520.00 \pm 6644.81$ & $142069.00 \pm 376.92$ & $423766.00 \pm 650.97$ & $2910539.00 \pm 1706.03$ \\
evtweight totalWeight & $12205790.00 \pm 3493.68$ & $50069.44 \pm 41.37$ & $10131636.20 \pm 1692.17$ & $330348.19 \pm 1069.50$ & $27442.09 \pm 46.73$ & $247914.33 \pm 156.59$ \\
trigE $==$ 1 OR trigM $==$ 1 & $12205790.00 \pm 3493.68$ & $50069.44 \pm 41.37$ & $10131636.20 \pm 1692.17$ & $330348.19 \pm 1069.50$ & $27442.09 \pm 46.73$ & $247914.33 \pm 156.59$ \\
Size(goodLepts) $==$ 2 & $7722052.00 \pm 2778.86$ & $28198.82 \pm 31.54$ & $7613297.59 \pm 1457.81$ & $1426.72 \pm 71.55$ & $11675.27 \pm 30.43$ & $120731.99 \pm 109.61$ \\
Pt(goodLepts) descend & $7722052.00 \pm 2778.86$ & $28198.82 \pm 31.54$ & $7613297.59 \pm 1457.81$ & $1426.72 \pm 71.55$ & $11675.27 \pm 30.43$ & $120731.99 \pm 109.61$ \\
Size(JET) == 0 & $4795074.00 \pm 2189.77$ & $8678.57 \pm 22.43$ & $4871553.27 \pm 1166.34$ & $666.96 \pm 48.69$ & $610.43 \pm 6.95$ & $1271.77 \pm 11.23$ \\
q(Lepton1)*q(Lepton2) $<$ 0 & $4780628.00 \pm 2186.46$ & $8286.75 \pm 17.36$ & $4853306.86 \pm 1164.26$ & $509.28 \pm 42.63$ & $602.94 \pm 6.90$ & $1255.02 \pm 11.15$ \\
Abs(pdgID(Lepton1)) $==$ Abs(pdgID(Lepton2)) & $4769375.00 \pm 2183.89$ & $4982.66 \pm 12.78$ & $4848133.49 \pm 1162.69$ & $211.55 \pm 25.88$ & $280.66 \pm 4.69$ & $576.36 \pm 7.53$ \\
zMassWindow $<$ 25 & $4626253.00 \pm 2150.87$ & $3166.77 \pm 9.60$ & $4731427.29 \pm 1148.20$ & $105.43 \pm 18.08$ & $94.59 \pm 2.72$ & $191.54 \pm 4.33$ \\ \hline
\end{tabular}
}
\label{tab:cutflow_Z}
\end{center}
\end{table}

\begin{table}[!ht]
\begin{center}
\caption{Event yields after all selections for each sample in the $Z \rightarrow \ell\ell$ channel.}
\begin{tabular}{|l|c|c|}
\hline
Sample & ADL/CutLang & Open Data Framework \\ \hline \hline
Data & $4613335.00 \pm 2147.87 $ & $4613335.00$ \\
Diboson & $3111.68 \pm 9.43 $ & $3111.68$ \\
$Z+jets$ & $4718396.13 \pm 1145.58 $ & $4717887.39$  \\
$W+jets$ & $101.03 \pm  17.73$ & $101.03$ \\
$t\bar{t}$ & $184.56 \pm 4.18 $ & $184.56$ \\ 
Single Top & $90.98 \pm 2.61$ & $90.98$ \\
\hline
\end{tabular}
\label{tab:yields_Zyields}
\end{center}
\end{table}

\clearpage

\subsection{Standard model W boson production in the leptonic final states} \label{section_Wanalysis}
This analysis focuses on the study of W boson decays in the leptonic channel. In the decay process, a lepton (electron or muon) and a neutrino (electron neutrino or muon neutrino) are produced. 
While leptons can be detected using calorimeters and tracking detectors, the neutrino cannot be observed directly. Instead, the presence of a neutrino is inferred experimentally through the missing transverse momentum, $p_T^{miss}$, which serves as an indirect measurement. 
The existence of neutrino makes this analysis slightly more complicated than the previous one.
Therefore, to reliably identify W boson events, the transverse mass, $m_T$, is used as a characteristic observable of the final state. The transverse mass is calculated using the transverse momentum of the charged lepton and the missing transverse momentum, based on their energy and angular properties. 
\paragraph{}
The transverse mass of the W boson, $M_T^W$, is computed as follows:

\begin{equation}
M_T^W = \sqrt{2 p_T^\ell p_T^{miss} (1 - \cos \Delta \phi)} \quad
\end{equation}

\noindent where $\Delta \phi$ represents the azimuthal angle between the charged lepton and the missing transverse momentum vector. After the same initialization block discussed in the previous analysis, the
event variables are defined using a structured approach accurately represent key physical quantities relevant to the analysis:
\begin{courier} \begin{lstlisting}
define leadLept : goodLepts[0]
define MTW : sqrt(2*Pt(leadLept)*MET*(1 - cos(dPhi(leadLept, METLV[0]))))
\end{lstlisting} \end{courier}

The first line defines the leading lepton referring to the single highest-$p_T$ lepton in the event. This lepton satisfies all selection criteria, including transverse momentum ($p_T >$ 35 GeV), tight identification, and isolation requirements. The $M_T^W$ formula for the $W$ boson is defined in the second line as a direct implementation of the previous definition. Both ADL and CutLang allow for the definition of complex formulas within the code using mathematical operations such as addition, multiplication, square roots, and trigonometric functions.

The event selection block {\tt WBosonAnalysis\_lnu} is implemented for the following conditions: single-electron (\texttt{trigE}) or single-muon (\texttt{trigM}) trigger must be fired; the $p_T^{miss}$ must exceed 30 GeV; exactly one lepton passing the selection criteria is required; and the $M_T^W$ must be greater than 60 GeV. The relevant block is given below:

\begin{courier} \begin{lstlisting}
region WBosonAnalysis_lnu
  preselections
  select trigE == 1 OR trigM == 1
  select Size(goodLepts) == 1
  select MET > 30.0  
  select MTW > 60.0
  bin "Wenu" Abs(pdgID(leadLept)) == 11
  bin "Wmunu" Abs(pdgID(leadLept)) == 13
\end{lstlisting} \end{courier}

\noindent Finally, events are categorized into two distinct electron and muon regions using the following blocks: 
\begin{courier} \begin{lstlisting}
region WBosonAnalysis_enu
  WBosonAnalysis_lnu
  select Abs(pdgID(leadLept)) == 11

region WBosonAnalysis_mnu
  WBosonAnalysis_lnu
  select Abs(pdgID(leadLept)) == 13
\end{lstlisting} \end{courier}

The {\tt WBosonAnalysis\_enu} region corresponds to the case where the W boson decays into an electron and an electron-neutrino, and the leading lepton is an electron, while the {\tt WBosonAnalysis\_mnu} region corresponds to the case where the W boson decays into a muon and a muon-neutrino, and the leading lepton is a muon. A region can also be partitioned into disjoint selections using the \texttt{bin} keyword, allowing the description of the analysis and the counting of events for multiple selections within a single \texttt{region} block. This approach is computationally efficient, as it removes the need to define many separate regions. {\tt WBosonAnalysis\_lnu} also illustrates how the \texttt{bin} keyword can be used to partition events into electron and muon channels within a same \texttt{region} block, by defining separate bins according to the lepton pdgID.

\begin{figure}[!ht]
\centering
\includegraphics[width=0.40\textwidth]{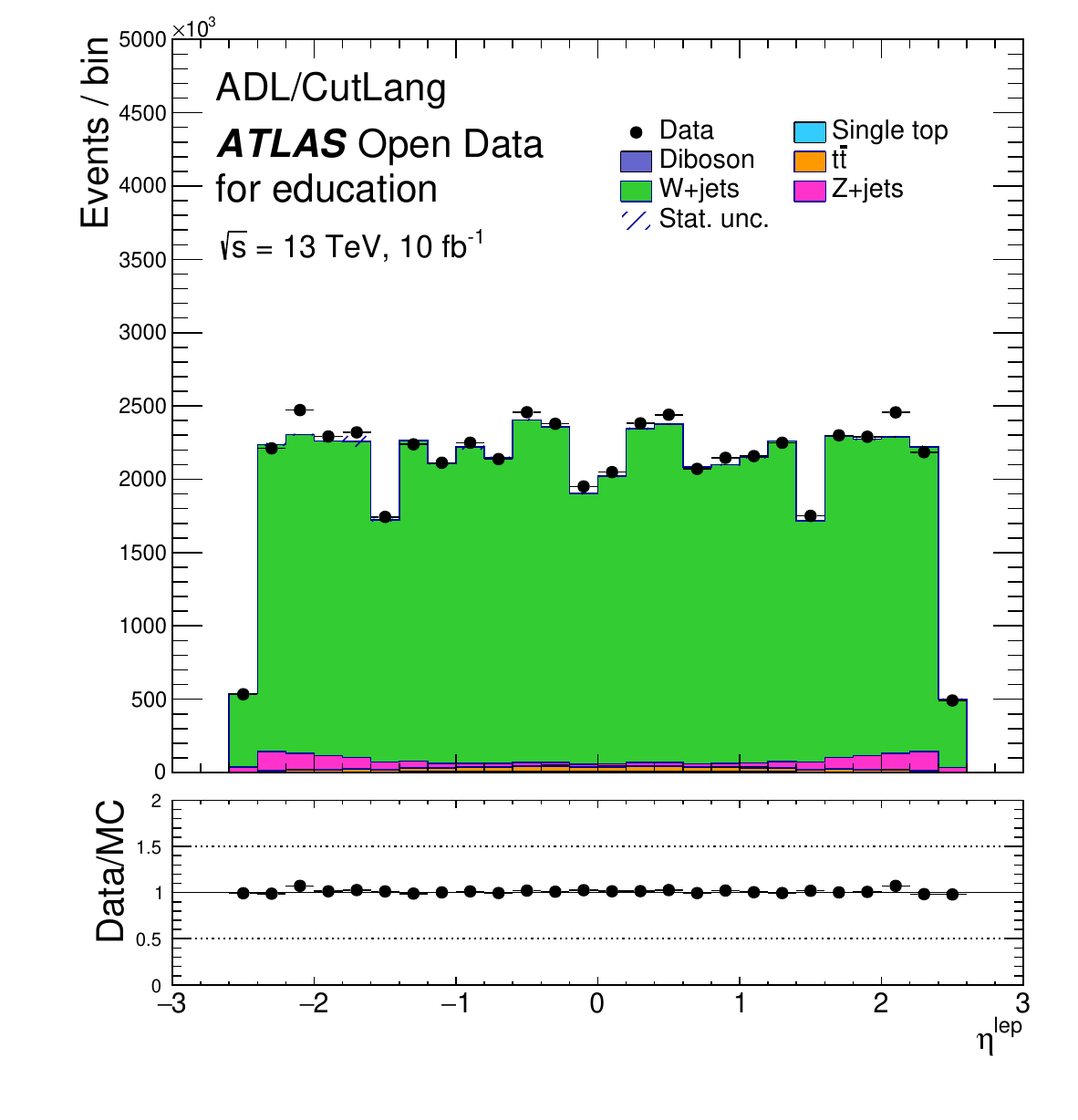}
\includegraphics[width=0.40\textwidth]{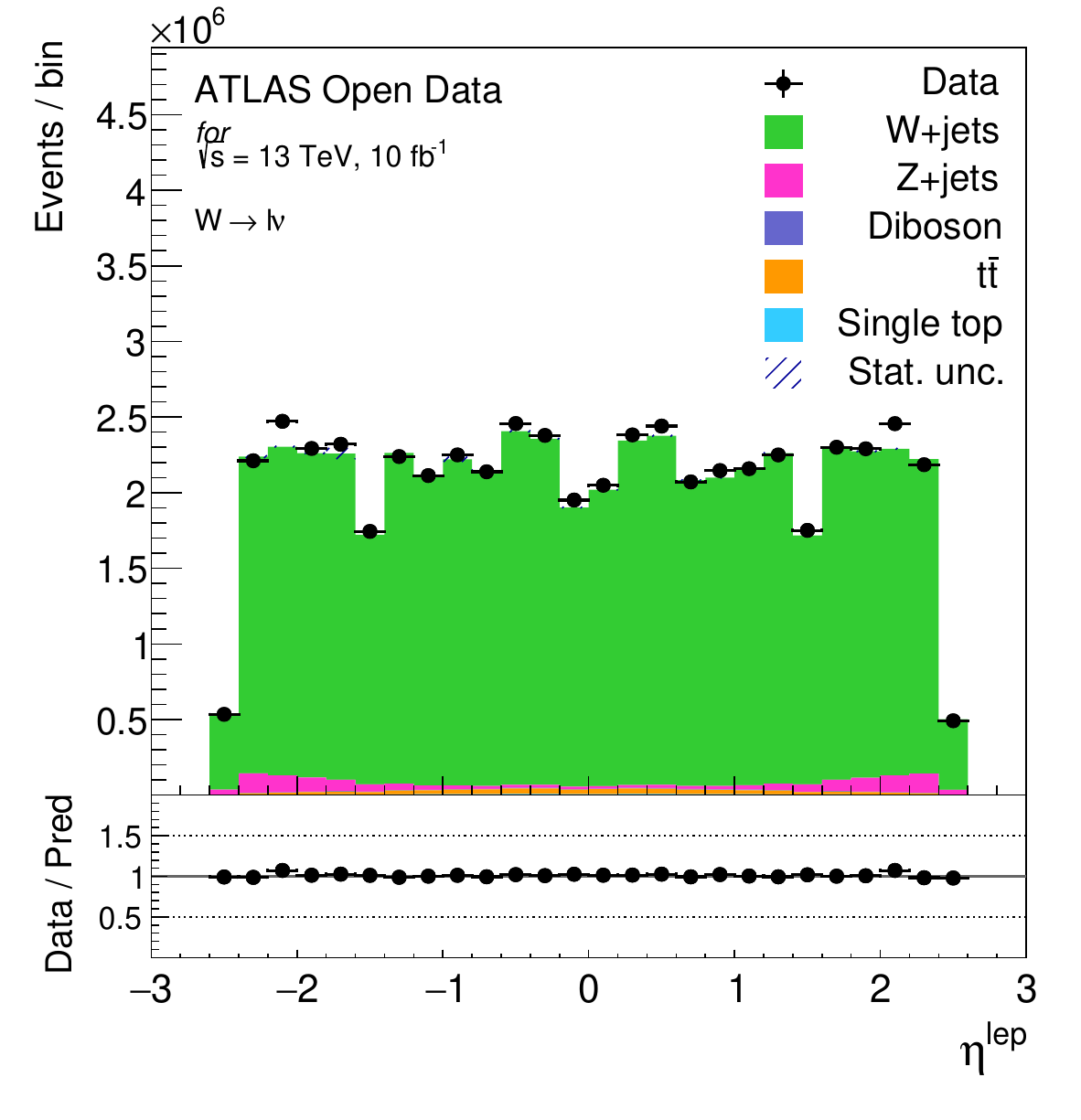}
\includegraphics[width=0.40\textwidth]{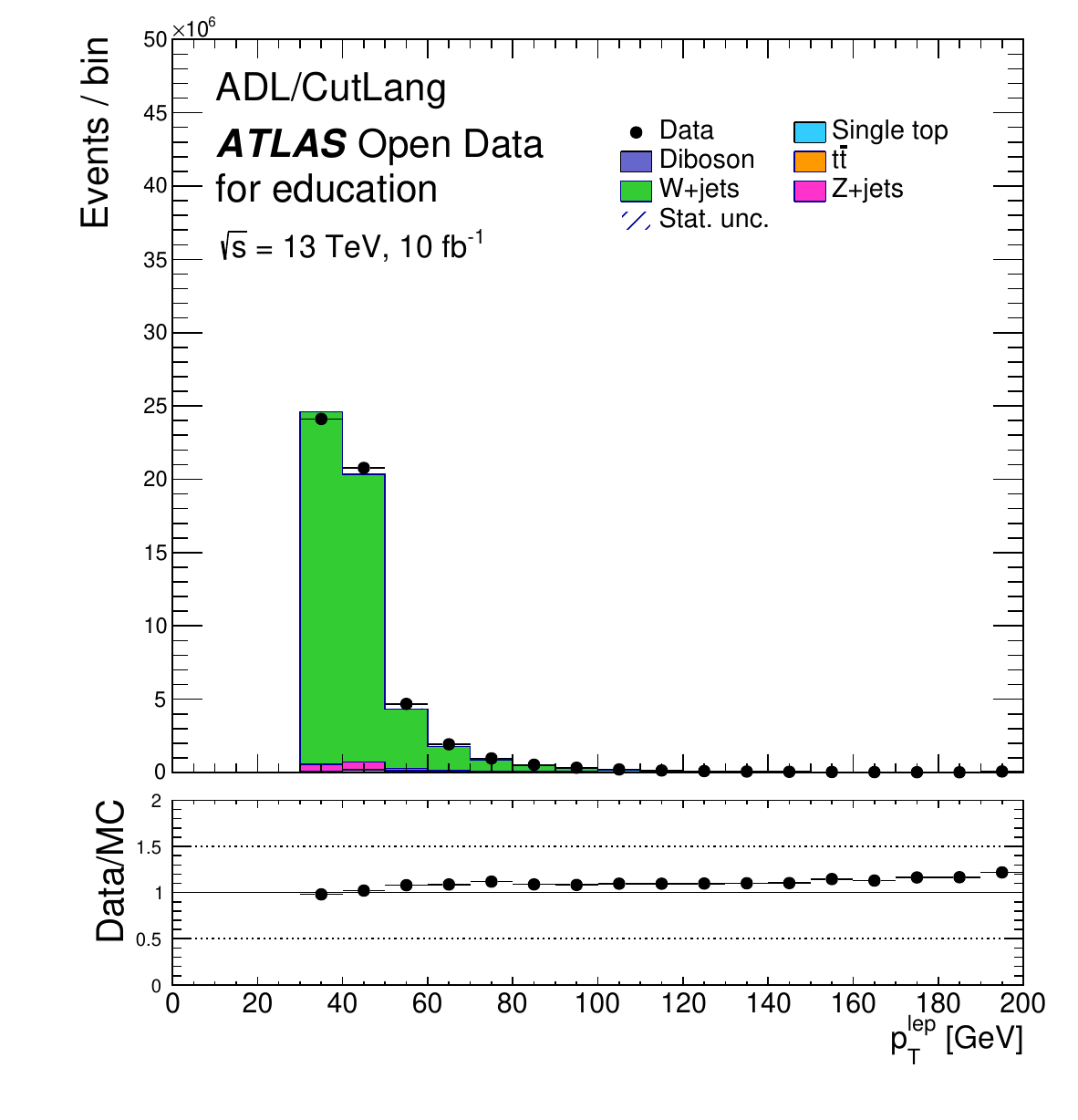}
\includegraphics[width=0.40\textwidth]{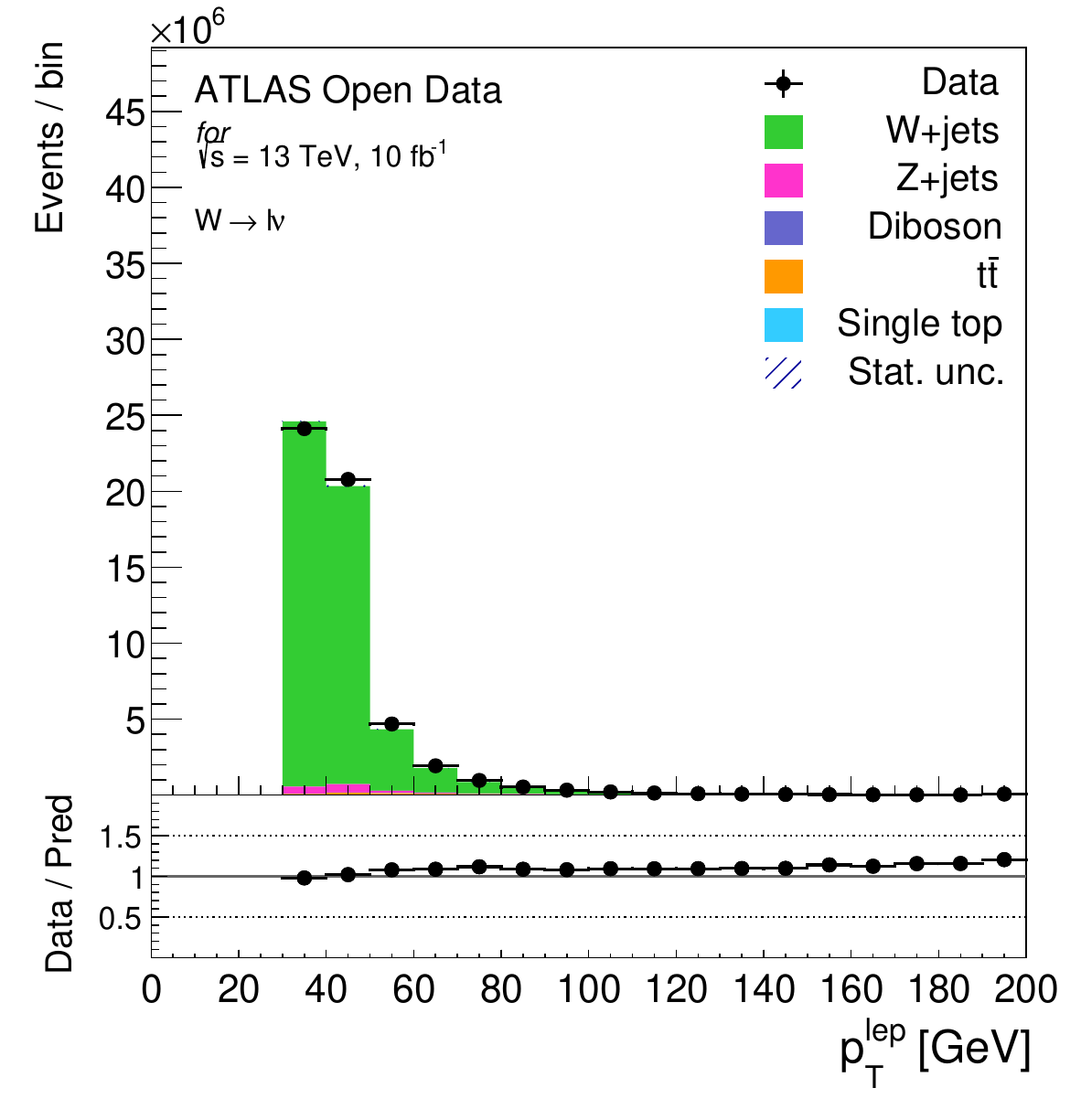}
\caption{Comparison of the lepton $\eta$ and $p_T$ distributions obtained with the ADL/CutLang (left) and Open Data (right) frameworks after the $W \rightarrow \ell\nu$ selections. The lower pad in each plot represents the Data/MC ratio. The last bin in each plot includes the overflow.}
\label{fig:CL_vs_OD_W_kinematics}
\end{figure}

\begin{figure}[!ht]
\centering
\includegraphics[width=0.40\textwidth]{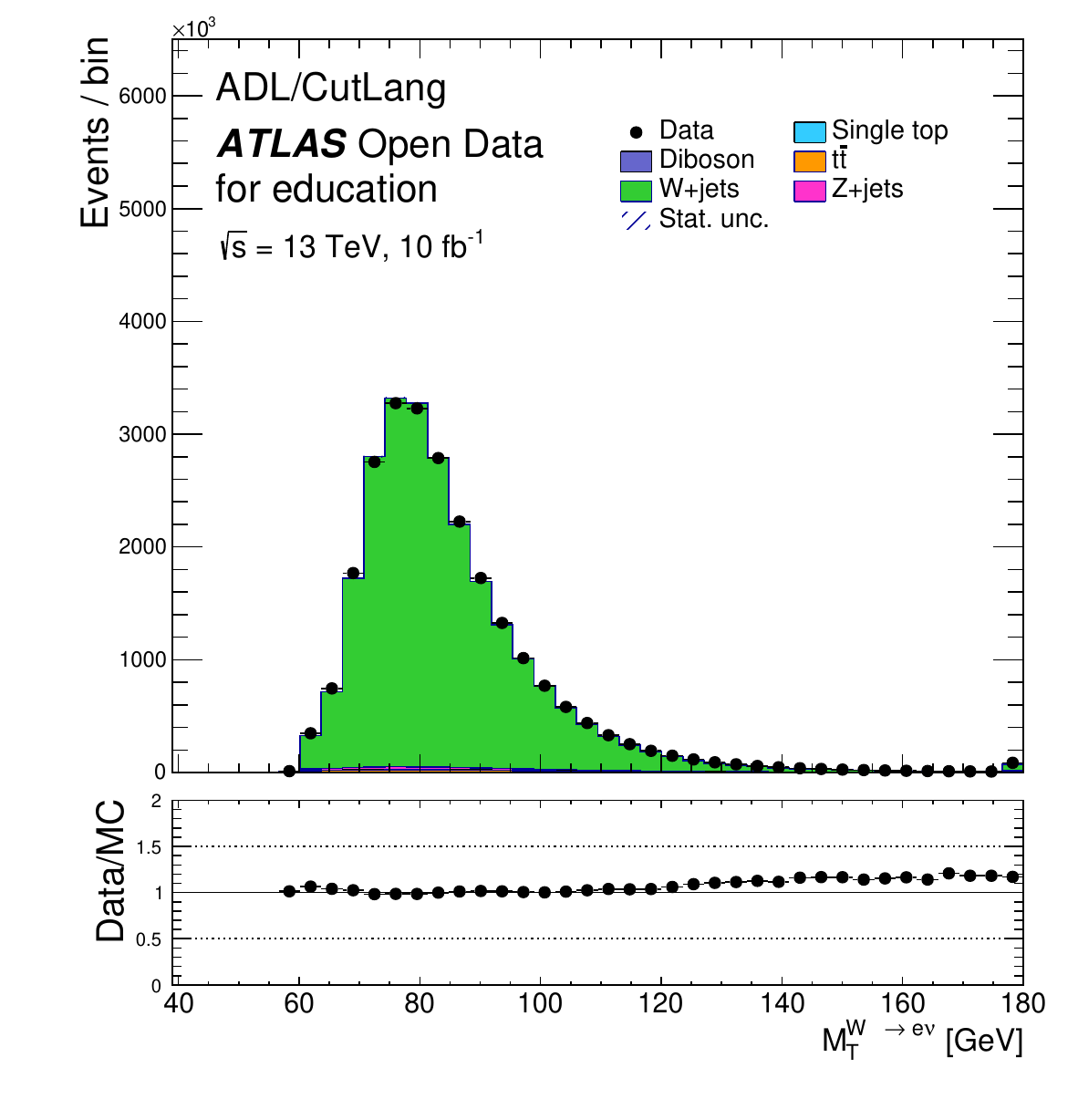}
\includegraphics[width=0.40\textwidth]{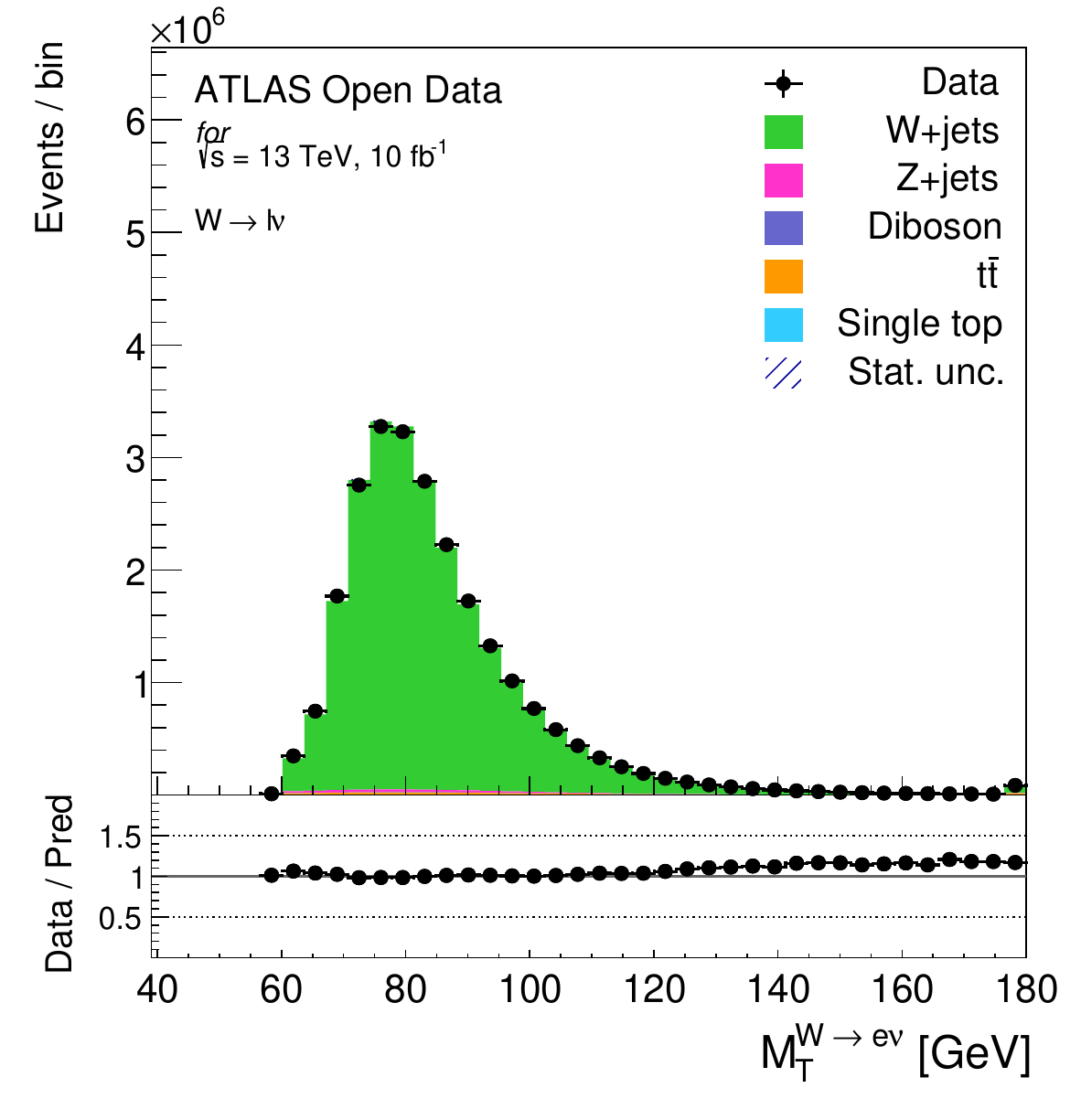}
\includegraphics[width=0.40\textwidth]{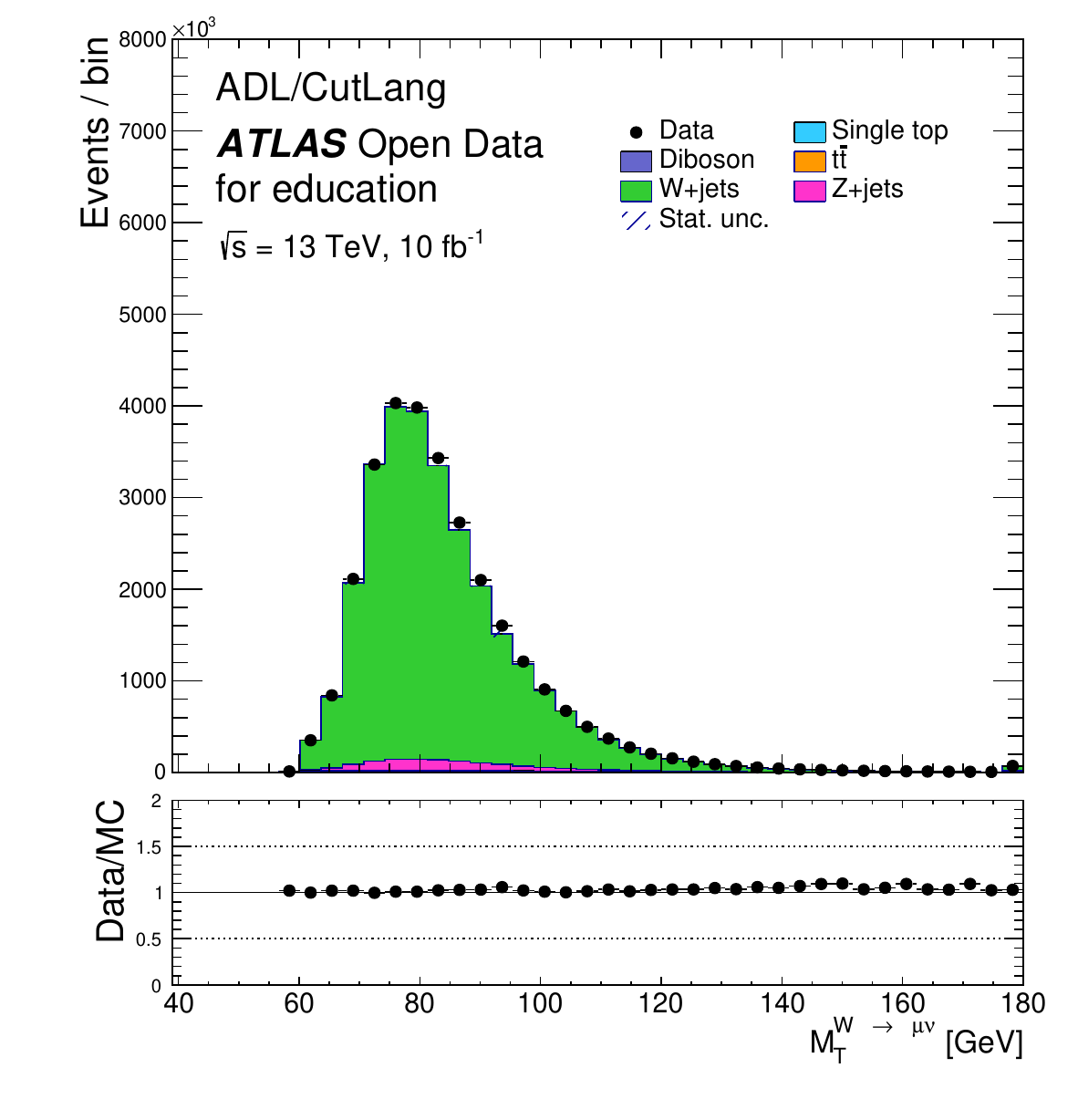}
\includegraphics[width=0.40\textwidth]{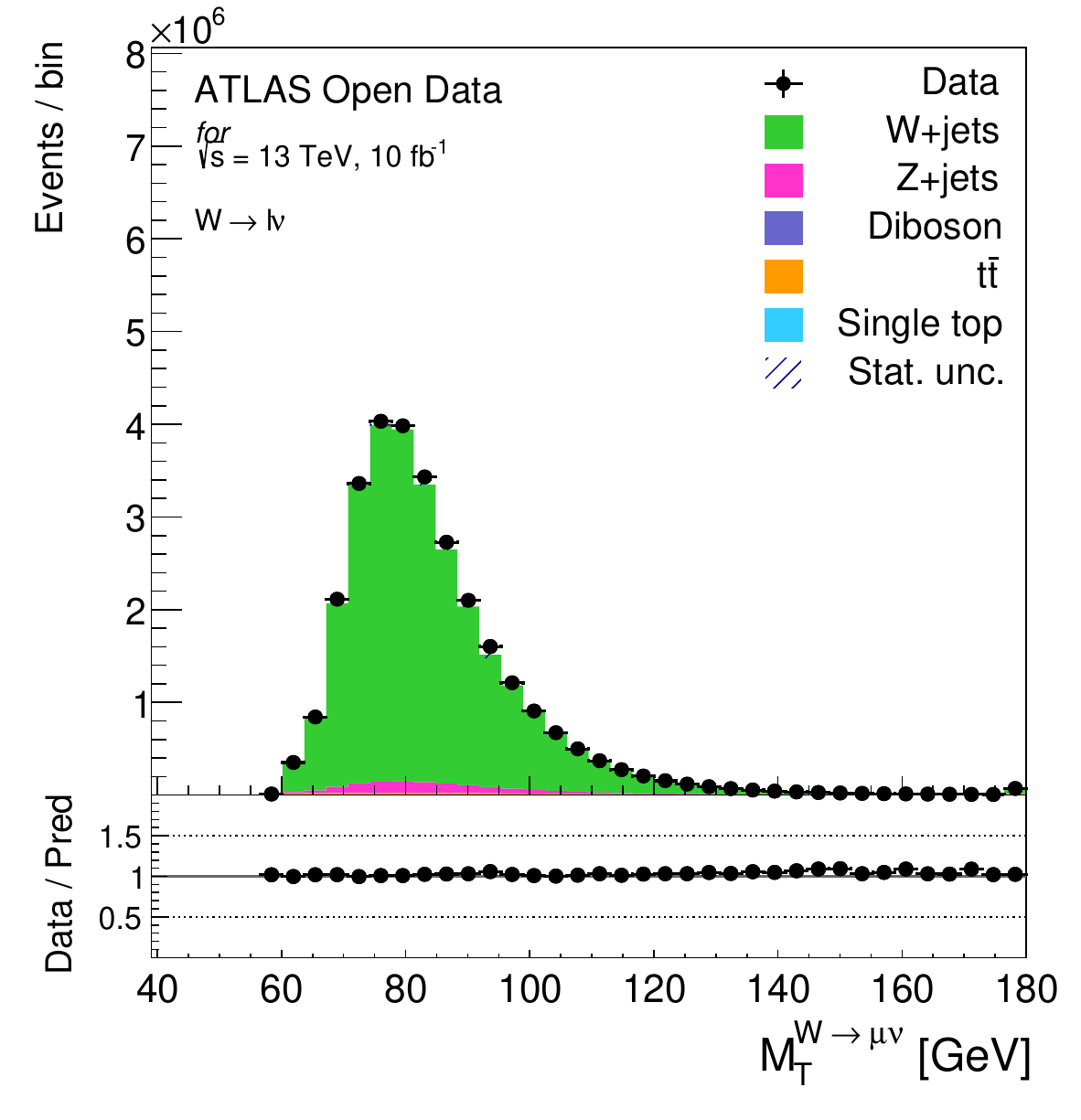}
\caption{Comparison of the $M_T^W$ distributions following the $W \rightarrow e\nu$ (top) and $W \rightarrow \mu\nu$ (bottom) selections obtained with the ADL/CutLang (left) and Open Data (right) frameworks. The lower pad in each plot represents the Data/MC agreement. The last bin in each plot includes the overflow.}
\label{fig:CL_vs_OD_W_masses}
\end{figure}

Figure \ref{fig:CL_vs_OD_W_kinematics} shows the $\eta$ and $p_T$ distributions of the single lepton in the event, and Figure \ref{fig:CL_vs_OD_W_masses} shows the $m_T^W$ distributions in the $W \rightarrow e\nu$ and $W \rightarrow \mu\nu$ channels. These figures show the distributions from the ADL/CutLang and Open Data frameworks, demonstrating consistent results between the two in terms of overall shape, data-to-MC event ratio, and normalization. The cutflow of events obtained with CutLang in the $W \rightarrow \ell\nu$ channel for each sample is shown in Table \ref{tab:cutflow_Wboson}. The corresponding event yields from both the ADL/CutLang and the Open Data frameworks, obtained using the integrals of the $W \rightarrow \ell\nu$, $W \rightarrow e\nu$, and $W \rightarrow \mu\nu$ distributions, are shown in Tables \ref{tab:yields_Wl_channel_yields}, \ref{tab:yields_We_channel_yields}, and \ref{tab:yields_Wmu_channel_yields}, respectively. The results are in excellent agreement, demonstrating consistency across the data and the individual simulation samples.

\begin{table}[!ht]
\begin{center}
\caption{Cutflow of event counts for each sample in the $W \rightarrow \ell \nu$ channel.}
\resizebox{\textwidth}{!}{
\begin{tabular}{|l|c|c|c|c|c|c|}
\hline
Selection & Data & $Z+jets$ & $W+jets$ & Single top & $t\bar{t}$ & Diboson \\ \hline 

ALL & $168533115.00 \pm 12982.03$ & $30246974.00 \pm 5499.72$ & $124384004.00 \pm 11152.76$ & $5398994.00 \pm 2323.57$ & $15747840.00 \pm 3968.35$ & $17855662.00 \pm 4225.60$ \\
Pt(goodLepts) descend & $168533115.00 \pm 12982.03$ & $30246974.00 \pm 5499.72$ & $124384004.00 \pm 11152.76$ & $5398994.00 \pm 2323.57$ & $15747840.00 \pm 3968.35$ & $17855662.00 \pm 4225.60$ \\
evtWeight totalWeight & $168533115.00 \pm 12982.03$ & $7340145.72 \pm 1530.00$ & $121782216.14 \pm 286254.05$ & $318964.86 \pm 162.19$ & $1362681.99 \pm 369.35$ & $224522.38 \pm 138.80$ \\
trigE == 1 OR trigM == 1 & $168533115.00 \pm 12982.03$ & $7340145.72 \pm 1530.00$ & $121782216.14 \pm 286254.05$ & $318964.86 \pm 162.19$ & $1362681.99 \pm 369.35$ & $224522.38 \pm 138.80$ \\
MET $>$ 30.0 & $102616082.00 \pm 10129.96$ & $2993377.76 \pm 973.25$ & $88774530.51 \pm 55047.27$ & $245769.05 \pm 142.36$ & $1112004.41 \pm 333.49$ & $161242.09 \pm 118.24$ \\
Size(goodLepts) == 1 & $58582884.00 \pm 7653.95$ & $1802239.31 \pm 745.73$ & $53906975.53 \pm 50002.49$ & $184541.97 \pm 123.59$ & $881717.20 \pm 297.40$ & $119285.69 \pm 103.86$ \\
MTW $>$ 60.0 & $54065955.00 \pm 7352.96$ & $1315323.71 \pm 629.18$ & $51241910.06 \pm 49680.28$ & $129510.68 \pm 103.48$ & $575972.55 \pm 240.35$ & $92791.24 \pm 92.94$ \\

\hline
\end{tabular}
}
\label{tab:cutflow_Wboson}
\end{center}
\end{table}

\begin{table}[!ht]
\begin{center}
\caption{Event yields after all selections in the lepton channel.}
\begin{tabular}{|l|c|c|}
\hline
Sample & ADL/CutLang & Opendata Framework \\ \hline \hline
Data & $54065955.00 \pm 7352.96$ & $54065954.00$ \\
Diboson & $92791.24 \pm 92.17  $ & $92791.26$ \\
$Z+jets$ & $1315323.71 \pm  628.08 $ & $1315374.30$\\
$W+jets$ & $51241910.06 \pm 49641.33$ & $51251174.89$ \\
$t\bar{t}$ & $575972.55 \pm 235.41  $ & $575973.01$ \\ 
Single top & $129510.68 \pm  102.29$ & $129510.86$ \\ \hline
\end{tabular}
\label{tab:yields_Wl_channel_yields}
\end{center}
\end{table}

\begin{table}[H]
\begin{center}
\caption{Event yields after all selections in the electron channel.}
\begin{tabular}{|l|c|c|}
\hline
Sample & ADL/CutLang & Opendata Framework \\ \hline \hline
Data & $24591458.00 \pm 4958.98 $ & $24591457.00$ \\
Diboson & $47149.62 \pm 62.60   $ & $47149.64$  \\
$Z + jets$ & $276520.90 \pm 291.09$ & $276524.45$ \\
$W + jets$ & $23736252.24 \pm 18699.39$ & $23736123.74$ \\
$t\bar{t}$ & $302045.81 \pm 171.12 $ & $302049.98$ \\ 
Single top & $ 66391.02 \pm 73.46 $ & $66391.14$ \\ \hline
\end{tabular}
\label{tab:yields_We_channel_yields}
\end{center}
\end{table}

\begin{table}[!h]
\begin{center}
\caption{Event yields after all selections in the muon channel.}
\begin{tabular}{|l|c|c|}
\hline
Sample & ADL/CutLang & Opendata Framework \\ \hline \hline
Data & $29474497.00 \pm 5429.04 $ & $29474496.00$ \\
Diboson & $45641.62 \pm 67.65  $ & $45641.62$ \\
$Z + jets$ & $ 1038802.48 \pm  556.55  $ & $1038849.85$ \\
$W + jets$ & $ 27505657.82 \pm 45984.72 $ & $27515051.15$ \\
$t\bar{t}$ & $ 273926.74 \pm 161.66$ & $273923.03$ \\ 
Single top & $63119.66 \pm 71.18$ & $63119.72$ \\ \hline
\end{tabular}
\label{tab:yields_Wmu_channel_yields}
\end{center}
\end{table}


\subsection{Standard model single top quark production in single lepton final state} \label{section_Stopanalysis}
This analysis describes the SM t-channel single top quark production in the $t \ + \ q \rightarrow Wb + q \rightarrow \ell\nu b + q$ decay. 
The novelty of this analysis is the inclusion of b-tagging for jets.
The object definition and the event selection algorithm of this analysis are loosely based on the ATLAS single-top t-channel production measurements at center-of-mass energies of $\sqrt{s}= 8$ TeV and $\sqrt{s}=13$ TeV \cite{atlas_singletop8tev, atlas_singletop13tev}. In addition to the standard object definitions described in Section \ref{section_objects}, leptons (electrons or muons) in the event required to have $p_T >$ 35 GeV and the b-tagged jets are required to pass the MV2c10 tagging algorithm at 70\% efficiency working point \cite{ATL-PHYS-PUB-2016-012}. The additional object definitions and the event variable definitions used for the analysis selections in ADL syntax, are given below: 

\begin{courier}
\begin{lstlisting}
object goodBJets
  take goodJets
  select jet_MV2c10(goodJets) > 0.8244273

object nonBJets
  take goodJets
  select jet_MV2c10(goodJets) < 0.8244273

define leadLept : goodLepts[0]
define BJet : goodBJets[0]
define nonBJet : nonBJets[0]
define dEtajj : Abs(Eta(BJet) - Eta(nonBJet))
define MTW : sqrt(2*Pt(leadLept)*MET*(1 - cos(dPhi(leadLept, METLV[0]))))
define HT : sum(Pt(goodEles)) + sum(Pt(goodMuos)) + sum(Pt(goodJets) + MET 

define lb : BJet leadLept
define mlb : m(lb)
\end{lstlisting}
\end{courier}

On this ADL snippet, b-tagged jets and non-b-tagged jets are well separated with two different object definitions and the $\eta$ separation between these two jets in the event are described using \texttt{define dEtajj} line. The standard transverse mass of the W boson, which is widely used in many analyses, particularly in Section \ref{section_Wanalysis}, is implemented in the \texttt{MTW} line.
As usual, \texttt{HT} is the scalar sum of $p_T^{\ell}$, $p_T^{jet}$ and the missing transverse energy, $E_T^{miss}$ in the event. The particle \texttt{lb} is obtained by adding the four-momenta of the b-tagged jet and the leading lepton in the event. Finally, \texttt{mlb} is the invariant mass of \texttt{lb} system, and it serves as an approximation to the reconstructed top-quark mass in the original analysis.

The event selection requires exactly one lepton and exactly two jets, of which one must be b-tagged. The $E_T^{miss}$, and the transverse mass of the $W$ boson, $M_T^W$, are required to be greater than 30 GeV and 60 GeV, respectively. The variable \texttt{mlb} is required to be less than 150 GeV, and $H_T$ must exceed 195 GeV in order to suppress the $W$+jets background contribution. Finally, the pseudorapidity of the non-b-tagged jet is required to satisfy $|\eta| > 1.5$, and the pseudorapidity separation between the b-tagged and non-b-tagged jets must be greater than 1.5 to further reduce the $t\bar{t}$ background. ADL representation of these event selections is shown below.

\clearpage

\begin{courier}
\begin{lstlisting}
region SingleTopAnalysis
  select ALL
  weight evtweight totalWeight
  select trigE == 1 OR trigM == 1
  select MET > 30
  select Size(goodLepts) == 1
  select Size(goodJets) == 2
  select Size(goodBJets) == 1
  select MTW > 60
  select mlb < 150
  select HT > 195
  select AbsEta(nonBJet) > 1.5
  select Abs(dEta(nonBJet, BJet)) > 1.5
\end{lstlisting}
\end{courier}

\begin{figure}[!ht]
\centering
\includegraphics[width=0.40\textwidth]{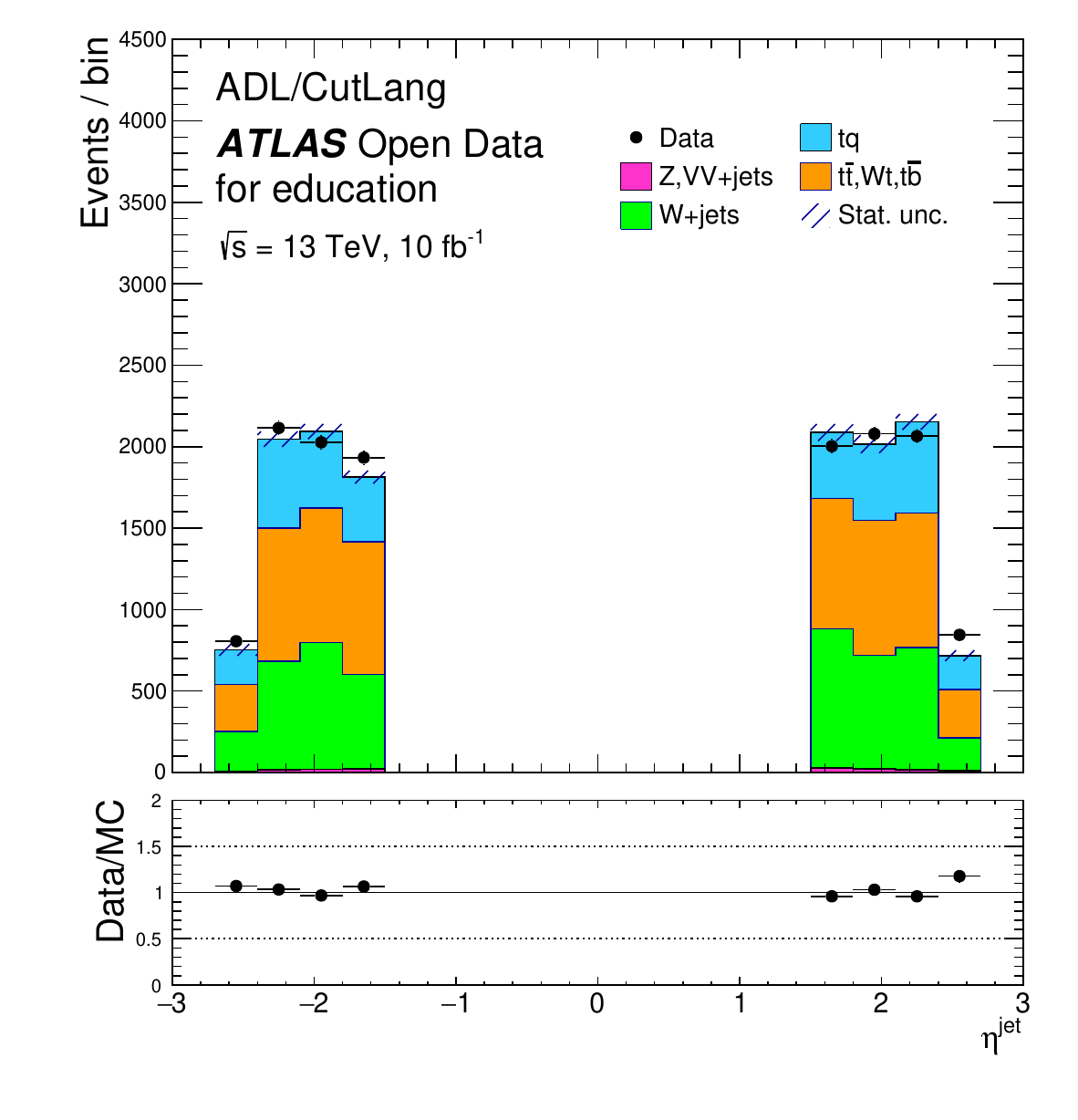}
\includegraphics[width=0.40\textwidth]{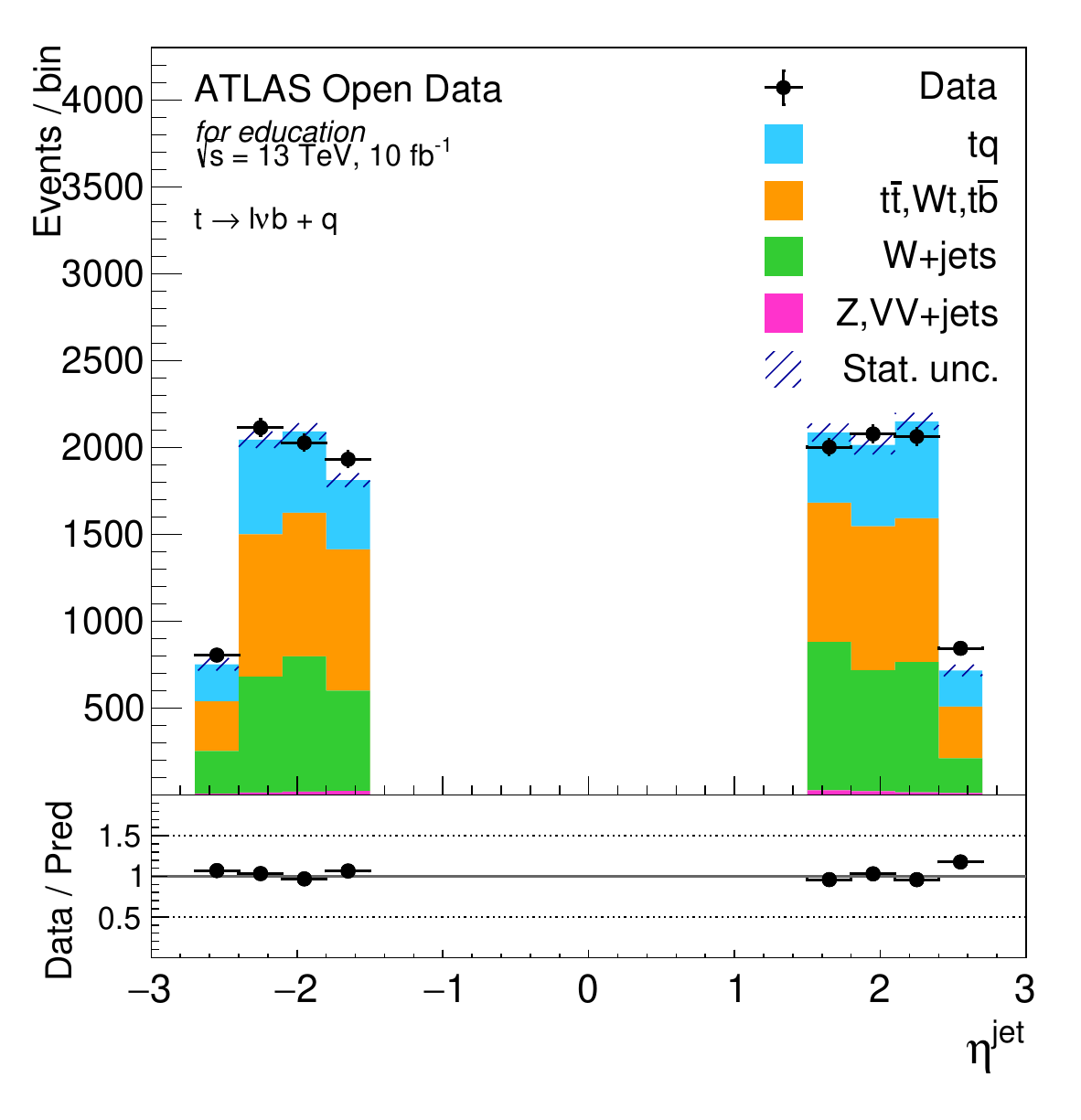}
\includegraphics[width=0.40\textwidth]{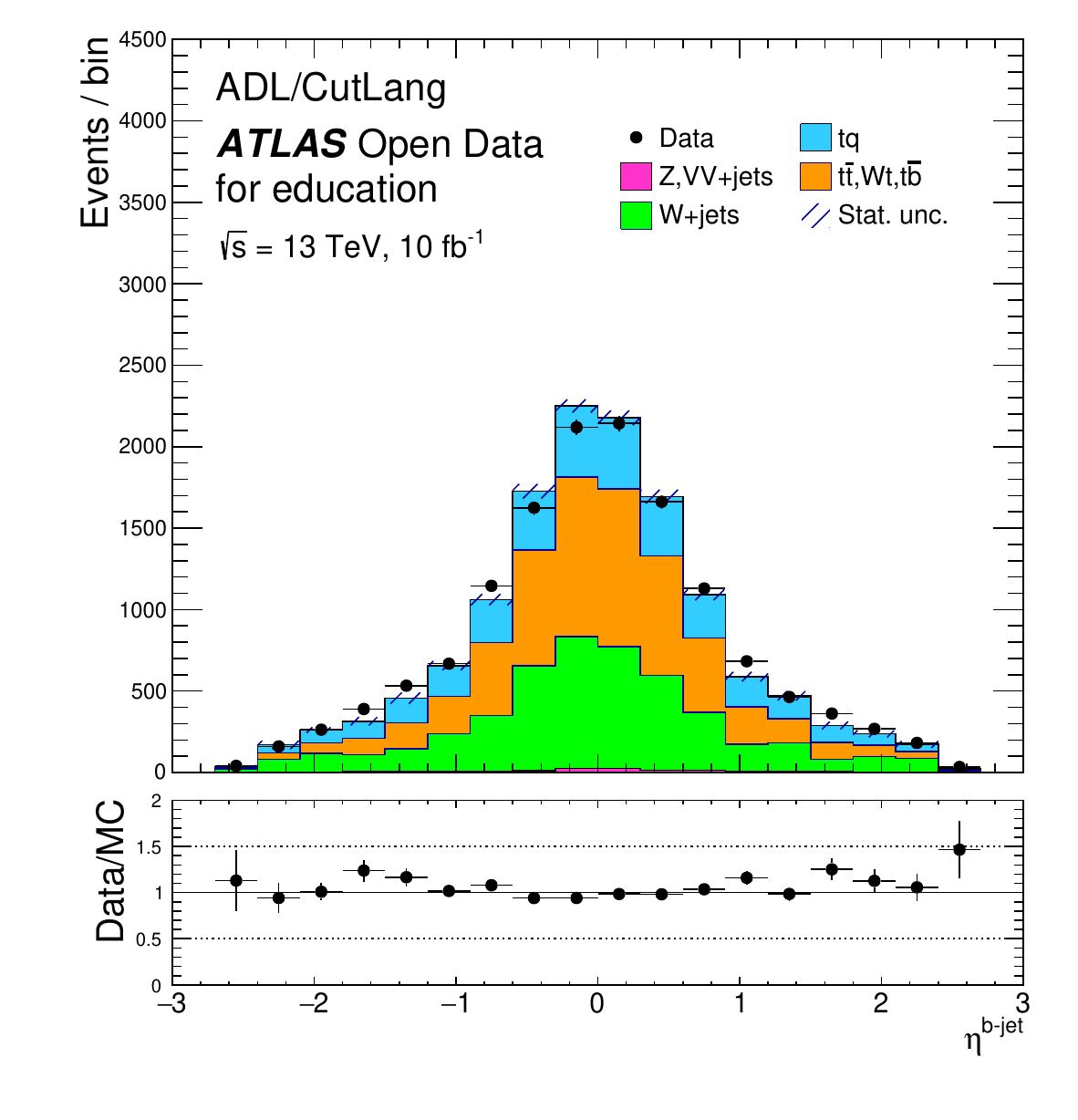}
\includegraphics[width=0.40\textwidth]{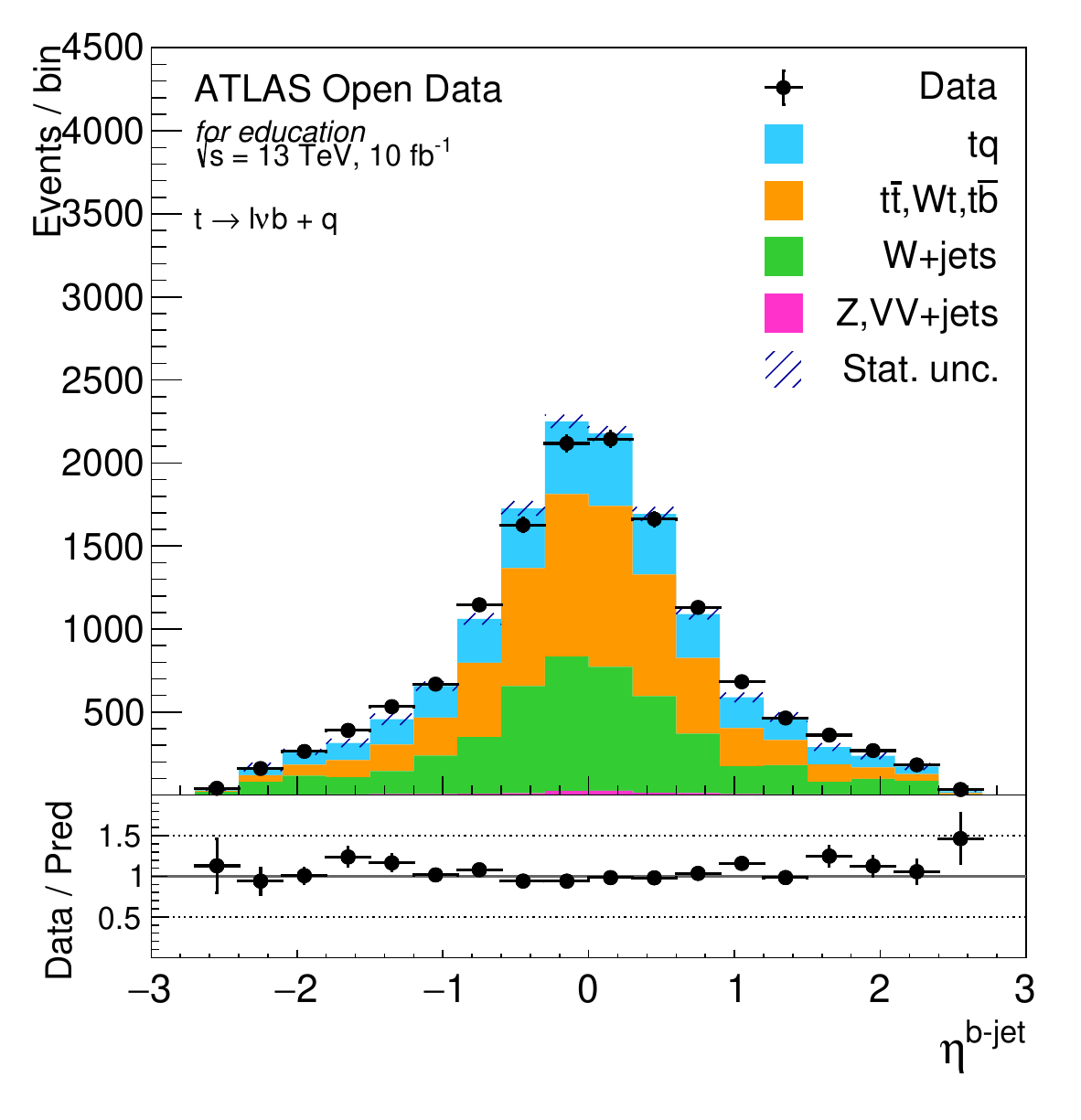}
\caption{Comparison of the $\eta$ distributions of the untagged jet (top) and the b-tagged jet (bottom), obtained with the ADL/CutLang (left) and Open Data (right) frameworks in the $t \ + \ q \rightarrow Wb + q \rightarrow \ell\nu b + q$ channel. The lower pad in each plot represents the Data/MC ratio. The last bin in each plot includes the overflow.}
\label{fig:CL_vs_OD_stop_kinematics}
\end{figure}

\begin{figure}[!ht]
\centering
\includegraphics[width=0.40\textwidth]{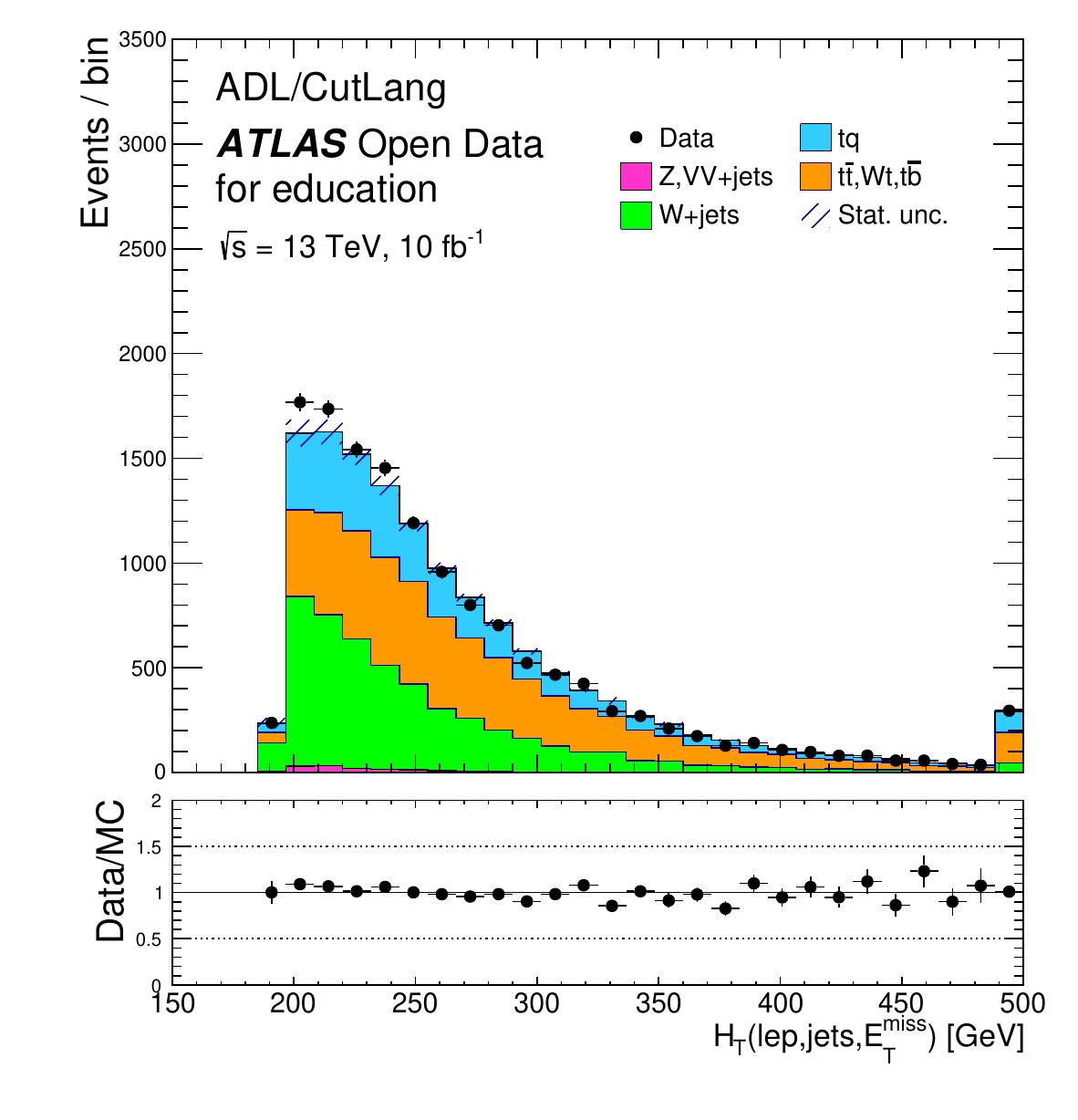}
\includegraphics[width=0.40\textwidth]{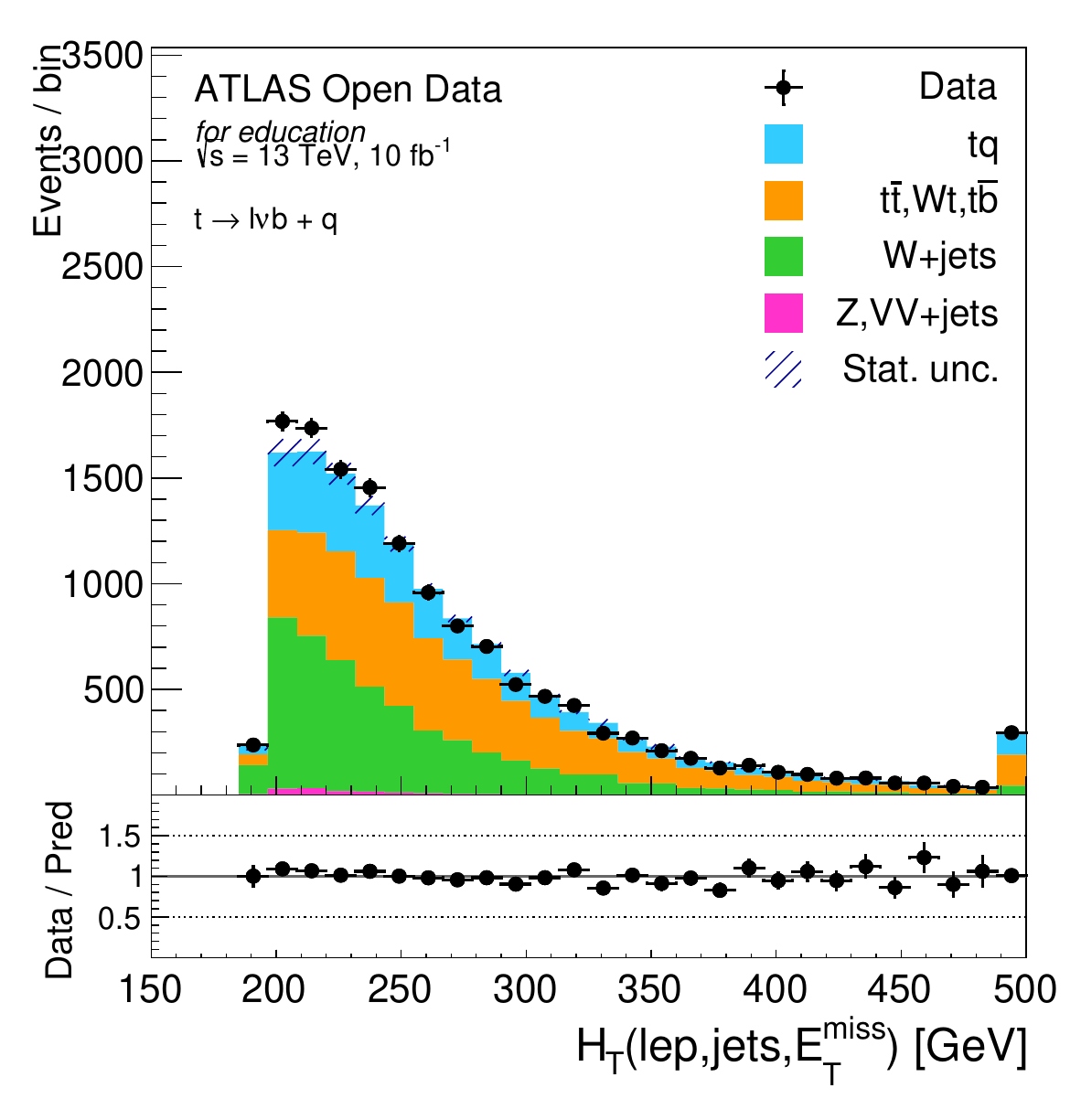}
\includegraphics[width=0.40\textwidth]{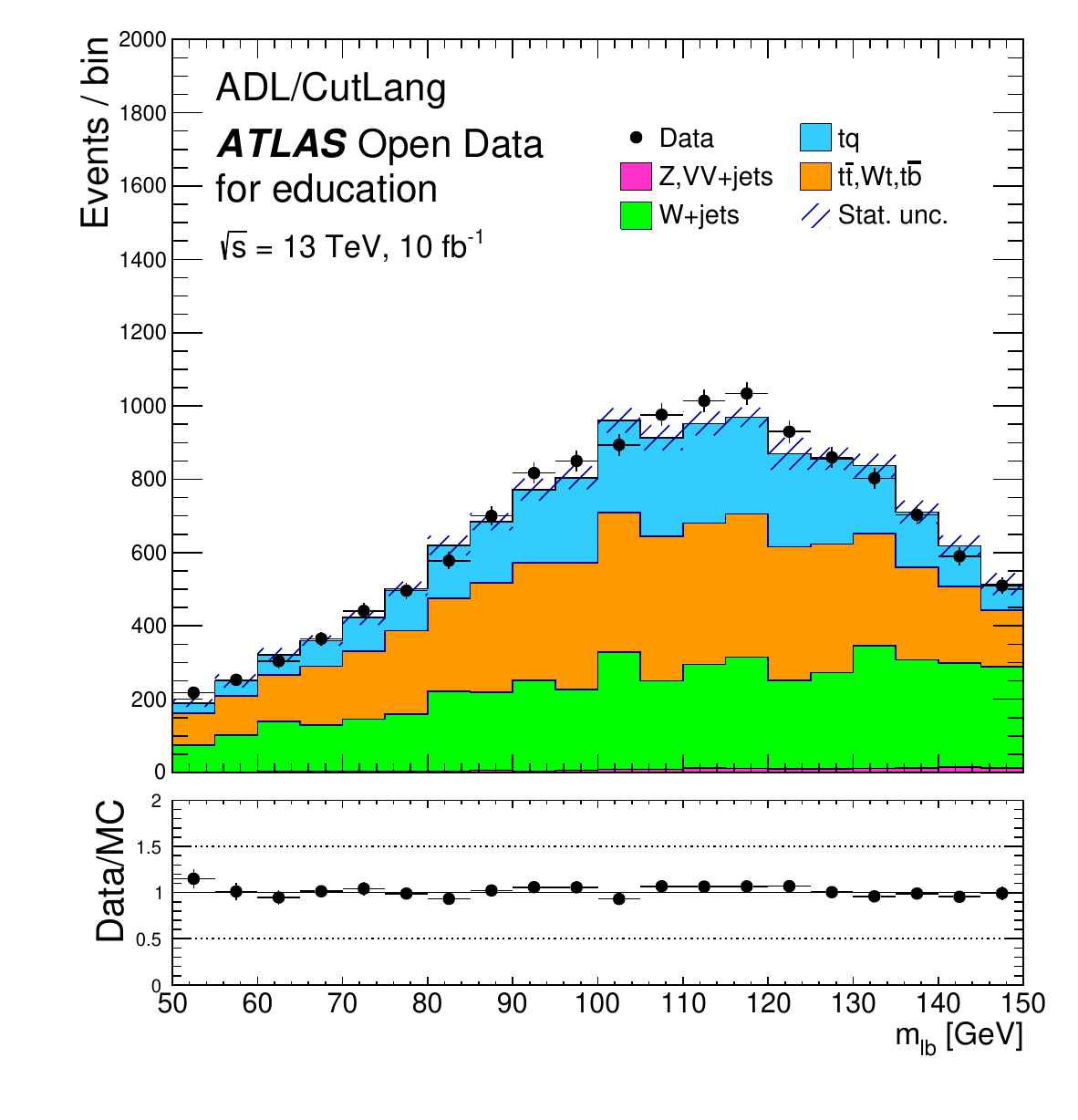}
\includegraphics[width=0.40\textwidth]{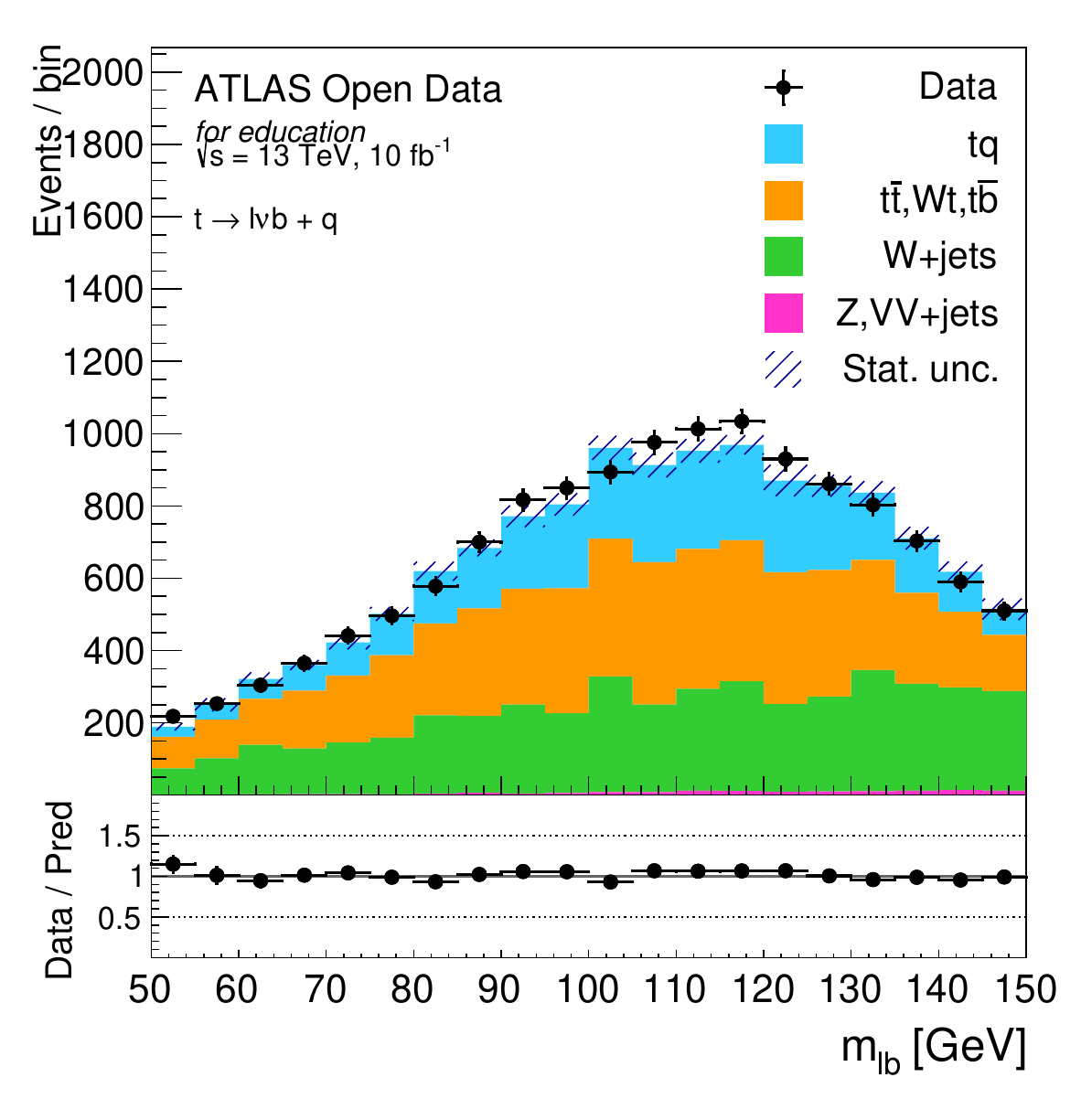}
\caption{Comparison of the $H_T$ (top) and the invariant mass of the lepton and b-tagged jet (bottom) distributions, obtained with the ADL/CutLang (left) and Open Data (right) frameworks in the $t \ + \ q \rightarrow Wb + q \rightarrow \ell\nu b + q$ channel. The lower pad in each plot represents the Data/MC ratio. The last bin in each plot includes the overflow.}
\label{fig:CL_vs_OD_stop_masses}
\end{figure}

The $\eta$ distributions after all selections for the untagged jet and the b-tagged jet are shown in Figure \ref{fig:CL_vs_OD_stop_kinematics}. Figure \ref{fig:CL_vs_OD_stop_masses} shows the $H_T$ distribution and the invariant mass of the lepton and b-tagged jet in the event. Both figures show the distributions from the ADL/CutLang and the Open Data frameworks on the left and right, respectively. In addition, cutflow tables summarizing the event counts after each selection step, as well as the final event yields for each sample, are presented in Tables \ref{tab:cutflow_stop} and \ref{tab:yields_stop}, respectively. For a fair comparison, the event yields for all samples are obtained using the integral of the $H_T$ distributions in both frameworks.\\

The distributions show that the results from both frameworks yield consistent Data/MC ratios in terms of event variables in both shape and normalization. In addition, the event yields show excellent agreement between the ADL/CutLang and Open Data frameworks. The simplicity of the ADL syntax and the ease of use of the CutLang framework is self evident.

\begin{table}[!hp]
\begin{center}
\caption{Cutflow of event counts for each sample in the single-top channel.}
\resizebox{\textwidth}{!}{
\begin{tabular}{|l|c|c|c|c|c|}
\hline
Selection & Data & $tq$ & ZVV & $t\bar{t}, Wt, tb$ & $W + jets$ \\ \hline 
ALL & $168533115.00 \pm 12982.03$ & $3032852.00 \pm 1741.51$ & $48102636.00 \pm 6935.61$ & $18113982.00 \pm 4256.05$ & $125060048.00 \pm 11183.02$ \\
evtweight totalWeight & $168533115.00 \pm 12982.03$ & $185685.44 \pm 127.58$ & $7564584.78 \pm 1536.49$ & $1497028.35 \pm 383.95$ & $121793301.42 \pm 286141.83$ \\
trigE $==$ 1 OR trigM $==$ 1 & $168533115.00 \pm 12982.03$ & $185685.44 \pm 127.58$ & $7564584.78 \pm 1536.49$ & $1497028.35 \pm 383.95$ & $121793301.42 \pm 286141.83$ \\
MET $>$ 30.0 & $102616082.00 \pm 10129.96$ & $140543.68 \pm 111.07$ & $3154773.85 \pm 980.63$ & $1218060.65 \pm 346.29$ & $88781958.24 \pm 55043.33$ \\
Size(goodLepts) $==$ 1 & $58582884.00 \pm 7653.95$ & $100261.53 \pm 94.18$ & $1921477.98 \pm 753.05$ & $966640.09 \pm 308.97$ & $53908408.88 \pm 49995.04$ \\
Size(goodJets) $==$ 2 & $2874858.00 \pm 1695.54$ & $33447.91 \pm 54.46$ & $138767.78 \pm 203.35$ & $210948.06 \pm 143.56$ & $2218750.58 \pm 8693.77$ \\
Size(goodBJets) $==$ 1 & $295206.00 \pm 543.33$ & $19930.20 \pm 42.02$ & $7206.56 \pm 51.71$ & $114061.97 \pm 106.22$ & $125036.53 \pm 1904.88$ \\
MTW $>$ 60.0 & $198941.00 \pm 446.03$ & $14167.12 \pm 35.32$ & $3950.17 \pm 33.65$ & $78435.36 \pm 88.03$ & $92755.56 \pm 1890.89$ \\
mlb $<$ 150.0 & $131784.00 \pm 363.02$ & $12837.51 \pm 33.41$ & $2405.32 \pm 26.18$ & $56663.63 \pm 74.78$ & $51885.51 \pm 1872.57$ \\
HT $>$ 195.0 & $97861.00 \pm 312.83$ & $10687.77 \pm 30.53$ & $1408.73 \pm 19.81$ & $51241.02 \pm 71.16$ & $32465.01 \pm 1839.77$ \\
AbsEta(nonBJet) $>$ 1.5 & $31789.00 \pm 178.29$ & $4924.88 \pm 20.72$ & $426.21 \pm 10.79$ & $15833.27 \pm 39.69$ & $10050.40 \pm 149.88$ \\
Abs( dEta(nonBJet, BJet) ) $>$ 1.5 & $13873.00 \pm 117.78$ & $3264.47 \pm 16.70$ & $146.19 \pm 6.41$ & $5499.49 \pm 23.37$ & $4334.50 \pm 113.49$ \\
\hline
\end{tabular}
}
\label{tab:cutflow_stop}
\end{center}
\end{table}

\begin{table}[!h]
\begin{center}
\caption{Event yields after all selections for the single-lepton channel.}
\begin{tabular}{|l|c|c|}
\hline
Sample & ADL/CutLang & Opendata Framework \\ \hline \hline
Data & $13873.00 \pm 117.78 $ & $13873.00$ \\
$tq$ &$ 3264.47 \pm 16.45 $ & $3264.27$  \\
$ZVV$ & $146.19 \pm 6.41 $ & $146.15$ \\
$t\bar{t},Wt,tb$& $5499.49 \pm 23.09$ & $5499.31$ \\
$W+jets$ & $4767.95 \pm 124.81$ & $4768.05$ \\ \hline
\end{tabular}
\label{tab:yields_stop}
\end{center}
\end{table}

\clearpage

\subsection{Top quark pair production in the single lepton final state} \label{section_ttbaranalysis}

The analysis targets the lepton+jets final state arising from top-quark pair (\( t\bar{t} \)) decays. In this process, each top quark decays into a \( W \) boson and a \( b \)-quark. One \( W \) boson decays leptonically (\( W \to \ell \nu_\ell \)), producing a charged lepton (\( \ell \)) and a neutrino (\( \nu_\ell \)), while the other decays hadronically (\( W \to q\bar{q}' \)), resulting in two jets. The final state therefore includes a lepton (\( e \) or \( \mu \)), $E_T^{miss}$ from the neutrino, and at least four jets, two of which are associated with \( b \)-quarks. 

The event reconstruction, as implemented using the algorithm in reference \cite{ATL-OREACH-PUB-2020-001}, focuses on the kinematic reconstruction of the fully hadronically decaying top quark and the leptonically decaying $W$ boson. The hadronic top-quark decay ($t \rightarrow W^+ b$) is reconstructed by identifying the $b$-jet and the jet pair associated with the hadronic W boson decay ($W \rightarrow q \bar{q}'$). The invariant mass of the three-jet system ($m_{jjj}$) provides a handle on the top-quark mass. Simultaneously, the leptonically decaying $W$ boson ($W \rightarrow \ell \nu_{\ell}$) is reconstructed using the charged lepton ($\ell$) and $E_T^{miss}$ from the neutrino.
This analysis differs from previously presented  studies in its implementation, particularly in utilization of Heaviside step  and the optimization functions.
Nearly the entire analysis algorithm is defined using event-level variables, making it a comprehensive and self-contained description.
For the reconstruction of the leptonically decaying W boson, the well-known and previously discussed transverse mass method is applied:
\begin{courier} \begin{lstlisting}
define leadLept : goodLepts[0]
define MTW : sqrt(2*Pt(leadLept)*MET*(1 - cos(dPhi(leadLept, METLV[0])))) 
\end{lstlisting} \end{courier} 

\paragraph{}
The variables defined and used to identify the hadronically decaying top-quark jet in a four-jet final-state event are given below. 
Three jets are selected from the {\tt goodJets} object class, denoted as {\tt goodJetA}, {\tt goodJetB} and {\tt goodJetC}.
Negative indices (-1,-2 and -3) are used to represent the unknown but distinct jet indices, which are determined for each event only after an optimization procedure.
The transverse momentum of the three jet system is denoted as \texttt{$p_T^{jjj}$}. 
The optimization procedure  identifies the jet combination with the highest $p_T$, thereby selecting the optimal three-jet system, and it will be discussed later. 

\begin{courier} \begin{lstlisting}
define goodJetA : goodJets[-1]
define goodJetB : goodJets[-2]
define goodJetC : goodJets[-3]
define Ptjjj : Pt(goodJetA + goodJetB + goodJetC)
define mTopjjj : m(goodJetA + goodJetB + goodJetC)
\end{lstlisting} \end{courier}
\paragraph{}
After identifying the three-jet system with the highest total $p_T$, the next step involves selecting the two-jet combination with the highest $p_T$. This two-jet combination is then used to reconstruct the mass of a hadronically decaying $W$ boson candidate. The necessary algorithms were implemented using the \texttt{HSTEP()} function to correctly identify and calculate the relevant quantities. This function  returns 0 when its conditional argument is false and 1 when the argument is true.
In the definitions below, therefore, only one of the three possible configurations becomes non-zero, and its corresponding mass is used to define the final variable {\tt mWjj}.

\begin{courier} \begin{lstlisting}
define Ptjj1 : Pt(goodJetA + goodJetB)
define Ptjj2 : Pt(goodJetB + goodJetC)
define Ptjj3 : Pt(goodJetA + goodJetC)

define PtjjMax : max(Ptjj1,Ptjj2,Ptjj3)
define Mjj1 : m(goodJetA + goodJetB)*HSTEP(Ptjj1 == PtjjMax)
define Mjj2 : m(goodJetB + goodJetC)*HSTEP(Ptjj2 == PtjjMax)
define Mjj3 : m(goodJetA + goodJetC)*HSTEP(Ptjj3 == PtjjMax)
define mWjj : Mjj1 + Mjj2 + Mjj3
\end{lstlisting} \end{courier}

The event selection process begins with the \texttt{preselections} region, which sets general conditions for the analysis. This block sets up the event weight for the MC events; ensures that the event contains more than three jets, significant $E_T^{miss}$, and valid leptons.
The $E_T^{miss}$ limit of 30 GeV is typical for events in which $W$ bosons decay leptonically.
Finally, the selected leptons are sorted in descending order of their transverse momenta.

\begin{courier} \begin{lstlisting}
region preselections
  select ALL
  weight evtWeight totalWeight
  select Size(JET) > 3
  select MET > 30
  sort Pt(goodLepts) descend
\end{lstlisting} \end{courier}

The region {\tt TTbarAnalysis} applies more specific cuts, narrowing down the event selection to those that match the expected characteristics of a top quark pair decay into a lepton+jets final state. The region is presented in the listing below.
A single-electron or single-muon trigger must be satisfied, and the event must contain exactly one good lepton.  
At least four good jets, with two being b-tagged, are required to identify the jets coming from top quark decays. 
The $M_T^W$ must be greater than 30 GeV. 
The line \texttt{select Ptjjj $\sim$= 99999} is part of the optimization process for identifying the best combination of three jets in a four-jet final state. The operator $\sim=$ instructs the system to find the best combination that yields a {\tt Ptjjj} as close as possible to the target value,  99999, which is chosen to be sufficiently large.

\begin{courier} \begin{lstlisting}
region TTbarAnalysis
  preselections
  select trigE == 1 OR trigM == 1
  select Size(goodLepts) == 1 
  select Size(goodJets) >= 4 
  select Size(goodBJets) >= 2 
  select MTW > 30 
  select Ptjjj ~= 99999 #finding largest Pt of 3-jet system
\end{lstlisting} \end{courier}

\begin{figure}[!ht]
\centering
\includegraphics[width=0.40\textwidth]{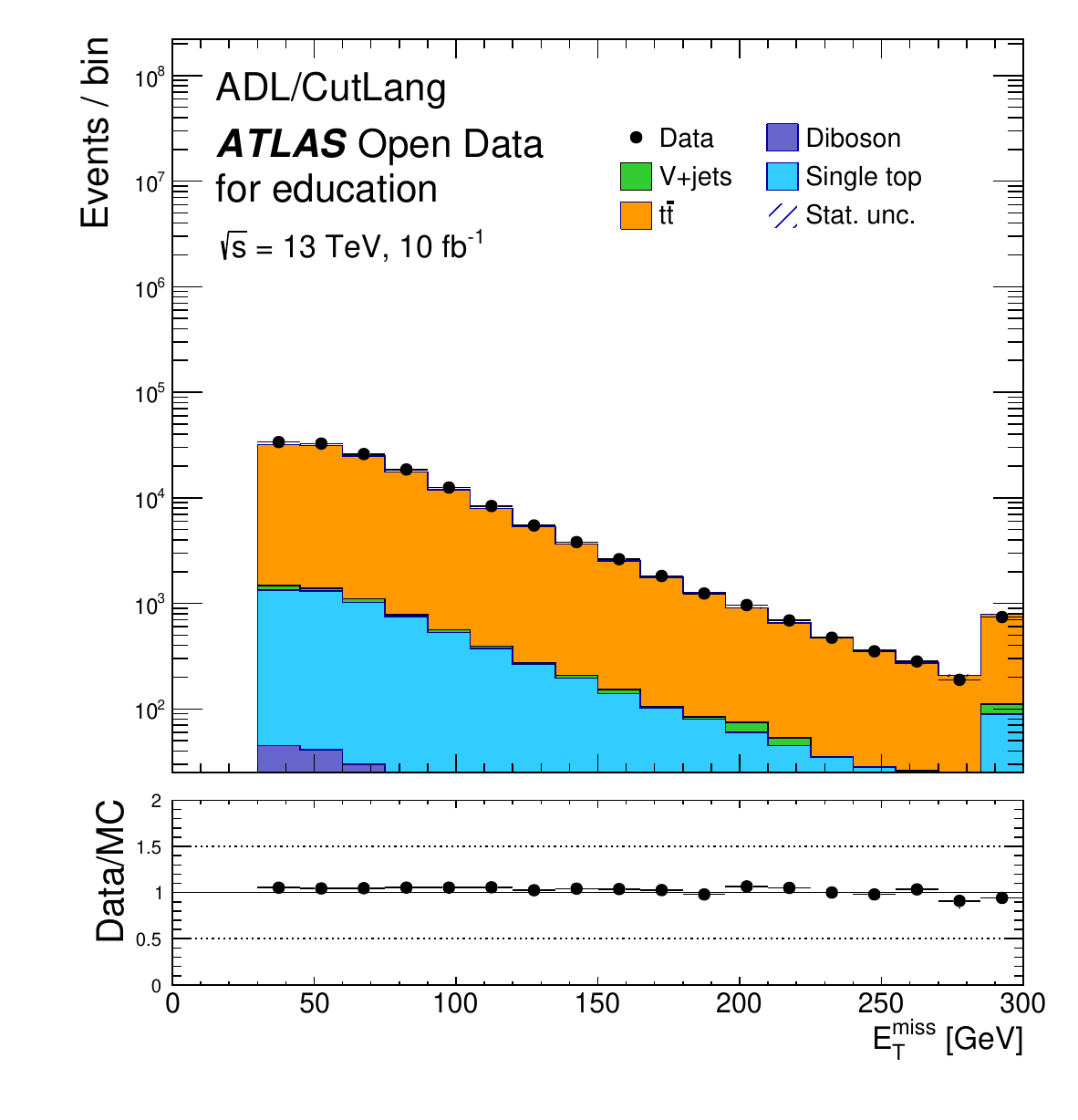}
\includegraphics[width=0.40\textwidth]{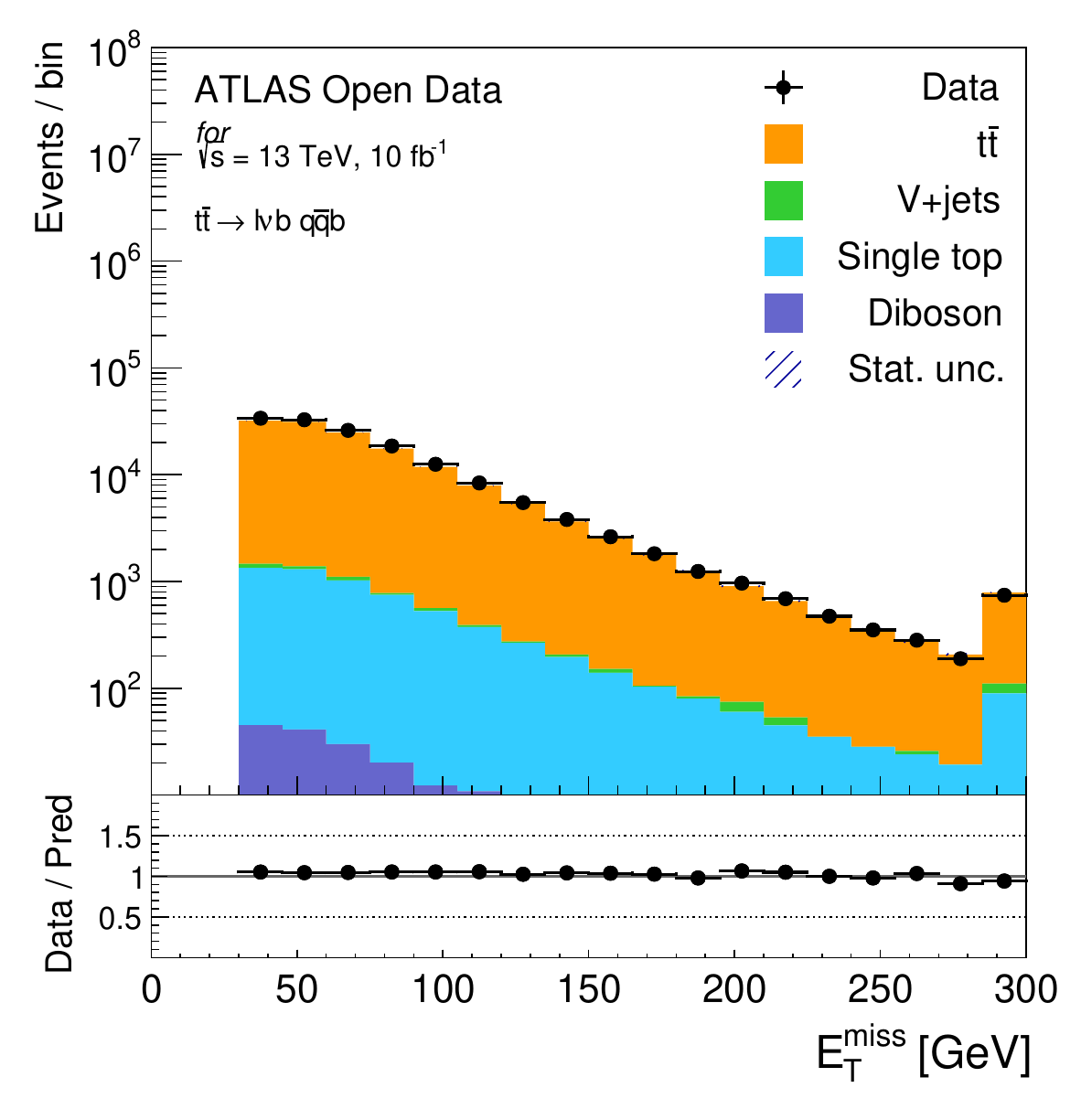}
\includegraphics[width=0.40\textwidth]{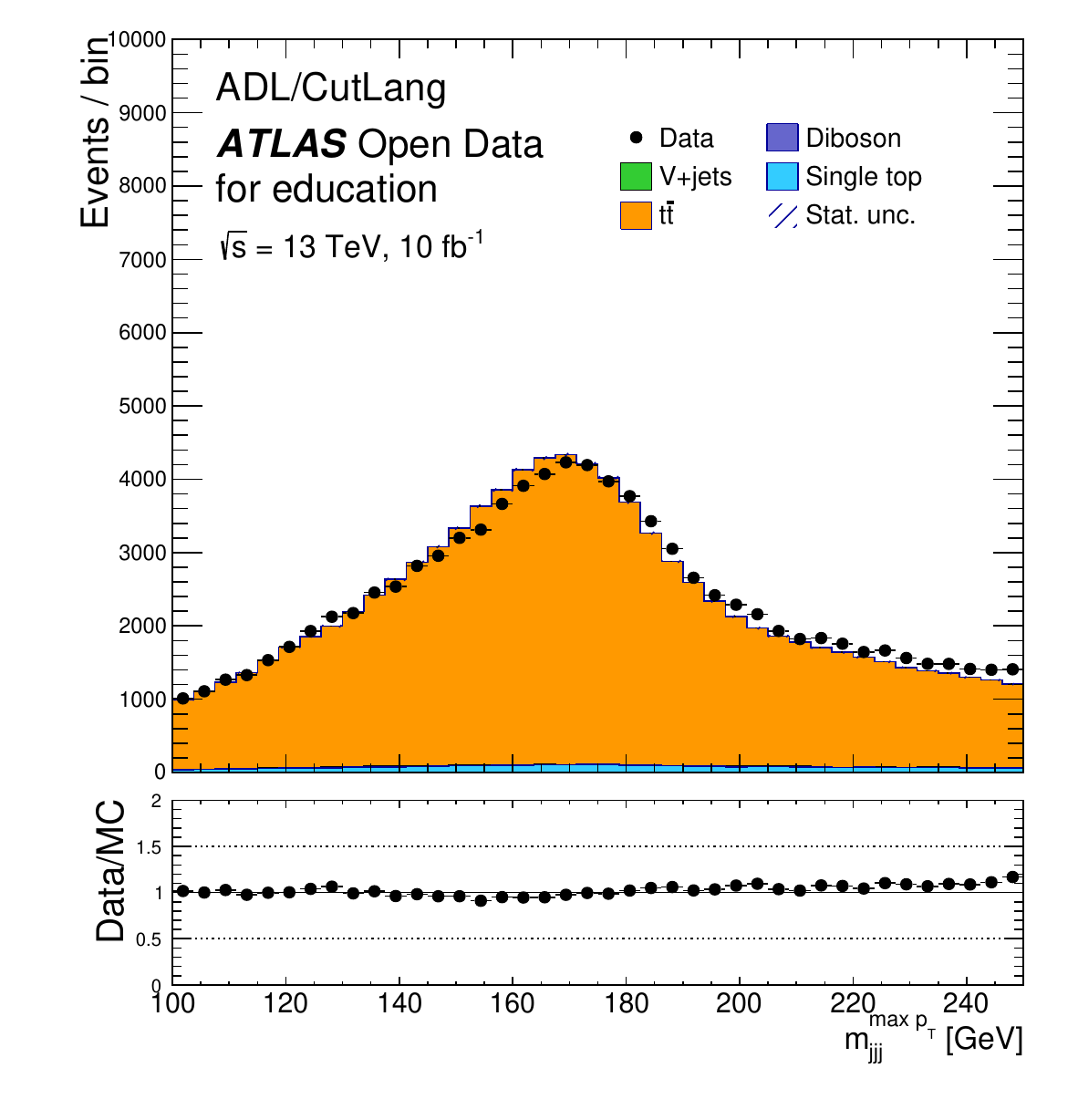}
\includegraphics[width=0.40\textwidth]{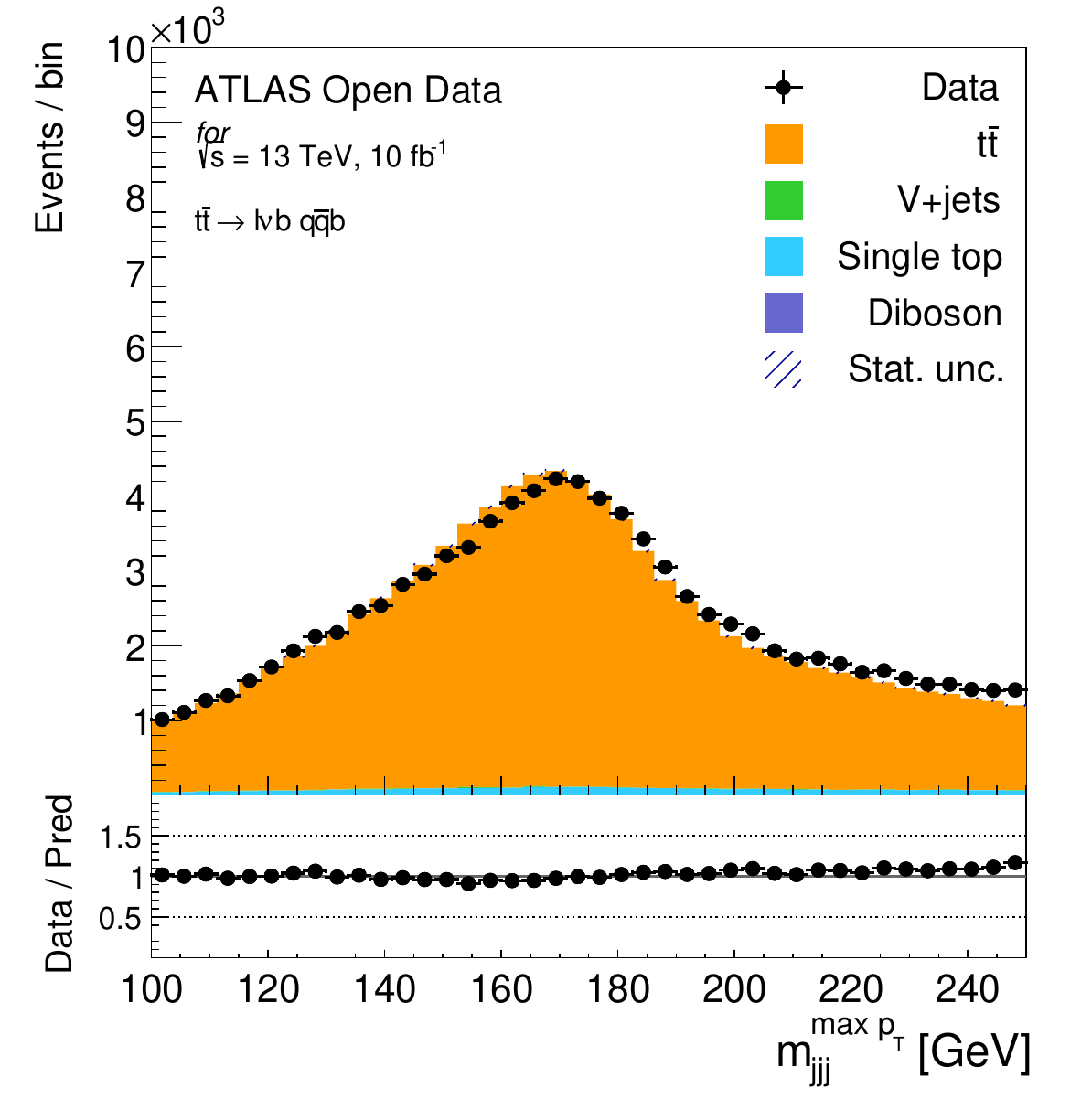}
\caption{Comparison of the $E_T^{miss}$ (top), and the invariant mass of three jets (bottom) distributions, obtained with the ADL/CutLang (left) and Open Data (right) frameworks in the $t\bar{t} \rightarrow W^+W^- b\bar{b} \rightarrow \ell \nu q\bar{q}' b\bar{b}$ channel. The lower pad in each plot represents the Data/MC ratio. The last bin in each plot includes the overflow.}
\label{fig:CL_vs_OD_TTbar_kinematics}
\end{figure}

A comparison of the ADL/CutLang and Open Data frameworks, showing the $E_T^{\text{miss}}$ and three-jet mass ($m_{jjj}$) distributions after all event selections in this region, can be seen in Figure~\ref{fig:CL_vs_OD_TTbar_kinematics}. Table \ref{tab:cutflow_TTbar} shows the cutflow of events after each selection in the ADL/CutLang framework, and Table \ref{tab:yields_TTbaryields} presents the event yields, obtained from the integral of the $E_T^{\mathrm{miss}}$ distribution, for a comparison between the ADL/CutLang and Open Data frameworks.\\

The two frameworks yield distributions with similar shapes, and data-to-MC ratios close to unity.
In addition, the event yields from ADL/CutLang are nearly identical to those from the Open Data framework, with differences of less than 1\%. This confirms that the same event selection algorithm can be successfully reproduced using ADL/CutLang framework.

\begin{table}[!ht]
\begin{center}
\caption{Cutflow of event counts for each sample in the $t\bar{t} \rightarrow W^+W^- b\bar{b} \rightarrow \ell \nu q\bar{q}' b\bar{b}$ channel.}
\resizebox{\textwidth}{!}{
\begin{tabular}{|l|c|c|c|c|c|}
\hline
Selection & Data & $t\bar{t}$ & $V+jets$ & Single top & Diboson \\ \hline 
ALL & $168533115.00 \pm 12982.03 $ & $15747840.00 \pm 3968.35 $ & $77715321.00 \pm 8815.63 $ & $5398994.00 \pm  2323.57 $ & $17855662.00 \pm4225.60$ \\
evtWeight totalWeight & $168533115.00 \pm 12982.03$ & $1361029.35 \pm 369.81$ & $123964861.46 \pm 20139.85$ & $319199.63 \pm 162.58$ & $224483.42 \pm 139.08$ \\
MET $>$ 30.0 & $102616082.00 \pm 10129.96$ & $1110623.03 \pm 333.89$ & $87453318.83 \pm 17117.00$ & $245953.66 \pm 142.70$ & $161221.98 \pm 118.48$ \\
Size(JET) $>$ 3 & $2705945.00 \pm 1644.98$ & $756028.28 \pm 275.36$ & $407749.02 \pm 1132.61$ & $66288.54 \pm 73.90$ & $27946.08 \pm 45.96$ \\
Pt(goodLepts) descend & $2705945.00 \pm 1644.98$ & $756028.28 \pm 275.36$ & $407749.02 \pm 1132.61$ & $66288.54 \pm 73.90$ & $27946.08 \pm 45.96$ \\
trigE $==$ 1 OR trigM $==$ 1 & $2705945.00 \pm 1644.98$ & $756028.28 \pm 275.36$ & $407749.02 \pm 1132.61$ & $66288.54 \pm 73.90$ & $27946.08 \pm 45.96$ \\
Size(goodLepts) $==$ 1 & $2075687.00 \pm 1440.72$ & $678955.32 \pm 261.14$ & $333729.85 \pm 1025.98$ & $59530.56 \pm 70.07$ & $25082.07 \pm 43.25$ \\
Size(goodJets) $\geq$ 4 & $693887.00 \pm 833.00$ & $397519.16 \pm 199.89$ & $36538.38 \pm 338.83$ & $25502.44 \pm 45.95$ & $9703.39 \pm 25.08$ \\
Size(goodBJets) $\geq$ 2 & $178807.00 \pm 422.86$ & $162208.51 \pm 130.98$ & $571.67 \pm 42.94$ & $7469.74 \pm 25.65$ & $233.09 \pm 3.58$ \\
MTW $>$ 30.0 & $150502.00 \pm 387.95$ & $136844.74 \pm 120.31$ & $461.10 \pm 38.88$ & $6218.46 \pm 23.42$ & $195.23 \pm 3.33$ \\
Ptjjj $\sim=$ 99999 & $150502.00 \pm 387.95$ & $136844.74 \pm 120.31$ & $461.10 \pm 38.88$ & $6218.46 \pm 23.42$ & $195.23 \pm 3.33$ \\
\hline
\end{tabular}
}
\label{tab:cutflow_TTbar}
\end{center}
\end{table}

\begin{table}[!ht]
\begin{center}
\caption{Event yields after $t\bar{t} \rightarrow W^+W^- b\bar{b} \rightarrow \ell \nu q\bar{q}' b\bar{b}$ channel selections. }
\begin{tabular}{|l|c|c|}
\hline
Sample & ADL/CutLang & Opendata Framework \\ \hline \hline
Data & $150502.00 \pm 387.95 $ & $150500.00$ \\
$t\bar{t}$ &$ 136844.62 \pm 120.08  $ & $136837.56$  \\
$V + jets$ & $461.10 \pm 38.11 $ & $461.10$ \\
Single top& $6218.46 \pm  23.30$ & $6218.13$ \\
Diboson & $195.23 \pm 3.29$ & $195.17$ \\ \hline
\end{tabular}
\label{tab:yields_TTbaryields}
\end{center}
\end{table}

\clearpage

\subsection{Standard model WZ diboson production in the three lepton final states} \label{section_WZ}
The analysis is designed to apply selection criteria for the leptonic decays of $W^{\pm} Z \to \ell \nu \ell \bar{\ell}$, where the Z boson decays into a pair of leptons, and the W boson decays into a charged lepton and a neutrino. In this process, $\ell$ and $\bar{\ell}$ can be either electrons or muons, resulting in three leptons in the final state.
\\
In the analysis, both the dilepton invariant mass, as defined in the {\tt ZBosonAnalysis}, and the transverse mass, as in the {\tt WBosonAnalysis}, are used as event variables. {\tt Lepton1} and {\tt Lepton2} correspond to the charged leptons originating from the decay of the $Z$ boson, while {\tt Lepton3} represents the charged lepton from the leptonic decay of the $W$ boson. These three leptons, selected from the {\tt goodLepts} collection, are used to reconstruct the events. {\tt Lepton1} and {\tt Lepton2} are paired together to form the {\tt Lepton12}, representing the $Z$ boson candidate, while {\tt Lepton3} is associated with the $W$ boson candidate. 
The transverse mass {\tt MTW} is then calculated using {\tt Lepton3} and the {\tt MET} to characterize the $W$ boson decay.

\begin{courier} \begin{lstlisting}
define Lepton1 : goodLepts[-1]
define Lepton2 : goodLepts[-2]
define Lepton3 : goodLepts[-3]
define Lepton12 : Lepton1 Lepton2
define MTW = sqrt(2*Pt(Lepton3)*MET*(1 - cos(dPhi(Lepton3, METLV[0]))))
\end{lstlisting} \end{courier}

The indices of the leptons forming the Z and W bosons are not known a priori; therefore, as in the previous analysis, negative indices are used to refer to these leptons. The negative indices indicate that the specific leptons will be selected after applying an optimization procedure in each event, allowing the combination of particles that best satisfies the reconstruction criteria for the W and Z boson decays. 
The indices of these leptons can be determined through a $\chi^2$ optimization algorithm within a specific goal. 
The optimization parameter, {\tt Chi2WZ} for the reconstruction of W and Z bosons is defined as follows:
\begin{courier} \begin{lstlisting}
define Chi2WZ : (m(Lepton12) - 91.18)^2 
               + (9999*pdgID(Lepton12))^2 
               + 0*MTW
\end{lstlisting} \end{courier}

The first term, \texttt{(m(Lepton12) - 91.18))$^2$ }, ensures that the invariant mass of the lepton pair is optimized to match the known mass of the $Z$ boson, 91.18 GeV. 
As the invariant mass of the lepton pair approaches the $Z$ boson mass, the contribution from this term decreases, indicating a better $Z$ boson candidate.

The second term, \texttt{pdgID(Lepton12)}, ensures that the lepton pair forming the Z boson candidate belongs to the same lepton family and constitutes a particle-antiparticle pair.
The $\tt pdgID$ function calculates the Particle Data Group (PDG) particle identification index of the pair as the sum of the individual pdgIDs: {\tt pdgID(Lepton12) = pdgID(Lepton1) + pdgID(Lepton2)}. For a valid Z boson decay, this value must be 0, indicating a pair of leptons with opposite electric charges and the same flavor (e.g., two electrons or two muons). A large factor of 9999 strongly emphasizes selecting such pairs, helping the algorithm focus on the correct lepton pair for the Z boson decay.

The last term, {\tt 0*MTW}, does not contribute to the {\tt Chi2WZ} value directly, but helps in identifying the third unknown lepton index, as it becomes the remaining lepton in the event.

Event selection for identifying WZ di-boson events is defined as follows:
\begin{courier}\begin{lstlisting}
region WZDiBosonAnalysis
  preselections
  select trigE == 1 OR trigM == 1
  select Size(goodLepts) == 3
  select Chi2WZ ~= 0
  select pdgID(Lepton12) == 0
  select Pt(goodLepts) > 25
  select Abs(m(Lepton12) - 91.18) < 10.0
  select MET > 30.0
  select MTW  > 30.0
\end{lstlisting} \end{courier}

\begin{figure}[!ht]
\centering
\includegraphics[width=0.40\textwidth]{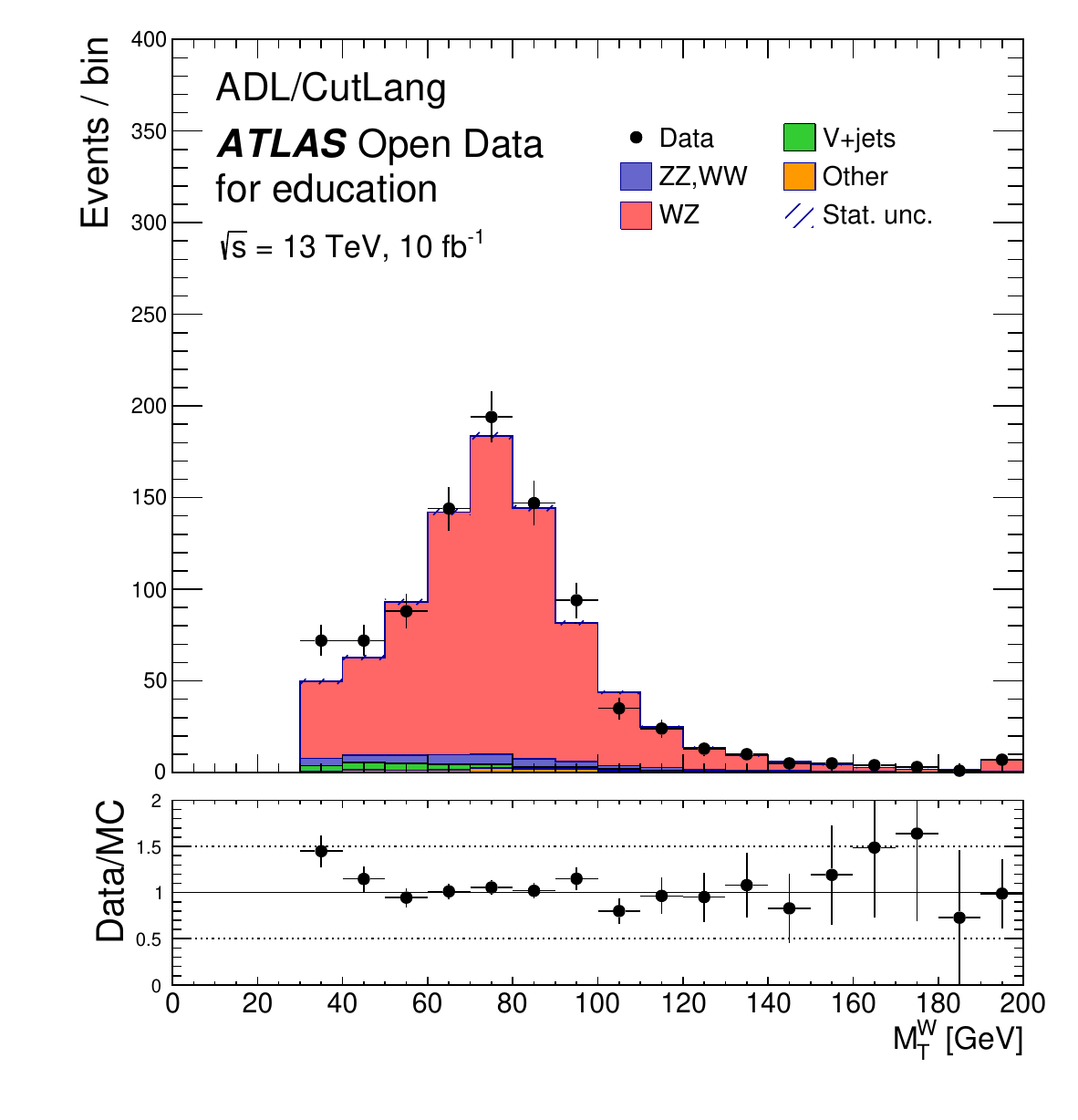}
\includegraphics[width=0.40\textwidth]{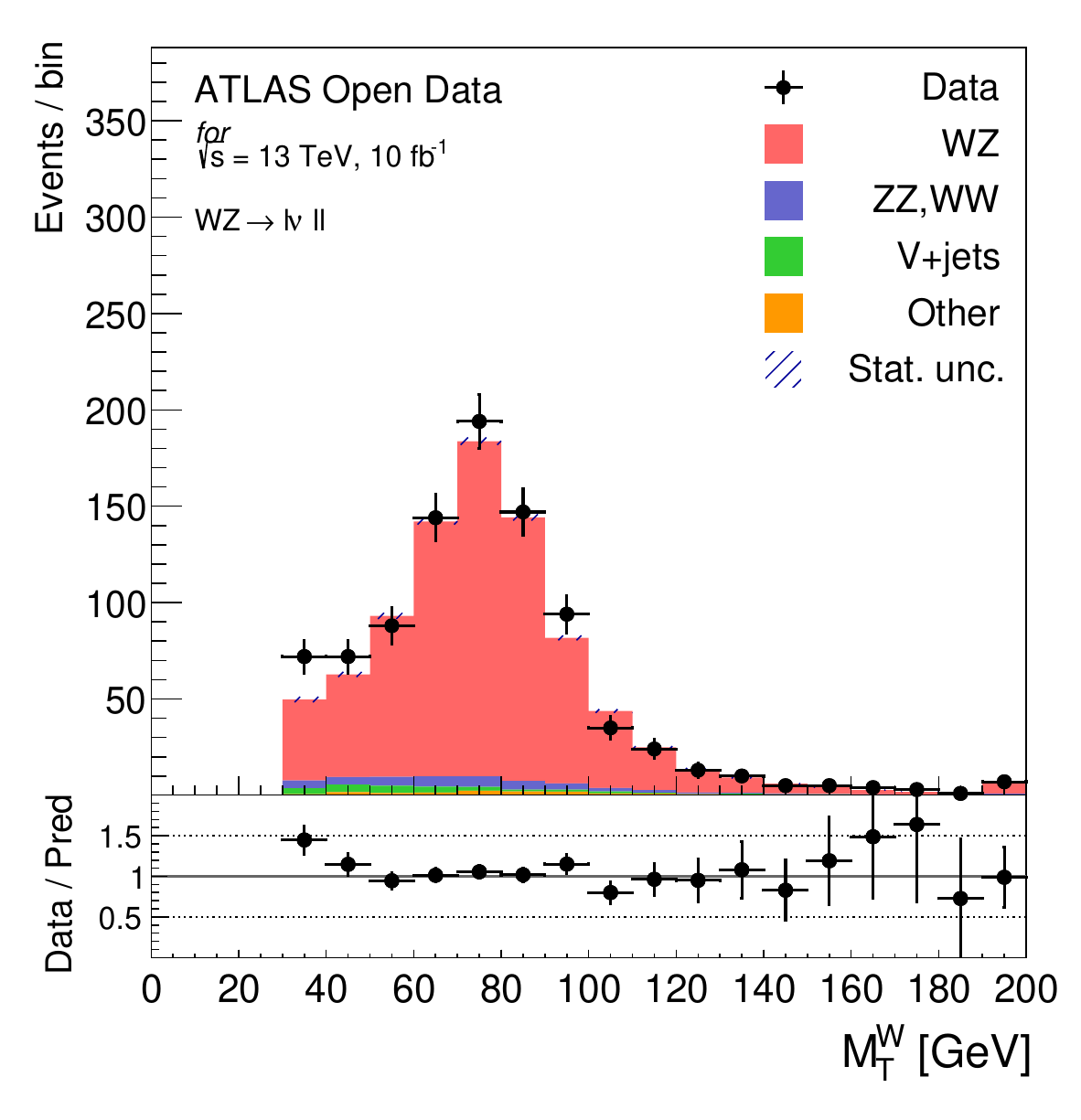}
\includegraphics[width=0.40\textwidth]{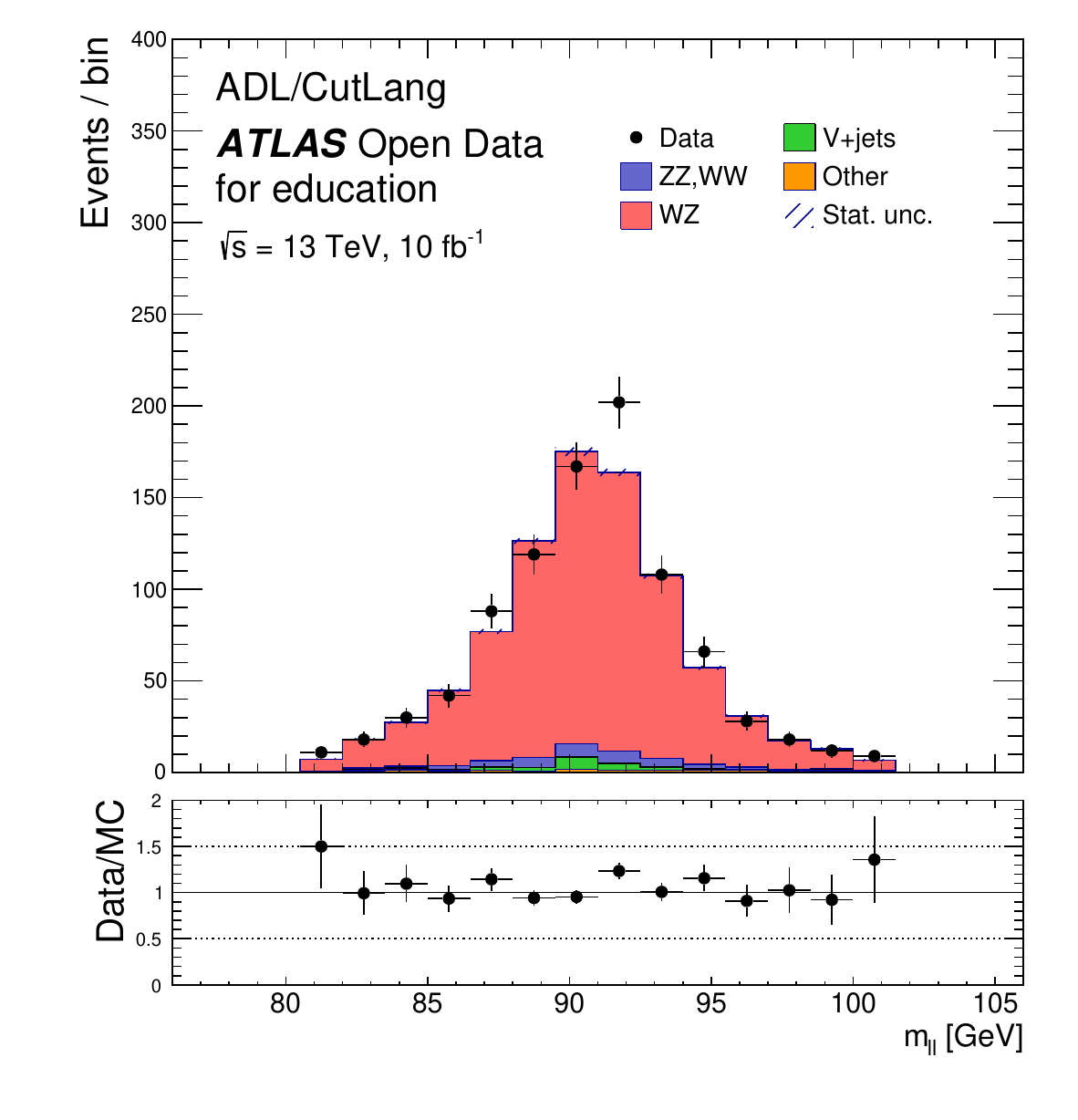}
\includegraphics[width=0.40\textwidth]{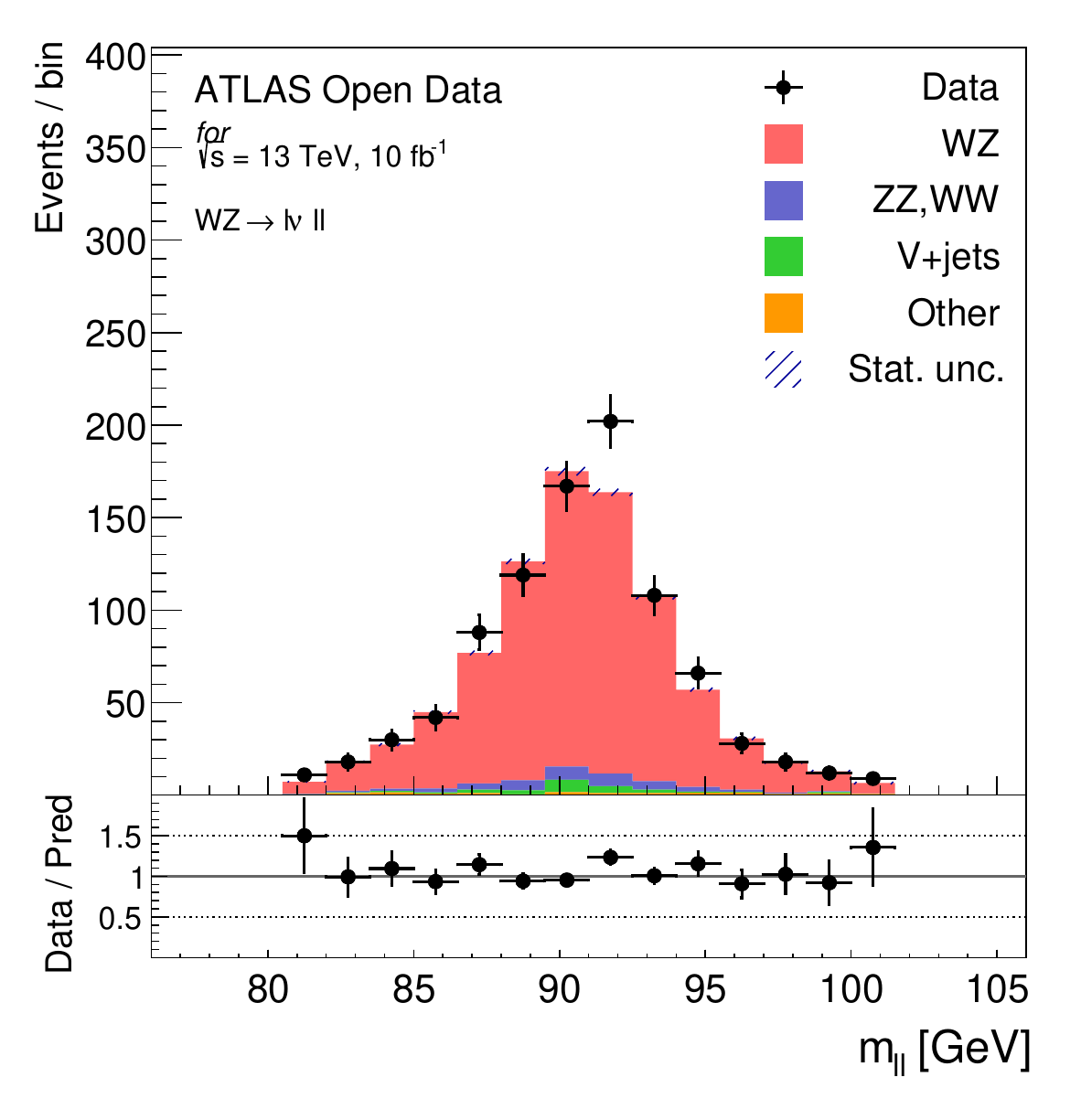}
\caption{Comparison of the transverse mass of the W boson ($M_T^W$, top) and the invariant mass of the lepton pair ($m_{l^+l^-}$, bottom) distributions obtained with the ADL/CutLang (left) and Open Data (right) frameworks after all event selections. The lower pad in each plot represents the Data/MC ratio. The last bin in each plot includes the overflow.}
\label{fig:CL_vs_OD_WZ_masses}
\end{figure}

The first three selections check that the event has been triggered by the relevant lepton type, that exactly three leptons are present in the event, and each of these leptons satisfies a loose lepton $p_T$ requirement and tight lepton identification criteria. Next, the optimization rule is applied next with \texttt{select Chi2WZ $\sim=$ 0}. By minimizing the $\chi^2$, the indices of all leptons are determined. Instead of imposing a limit on the $\chi^2$, the following selection ensures that the lepton pair forming the Z boson candidate has opposite sign  and same flavor.
The final selection focuses on the $W$ and $Z$ bosons, applying slightly looser selection criteria compared to previous analyses. Figure \ref{fig:CL_vs_OD_WZ_masses} shows the transverse mass of the $W$ candidate and the mass of the $Z$ candidate ($m_{\ell^+\ell^-}$) distributions from ADL/CutLang and Open Data frameworks. Additionally, a cutflow of event counts and the event yields for each sample are presented in Tables \ref{tab:cutflow_WZ} and \ref{tab:yields_WZyields}, respectively. The integral of $m_{\ell^+\ell^-}$ distribution was used to obtain the event yields from both frameworks. 
These results demonstrate that ADL/CutLang is in excellent agreement with the Open Data framework in terms of weighted event yields, distribution shapes, and the data to MonteCarlo ratio, confirming that the same analysis algorithm can be successfully reproduced using ADL.

\begin{table}[!ht]
\begin{center}
\caption{Cutflow of event counts for each sample in the $W^{\pm}Z \rightarrow \ell\nu \ell \ell'$ channel.}
\resizebox{\textwidth}{!}{
\begin{tabular}{|l|c|c|c|c|c|}
\hline
Selection & Data & WZ & ZZ,WW & $V+jets$ & Other \\ \hline 

ALL & $73490.00 \pm 271.09$ & $1415266.00 \pm 1189.65$ & $1350029.00 \pm 1161.91$ & $241168.00 \pm 491.09$ & $83680.00 \pm 289.27$ \\
evtweigt totalWeight & $73490.00 \pm 271.09$ & $2918.90 \pm 10.48$ & $904.83 \pm 3.16$ & $51474.74 \pm 144.54$ & $6836.74 \pm 25.58$ \\
trigE == 1 OR trigM == 1 & $73490.00 \pm 271.09$ & $2918.90 \pm 10.48$ & $904.83 \pm 3.16$ & $51474.74 \pm 144.54$ & $6836.74 \pm 25.58$ \\
MET $>$ 30.0 & $32720.00 \pm 180.89$ & $2148.94 \pm 9.70$ & $342.31 \pm 2.63$ & $16045.65 \pm 91.26$ & $5938.14 \pm 23.84$ \\
Size(goodLepts) $==$ 3 & $1466.00 \pm 38.29$ & $1050.73 \pm 3.96$ & $69.13 \pm 0.55$ & $73.91 \pm 4.71$ & $109.63 \pm 3.22$ \\
Pt(goodLepts) descend & $1466.00 \pm 38.29$ & $1050.73 \pm 3.96$ & $69.13 \pm 0.55$ & $73.91 \pm 4.71$ & $109.63 \pm 3.22$ \\
Chi2WZ $\sim =$ 0 & $1466.00 \pm 38.29$ & $1050.73 \pm 3.96$ & $69.13 \pm 0.55$ & $73.91 \pm 4.71$ & $109.63 \pm 3.22$ \\
pdgID(Lepton12) $==$ 0 & $1418.00 \pm 37.66$ & $1043.29 \pm 3.94$ & $68.45 \pm 0.55$ & $73.33 \pm 4.70$ & $79.84 \pm 2.74$ \\
Pt(goodLepts) $>$ 25.0 & $1418.00 \pm 37.66$ & $1043.29 \pm 3.94$ & $68.45 \pm 0.55$ & $73.33 \pm 4.70$ & $79.84 \pm 2.74$ \\
Abs(m(Lepton12) - 91.18) $<$ 10.0 & $1086.00 \pm 32.95$ & $902.54 \pm 3.68$ & $55.30 \pm 0.50$ & $57.03 \pm 4.11$ & $17.84 \pm 1.30$ \\
MTW $>$ 30.0 & $918.00 \pm 30.30$ & $799.43 \pm 3.46$ & $37.33 \pm 0.35$ & $19.91 \pm 2.36$ & $15.41 \pm 1.20$ \\

\hline
\end{tabular}
}
\label{tab:cutflow_WZ}
\end{center}
\end{table}

\begin{table}[H]
\begin{center}
\caption{Event yields after all selections for each sample in the $W^{\pm}Z \rightarrow \ell\nu \ell \ell'$ channel.}
\begin{tabular}{|l|c|c|}
\hline
Sample & ADL/CutLang & Opendata Framework \\ \hline \hline
Data & $918.00 \pm 30.30 $ & 918 \\
$WZ$ & $799.43 \pm  3.45$ & 799.43 \\
$ZZ,WW$ & $37.33 \pm  0.35$ & 37.33 \\
$V+jets$ & $19.91 \pm  2.36$ & 19.91 \\
Other & $15.41 \pm 1.20$ & 15.41 \\  
\hline
\end{tabular}
\label{tab:yields_WZyields}
\end{center}
\end{table}

\subsection{Standard model Higgs boson production in the 
\texorpdfstring{$H \rightarrow ZZ^*$}{H → ZZ*} decay}  \label{section_HZZ}
Standard Model Higgs boson production in the four lepton final states is another analysis that benefits from CutLang's optimization function. In this search, $H \rightarrow ZZ^* \rightarrow 4\ell$ ($\ell = e$ or $ \mu$) decay channel is investigated. \newline\newline
This analysis requires two SFOS (same flavor, opposite sign) lepton pairs to reconstruct the Higgs boson. To select the best Z candidates, the mass optimization terms for leading and subleading SFOS pairs are defined as: $\chi^2_{Z_1} = |m_{\ell\ell^1} - m_Z|$  and $\chi^2_{Z_2} = |m_{\ell\ell^2} - m_Z|$, where $m_{\ell\ell}$ represents the invariant mass of the reconstructed Z bosons, and $m_Z$ is the nominal $Z$ boson mass. The steps of reconstructing the best $Z$ candidates are summarized as follows:

\begin{itemize}
    \item Any combination of four good leptons is selected. The four-momenta of two SFOS leptons are used to reconstruct $Z_1$ candidate, and the remaining two are used to reconstruct $Z_2$ candidate. 
    \item $\chi^2_{Z_1}$ and $\chi^2_{Z_2}$ expression are calculated for every combination. The nominal Z boson mass is set as 91.18 GeV.
    \item The combination with the minimum $\chi^2_{Z_1}$ is selected as the best $Z_1$ candidate.
    \item After identifying the best $Z_1$, the combination of the remaining leptons giving the minimum $\chi^2_{Z_2}$ is selected as the best $Z_2$ candidate.
\end{itemize}

Due to the nature of ADL, the mass reconstruction algorithm is applied in a region block along with the other event selections. In addition to the standard object definitions used in previous analyses, the event variable definitions in ADL syntax for HZZ analysis are shown below:

\begin{courier} \begin{lstlisting}
define Lepton1 : goodLepts[-1]
define Lepton2 : goodLepts[-2]
define Lepton3 : goodLepts[-3]
define Lepton4 : goodLepts[-4]

define Z1 : Lepton1 Lepton2
define Z2 : Lepton3 Lepton4

define ChiZ1 : (m(Z1) - 91.18)^2 + (9999*pdgID(Z1))^2 
define ChiZ2 : (m(Z2) - 91.18)^2 + (9999*pdgID(Z2))^2  

\end{lstlisting} \end{courier}

In the ADL syntax of this analysis, \texttt{9999*pdgID(Z1))} and \texttt{9999*pdgID(Z2))} terms simply force the BestZ definitions to have large values if they do not satisfy SFOS requirement during the optimization process. In this way, only the best possible Z boson candidates are selected by CutLang's optimization mechanism, which is executed with "$\sim =$" operator.

It is important to note that CutLang does not perform implicit event selections. Therefore, all operations related to event selection must be explicitly defined within the region blocks. In cases where no suitable SFOS combination exists in an event, CutLang still computes and retains the combination with the minimum $\chi^2$ value. Such incorrect combinations must therefore be removed through appropriate selection criteria at the region level. One example is the use of explicit PDG ID requirements for reconstructed $Z$ candidates. In the region block, the relevant selection is applied as \texttt{select pdgID(Z1) == 0 AND pdgID(Z2) == 0}, ensuring that both $Z$ candidates are formed from valid SFOS lepton pairs. This effectively selects events corresponding to $ee,ee$, $ee,\mu\mu$, or $\mu\mu,\mu\mu$ final states. The “pdgID cut” selection in Table \ref{tab:cutflow_HZZ} refers to this operation. The event selections used in this analysis are shown below in ADL syntax:

\begin{courier} \begin{lstlisting}
region preselections
  select ALL
  weight evtWeight totalWeight
  select trigE == 1 OR trigM == 1
  select Size(goodLepts) == 4
  Sort Pt(goodLepts) descend

region HZZAnalysis
  preselections
  select Pt(Lepton1) > 25
  select Pt(Lepton2) > 15
  select Pt(Lepton3) > 10
  select ChiZ1 ~= 0
  select ChiZ2 ~= 0
  select pdgID(Z1) == 0 AND pdgID(Z2) == 0
\end{lstlisting} \end{courier}

\begin{figure}[!ht]
\centering
\includegraphics[width=0.40\textwidth]{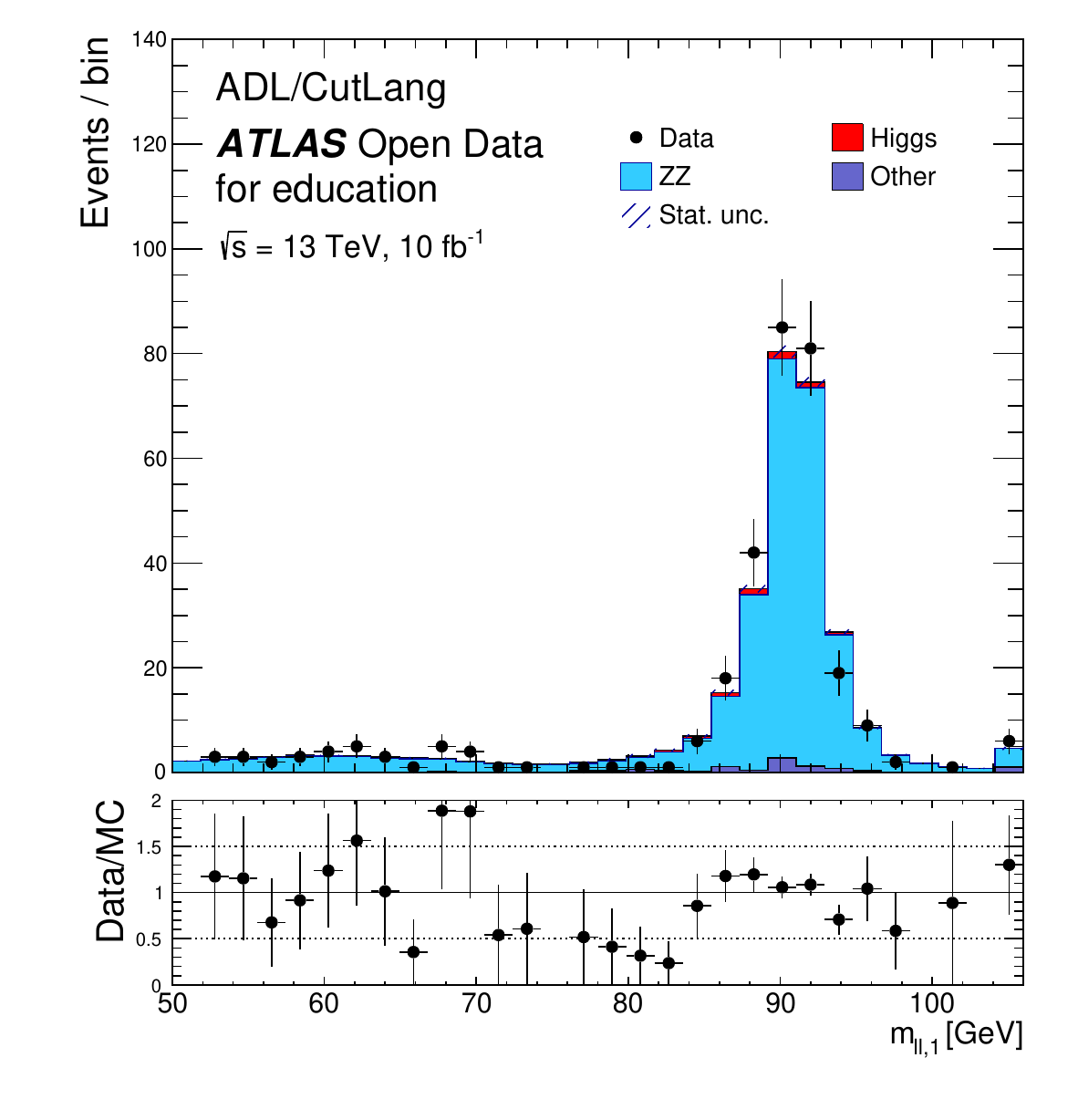}
\includegraphics[width=0.40\textwidth]{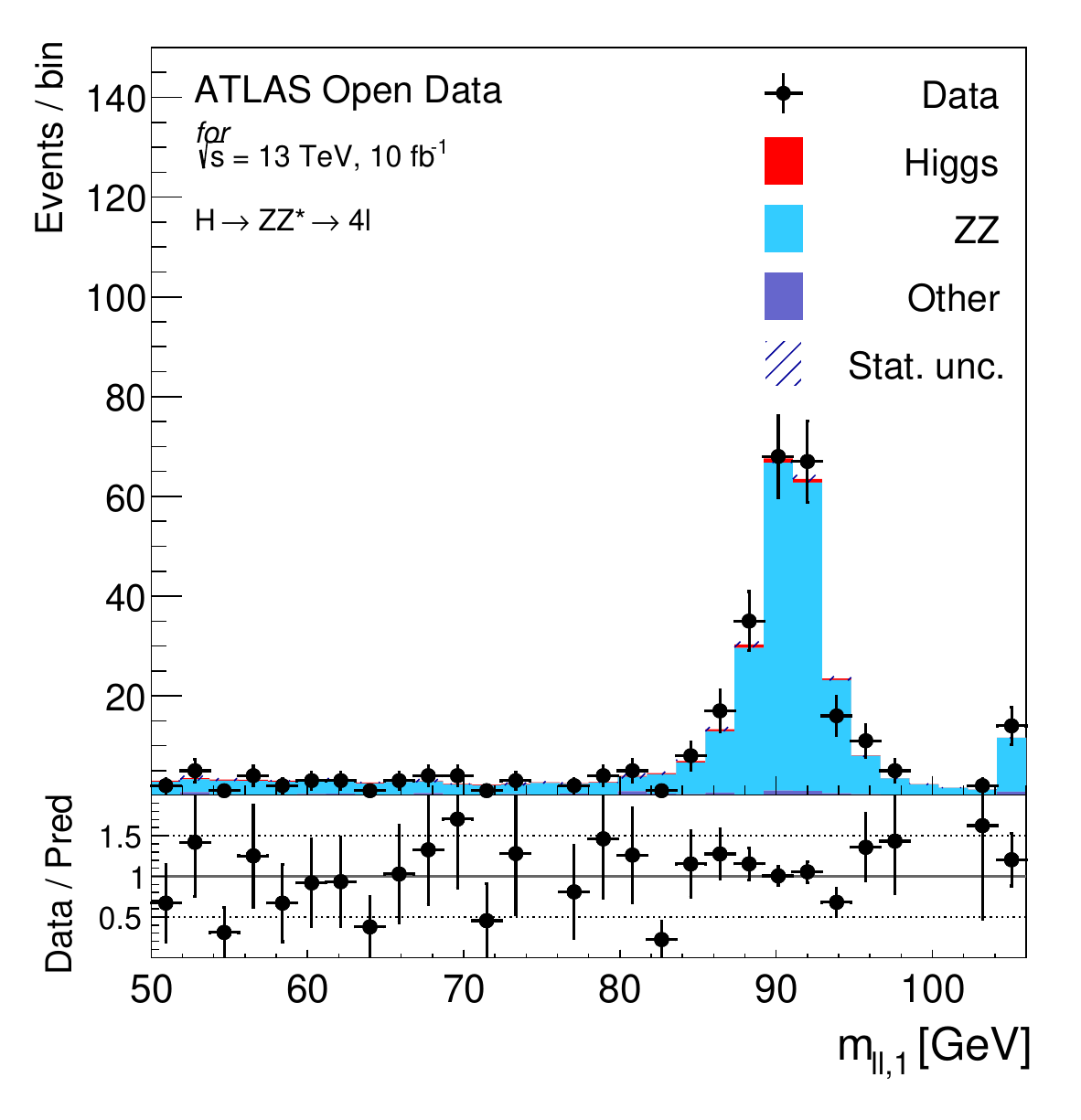}
\includegraphics[width=0.40\textwidth]{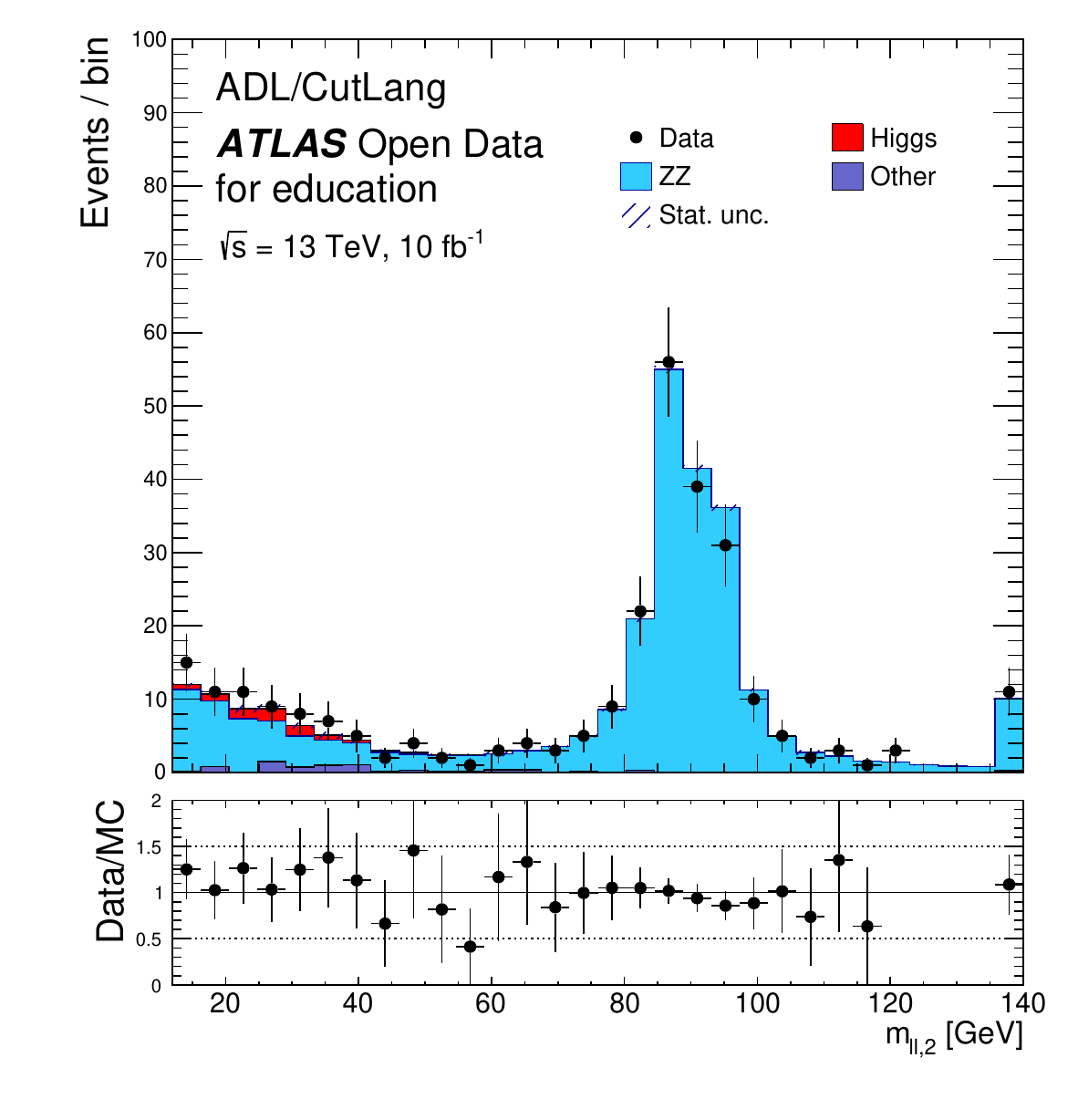}
\includegraphics[width=0.40\textwidth]{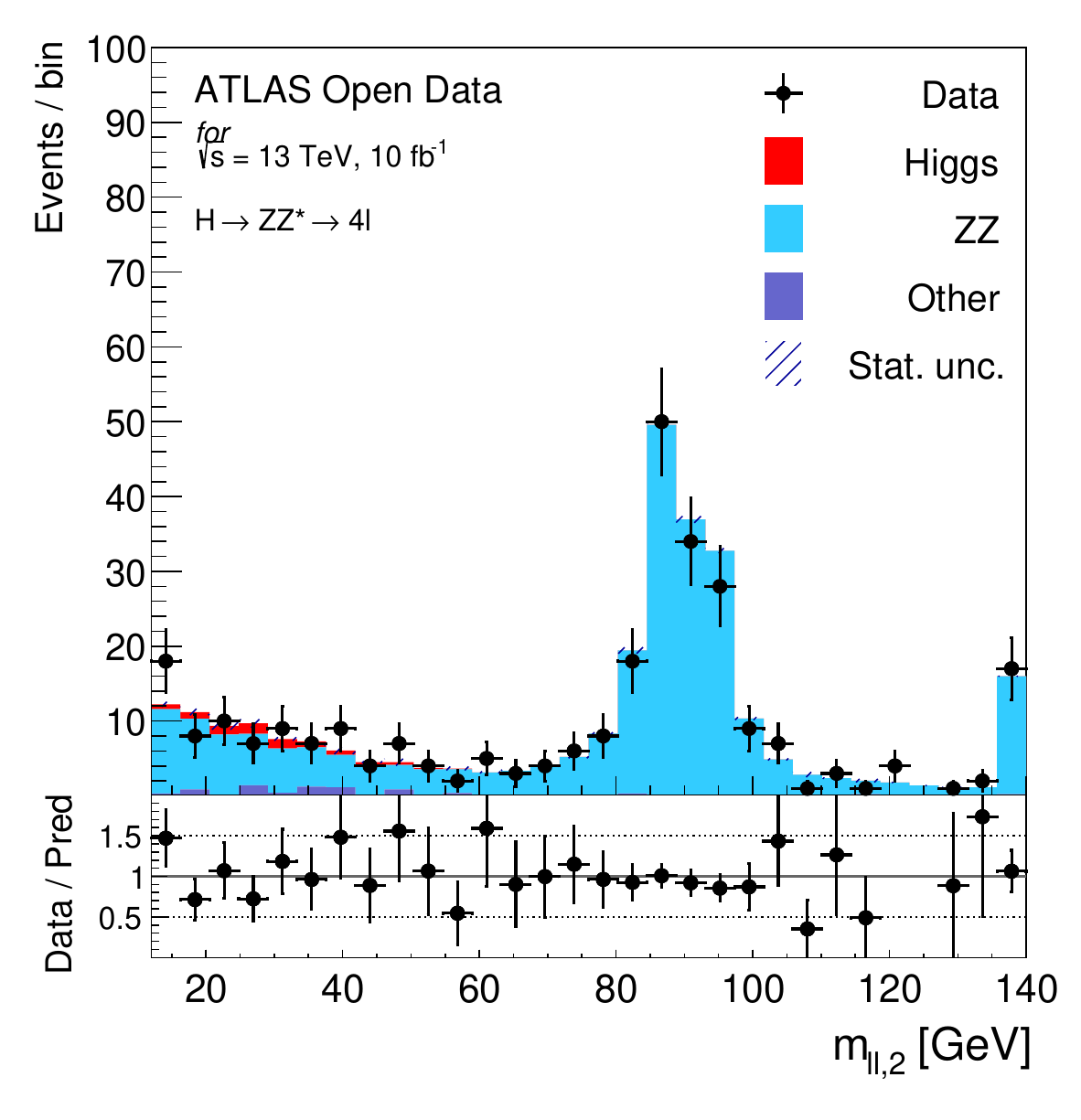}
\includegraphics[width=0.40\textwidth]{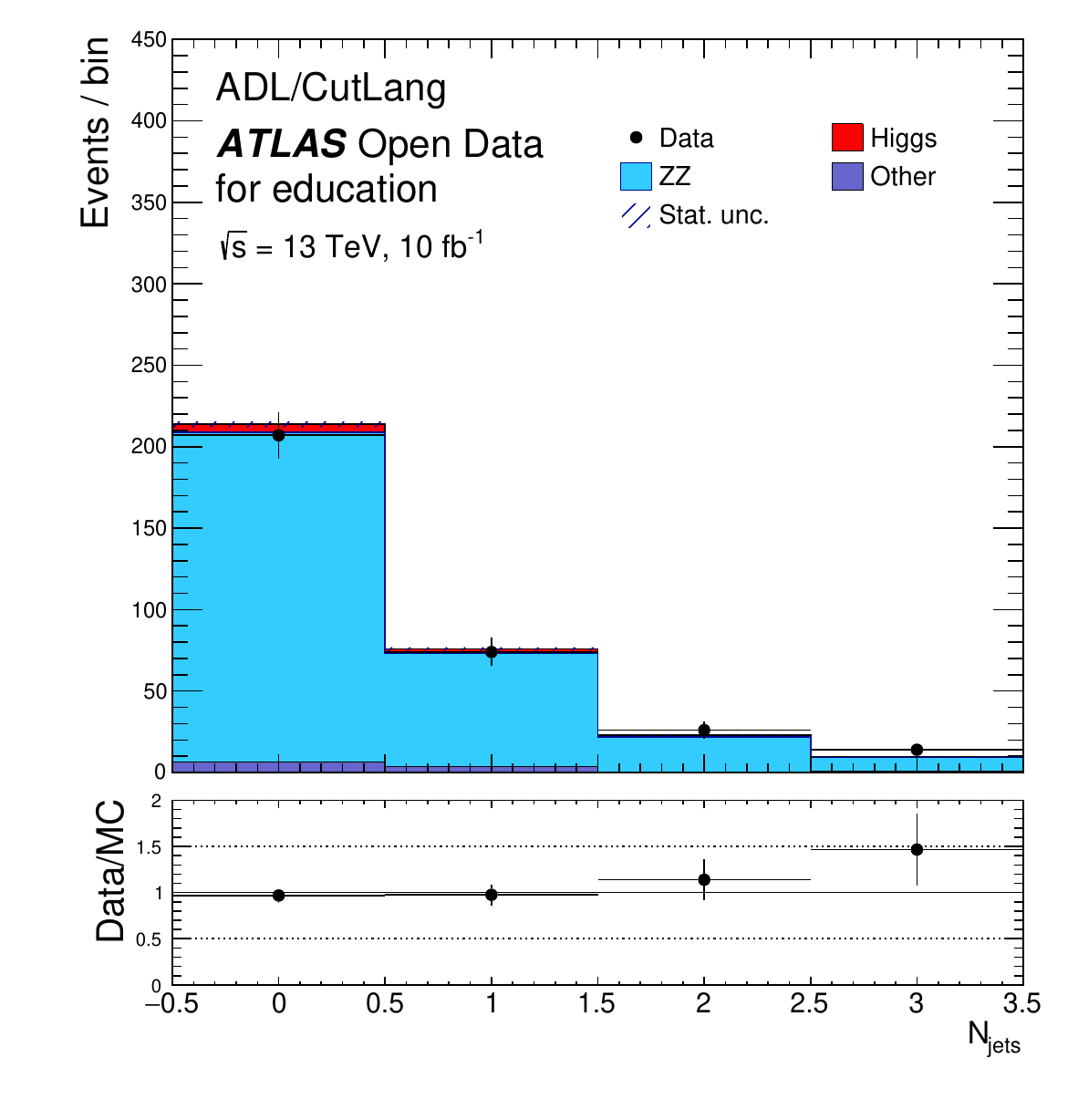}
\includegraphics[width=0.40\textwidth]{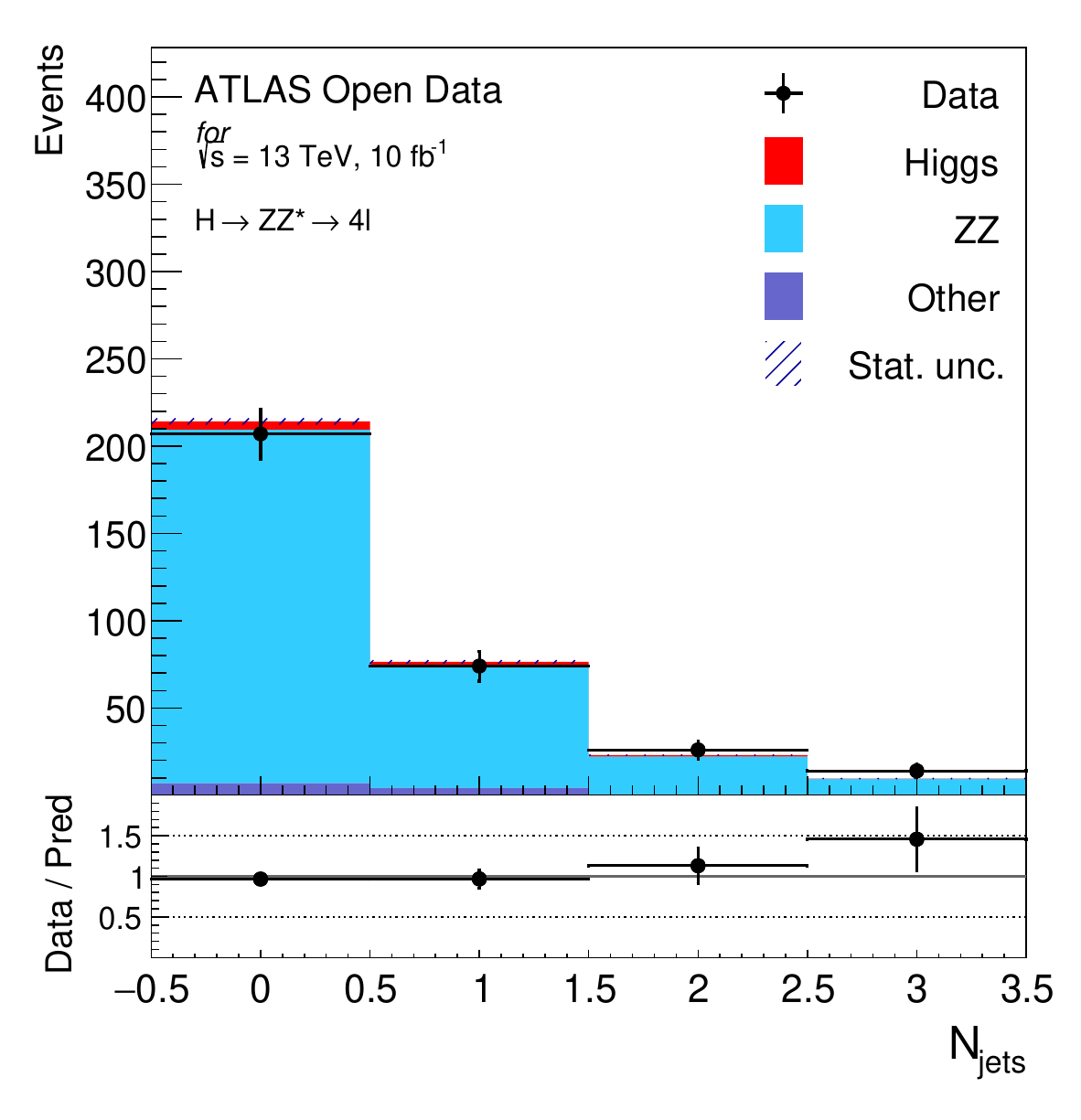}
\caption{Comparison of $m_{Z_1}$ (top) and $m_{Z_2}$ (middle) and the jet multiplicity (bottom) distributions obtained with the ADL/CutLang (left) and Open Data (right) frameworks after all event selections in the $H \rightarrow ZZ^* \rightarrow \ell^+\ell^- \ell'^+ \ell'^-$ channel. The lower pad in each plot represents the Data/MC ratio and the last bin in each plot includes the overflow.}
\label{fig:CL_vs_OD_HZZ_masses}
\end{figure}

The reconstructed $Z$ boson candidate masses and the number of jets in the event distributions from ADL/CutLang and the Open Data framework are shown in Figure \ref{fig:CL_vs_OD_HZZ_masses}. The mass distributions show slight differences between the two frameworks in terms of shape, and data-to-MC ratio. In particular, ADL/CutLang provides more events near the $Z$ boson mass window for both $Z_1$ and $Z_2$ candidates. This is because the optimization mechanism in CutLang is able to identify more candidates close to the nominal $Z$ boson mass compared to the Open Data framework. In contrast, the multiple independent \texttt{if} conditions used in the Open Data framework overwrite the earlier combinations during minimization, causing a portion of the final selected $Z$ masses to correspond to a non-optimal combinations. 
This example demonstrates the advantages of reusing a well-tested and optimized analysis software framework, enabling the analyst to focus on the physics while leaving computational details to the framework.

The Tables \ref{tab:cutflow_HZZ} and \ref{tab:yields_HZZ} present the event cutflow from CutLang and the comparison of event yields between the two frameworks, respectively. For a fair comparison of the weighted event yields after all selections, the integral of the jet multiplicity distribution is used in both frameworks. The results show that ADL/CutLang and the Open Data framework are in good agreement at the event selection level, with the remaining differences arising only in the distribution shapes due to the different mass minimization approaches.

\begin{table}[!ht]
\begin{center}
\caption{Cutflow of event counts for each sample in the $H \rightarrow ZZ^* \rightarrow \ell^+\ell^- \ell'^+ \ell'^-$ channel.}
\resizebox{0.85\textwidth}{!}{
\begin{tabular}{|l|c|c|c|c|}
\hline
Selection & Data & Higgs & ZZ & Other \\ \hline \hline
ALL & $832.0 \pm 28.8444$ & $385706.0 \pm 621.0523$ & $554670.0 \pm 744.7617$ & $12134.0 \pm 110.1544$ \\
evtWeight totalWeight & $832.0 \pm 28.844$ & $10.0118 \pm 0.0239$ & $282.3140 \pm 0.9944$ & $420.6022 \pm 9.9079$ \\
trigE OR trigM & $832.0 \pm 28.844$ & $10.0118 \pm 0.0239$ & $282.3140 \pm 0.9944$ & $420.6022 \pm 9.9079$ \\
Size(goodLepts) $==$ 4 & $416.0 \pm 20.3961$ & $8.9235 \pm 0.0226$ & $253.9507 \pm 0.9400$ & $41.3730 \pm 3.0085$ \\
Pt(goodLepts) descend & $416.0 \pm 20.3961$ & $8.9235 \pm 0.0226$ & $253.9507 \pm 0.9400$ & $41.3730 \pm 3.0085$ \\
Pt(Lepton1) $>$ 25 & $416.0 \pm 20.3961$ & $8.9235 \pm 0.0226$ & $253.9507 \pm 0.9400$ & $41.3730 \pm 3.0085$ \\
Pt(Lepton2) $>$ 15 & $409.0 \pm 20.2237$ & $8.8587 \pm 0.0225$ & $249.8708 \pm 0.9341$ & $40.2460 \pm 2.9929$ \\
Pt(Lepton3) $>$ 10 & $380.0 \pm 19.4936$ & $8.6819 \pm 0.0223$ & $242.5931 \pm 0.9234$ & $29.8562 \pm 2.5831$ \\
$\chi^2_{Z_1} \sim= 0$ & $380.0 \pm 19.4936$ & $8.6819 \pm 0.0223$ & $29.8562 \pm 2.5831$ & $242.5931 \pm 0.9234$ \\
$\chi^2_{Z_2} \sim= 0$ & $380.0 \pm 19.4936$ & $8.6819 \pm 0.0223$ & $29.8562 \pm 2.5831$ & $242.5931 \pm 0.9234$ \\
pdgID cut & $321.0 \pm 17.9165$ & $8.4996 \pm 0.0221$ & $11.1041 \pm 1.6848$ & $232.5621 \pm 0.9066$ \\

\hline
\end{tabular}
}
\label{tab:cutflow_HZZ}
\end{center}
\end{table}

\begin{table}[H]
\begin{center}
\caption{Event yields after all the selections for each sample in the $H \rightarrow ZZ^* \rightarrow \ell^+\ell^- \ell'^+ \ell'^-$ channel.}
\begin{tabular}{|l|c|c|}
\hline
Sample & ADL/CutLang & Opendata Framework \\ \hline \hline
Data & $321.0 \pm 17.92 $ & $321.0$ \\
Higgs & $ 8.50 \pm 0.02 $ & $8.50$ \\
ZZ & $ 302.38 \pm 1.18 $ & $302.38$ \\
Other & $ 11.10 \pm 1.68 $ & $12.29$ \\ \hline
\end{tabular}
\label{tab:yields_HZZ}
\end{center}
\end{table}


\subsection{SM \texorpdfstring{$ZZ$}{ZZ} diboson Production}\label{section_ZZ}
The analysis of $ZZ$ diboson production is crucial for testing the electroweak sector of the Standard Model. This process specifically involves the direct production of $ZZ$ boson pairs in proton-proton collisions, characterized by the decay of each $Z$ boson into lepton pairs. 

The analysis focuses on applying selection criteria to $ZZ$ diboson events in which both $Z$ bosons decay leptonically. For this analysis, a reconstruction procedure similar to that described in Section \ref{section_HZZ} is followed. Overall, the object and event variable definitions are the same as in the previous analysis, with one addition: $ZZ$ diboson analysis includes an additional event variable, defined as \texttt{define ZZsum : abs(m(Z1) - 91.18) + abs(m(Z2) - 91.18)} in the ADL syntax. This definition represents the sum of two $Z$ mass window deviations, and the actual cut on this variable is explicitly given in Table \ref{tab:cutflow_ZZ}. The event selections for this analysis in ADL syntax are shown below. They use the same \texttt{preselections} definitions in \ref{section_HZZ}.

\begin{courier} 
\begin{lstlisting}
region ZZDibosonAnalysis
  preselections
  select Chi2Z1 ~= 0
  select Chi2Z2 ~= 0
  select pdgID(Z1) == 0 AND pdgID(Z2) == 0
  select ZZsum < 50
\end{lstlisting} 
\end{courier}

\begin{figure}[!ht]
\centering
\includegraphics[width=0.40\textwidth]{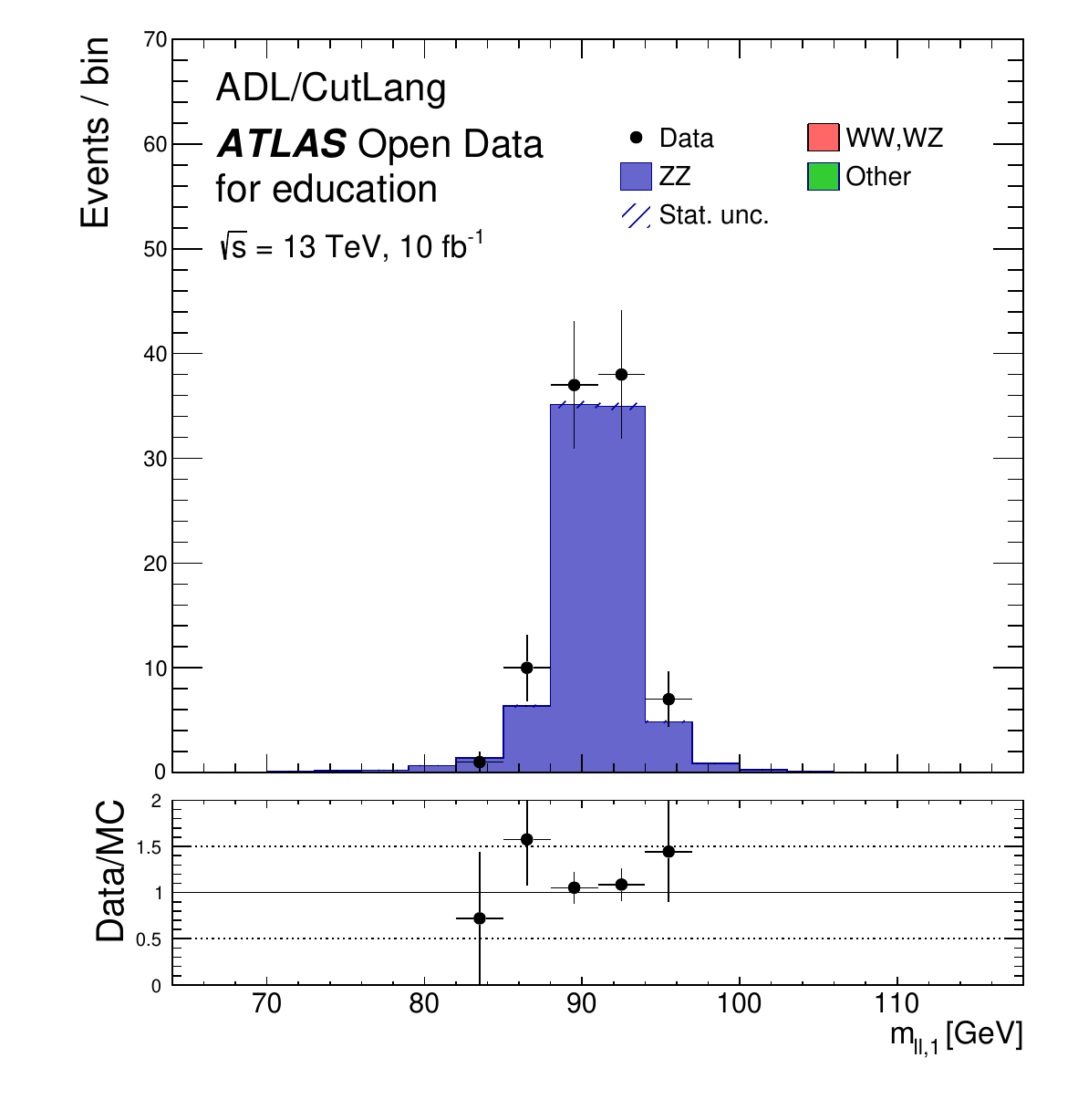}
\includegraphics[width=0.40\textwidth]{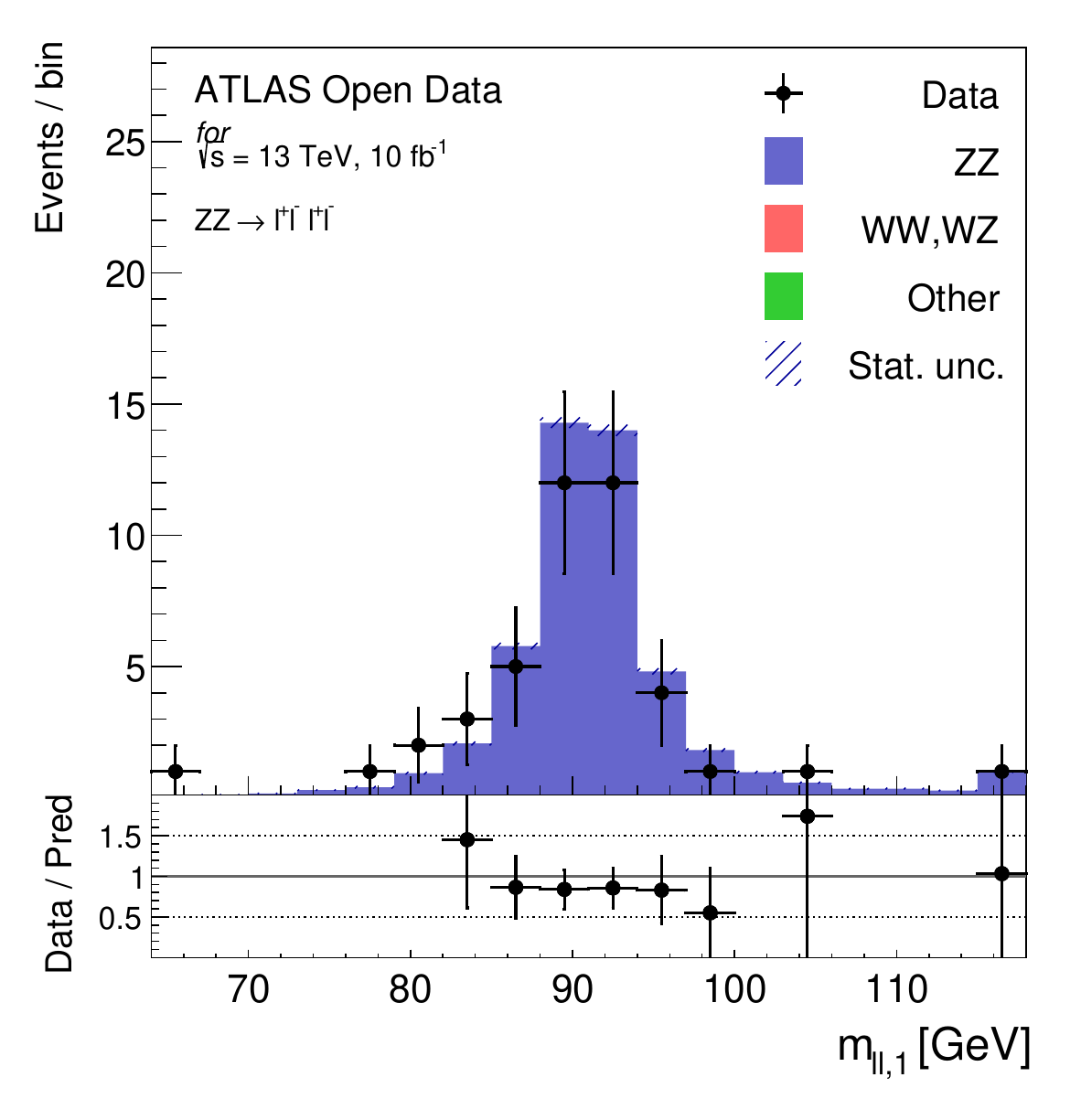}
\includegraphics[width=0.40\textwidth]{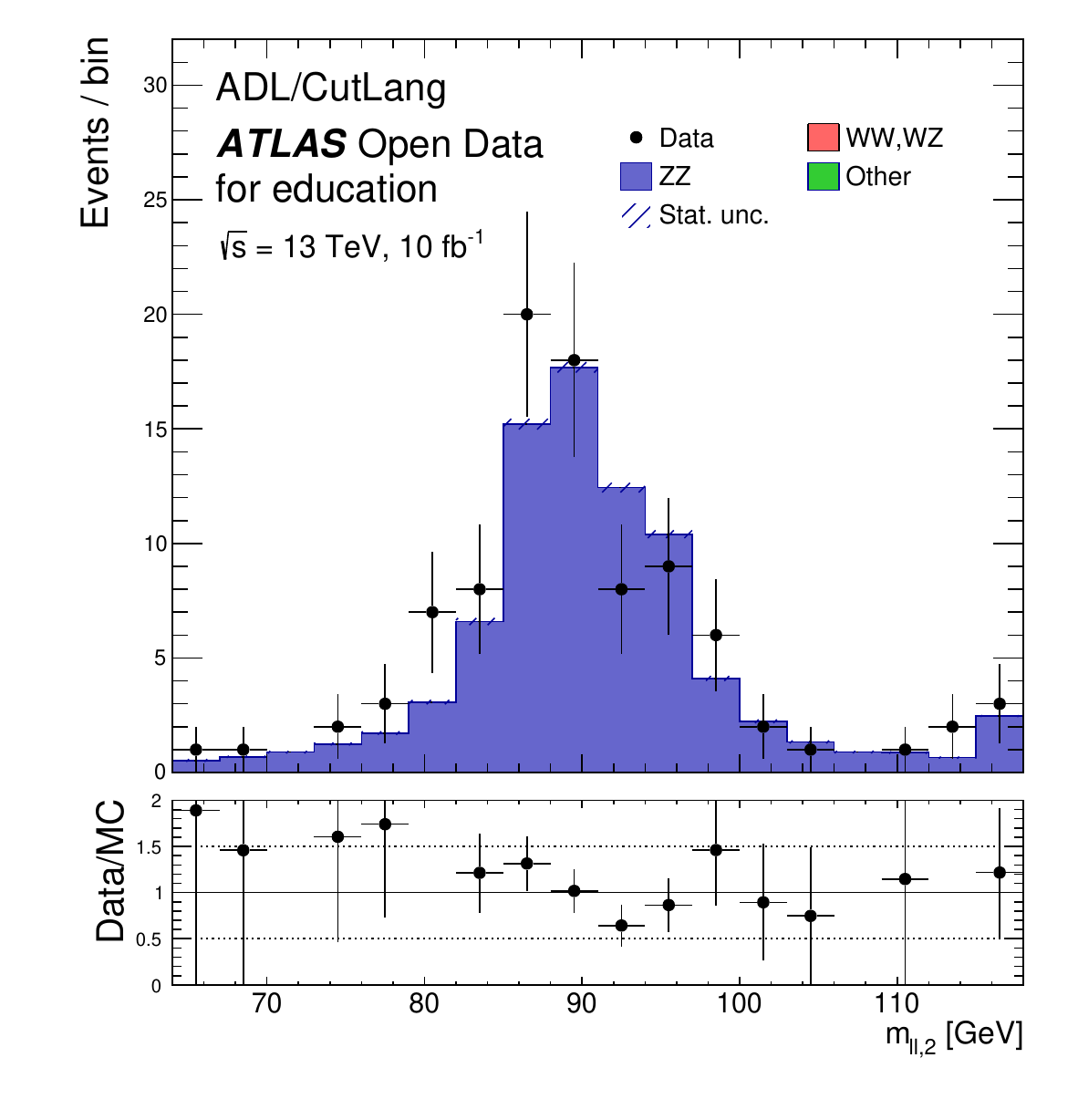}
\includegraphics[width=0.40\textwidth]{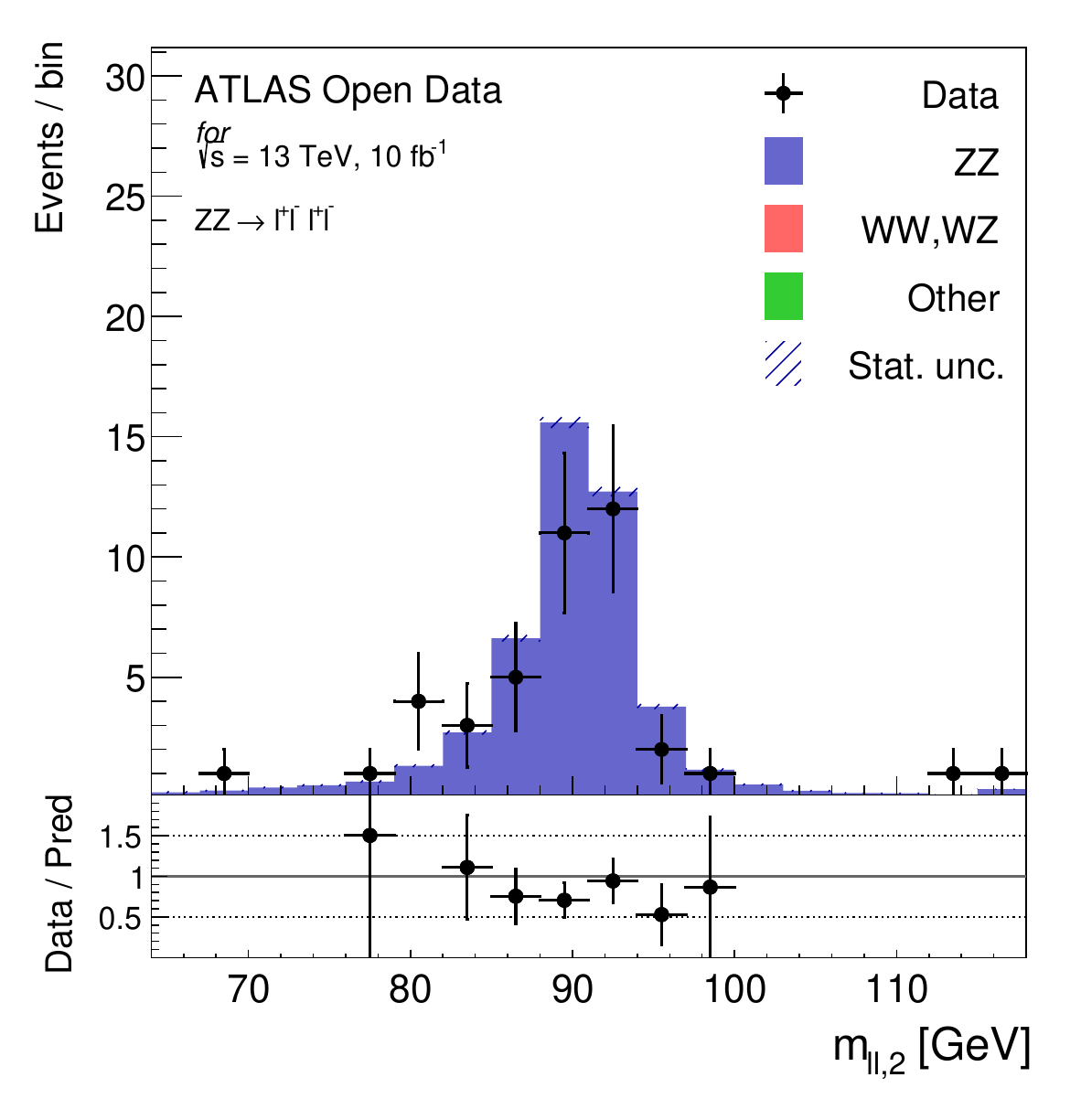}
\caption{Comparison of $m_{Z1}$ (top) and $m_{Z2}$ (bottom) distributions obtained with the ADL/CutLang (left) and the Open Data (right) frameworks in the $ZZ \rightarrow \ell^+\ell^- \ell'^+ \ell'^-$ channel. The lower pad in each plot represents the Data/MC ratio. The last bin in each plot includes the overflow.}
\label{fig:CL_vs_OD_ZZ_masses}
\end{figure}

The mass distributions of the best reconstructed $Z$ candidates from ADL/CutLang and the Open Data framework are shown in Figure \ref{fig:CL_vs_OD_ZZ_masses}. In general, this analysis exhibits slightly poorer data-to-MC agreement due to limited statistics. However, the results from ADL/CutLang show that about two times more events near the $Z$ boson mass with a better Data/MC ratio. 
As mentioned previously,  CutLang's optimization mechanism is able to select more $Z$ boson candidates than the Open Data framework's implementation with multiple independent \texttt{if} conditions, similar to the $H \rightarrow ZZ^*$ analysis. Table \ref{tab:cutflow_ZZ} shows the event cutflow from ADL/CutLang framework for each sample. The event yields, presented in Table \ref{tab:yields_ZZ} and obtained using the integral of the $m_{Z_1}$ distribution in both frameworks, show good agreement between ADL/CutLang and the Open Data framework, with differences of less than 5\%. These differences arise from the different mass minimization approaches used in the two frameworks, as discussed in the $H \rightarrow ZZ^*$ analysis. In particular, variations in the reconstruction of lepton pairs forming the $Z$ candidates affect the number of events passing the \texttt{ZZsum} $< 50$ GeV selection.

\begin{table}[!ht]
\begin{center}
\caption{Cutflow of event counts for each sample in the $ZZ \rightarrow \ell^+\ell^- \ell'^+ \ell'^-$ channel.}
\resizebox{\textwidth}{!}{
\begin{tabular}{|l|c|c|c|c|}
\hline
Selection & Data & $ZZ$  &  $WW,WZ$  & Other \\ \hline \hline
ALL &
$832.00 \pm 28.84$ &
$554670.00 \pm 744.76$ &
$9400.00 \pm 96.95$ &
$2734.00 \pm 52.29$ \\

evtWeight totalWeight &
$832.00 \pm 28.84$ &
$282.31 \pm 744.76$ &
$16.14 \pm 0.44$ &
$404.46 \pm 9.90$ \\

trigE == 1 OR trigM == 1 &
$832.00 \pm 28.84$ &
$282.31 \pm 744.76$ &
$16.14 \pm 0.44$ &
$404.46 \pm 9.90$ \\



Size(goodLepts) == 4 &
$109.00 \pm 10.44$ &
$94.81 \pm 0.55$ &
$0.11 \pm 0.03$ &
$0.00 \pm 0.00$ \\

Pt(goodLepts) descend &
$109.00 \pm 10.44$ &
$94.81 \pm 0.55$ &
$0.11 \pm 0.03$ &
$0.00 \pm 0.00$ \\

Chi2Z1 $\sim = 0$ &
$109.00 \pm 10.44$ &
$94.81 \pm 0.55$ &
$0.11 \pm 0.03$ &
$0.00 \pm 0.00$ \\

Chi2Z2 $\sim = 0$ &
$109.00 \pm 10.44$ &
$94.81 \pm 0.55$ &
$0.11 \pm 0.03$ &
$0.00 \pm 0.00$ \\

pdgID cut &
$103.00 \pm 10.15$ &
$94.81 \pm 0.55$ &
$0.11 \pm 0.03$ &
$0.00 \pm 0.00$ \\

ZZsum $< 50$ GeV &
$93.00 \pm 9.64$ &
$85.22 \pm 0.52$ &
$0.02 \pm 0.02$ &
$0.00 \pm 0.00$ \\
\hline
\end{tabular}
}
\label{tab:cutflow_ZZ}
\end{center}
\end{table}

\begin{table}[!ht]
\begin{center}
\caption{Event yields after all the selections for each sample in the $ZZ \rightarrow \ell^+\ell^- \ell'^+ \ell'^-$ channel.}
\begin{tabular}{|l|c|c|}
\hline
Sample & ADL/CutLang & Opendata Framework \\ \hline \hline
Data & $93.00 \pm 9.64 $ & $96.00$ \\
ZZ & $ 85.22 \pm 0.52 $ & $88.90$ \\
WWZ & $ 0.02 \pm 0.03 $ &  $0.02$ \\
Other & $ 0.00 \pm 0.00 $ & $0.00$ \\ \hline
\end{tabular}
\label{tab:yields_ZZ}
\end{center}
\end{table}

\clearpage

\subsection{Search for direct production of pairs of sleptons \texorpdfstring{$\tilde{\ell}$}{\~{l}}} \label{section_susy}
The analysis focuses on a search for the direct production of pairs of slepton pairs (denoted as $\tilde{\ell} \tilde{\ell}$), which are the superpartners of the SM leptons in SUSY theories. In this process, each slepton decays into the lightest neutralino ($\chi^0_1$) and the corresponding SM lepton. The slepton and neutralino masses are assumed to be 600 GeV and 300 GeV, respectively. The final state therefore consists of two leptons and the MET, which arises from the undetected neutralinos.

The event variables are defined as follows. The stransverse mass $M_{T2}$ is computed, which is derived from the transverse momentum of the leptons and the MET, and serves as a key variable for this analysis. 
\begin{courier} \begin{lstlisting}    
define leadLept    : signalLepts[0]
define subleadLept : signalLepts[1]
define mLL         : m(leadLept subleadLept)
define MT2         : fMT2(leadLept, subleadLept, METLV[0])
\end{lstlisting} \end{courier}


For the analysis, three regions are defined: \texttt{preselections}, \texttt{SR2loose}, and \texttt{SR2tight}:

\noindent The \texttt{preselections} is the first step in the analysis, where basic condition criteria are applied to select relevant events. The event must pass the lepton trigger, ensuring that it contains at least one lepton; while the analysis ultimately requires exactly two leptons in the final state. The leading lepton (\texttt{leadLept}) ($p_T > 25$ GeV) and the subleading lepton (\texttt{subleadLept}) ($p_T > 20$ GeV) must satisfy minimum transverse momentum requirements, ensuring that both leptons are sufficiently energetic for further analysis. Events containing jets assigned to \texttt{JetsA}, \texttt{JetsB}, \texttt{JetsC}, or \texttt{JetsD} are rejected, as specified in the ADL code block of the \texttt{preselections} region. The object selection and overlap removal procedures are analysis-specific, and their full definitions are provided in the ADL file in Ref.~\cite{ADL4LHC_ATLAS13TeV}.

\texttt{SR2loose}, and \texttt{SR2tight} regions are subsequently defined to refine the event selection with increasing levels of strictness.

\begin{courier} \begin{lstlisting}
region preselections
  select ALL
  weight evtweight totalWeight
  select trigE == 1 OR trigM == 1
  select Size(signalLepts) == 2
  select Pt(leadLept) > 25
  select Pt(subleadLept) > 20
  select Size(JetsA) == 0
  select Size(JetsB) == 0
  select Size(JetsC) == 0
  select Size(JetsD) == 0

region SR2loose
  preselections
  select pdgID(leadLept) + pdgID(subleadLept) == 0
  select MT2 > 100
  select mLL > 111

region SR2tight
  preselections
  select pdgID(leadLept) + pdgID(subleadLept) == 0
  select MT2 > 130
  select mLL > 300
\end{lstlisting} \end{courier}

\begin{figure}[H]
\centering
\includegraphics[width=0.40\textwidth]{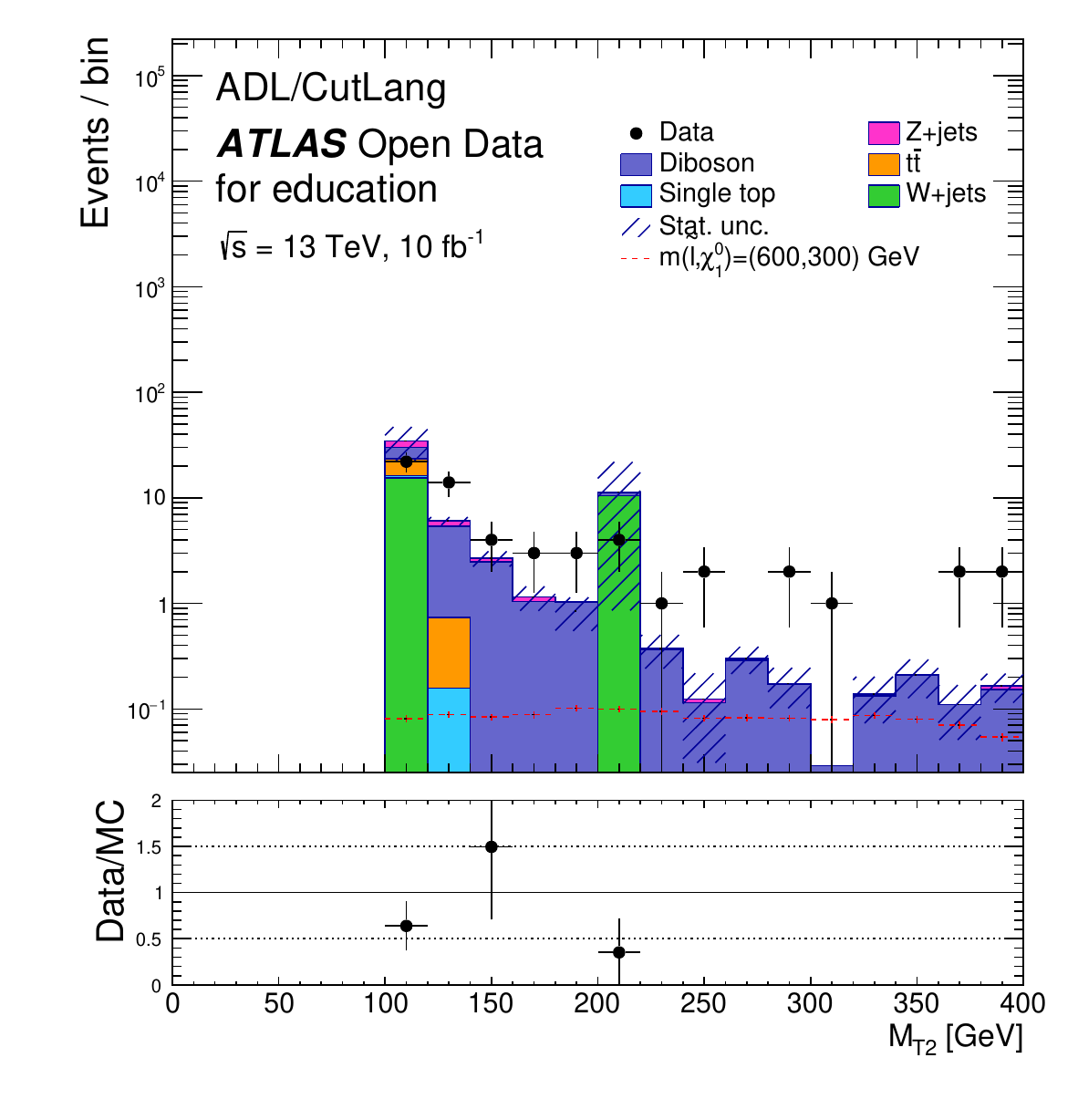}
\includegraphics[width=0.40\textwidth]{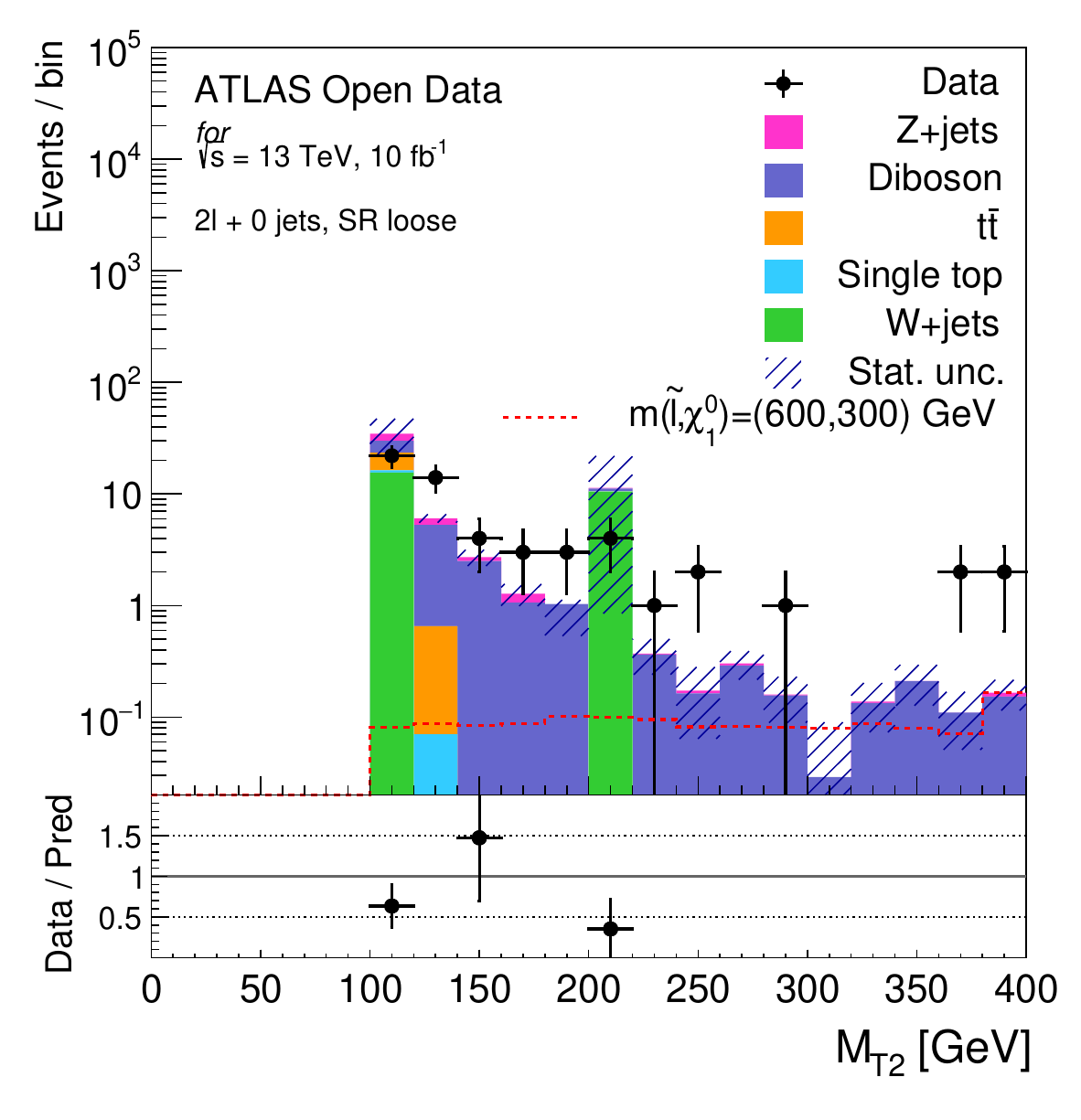}
\includegraphics[width=0.40\textwidth]{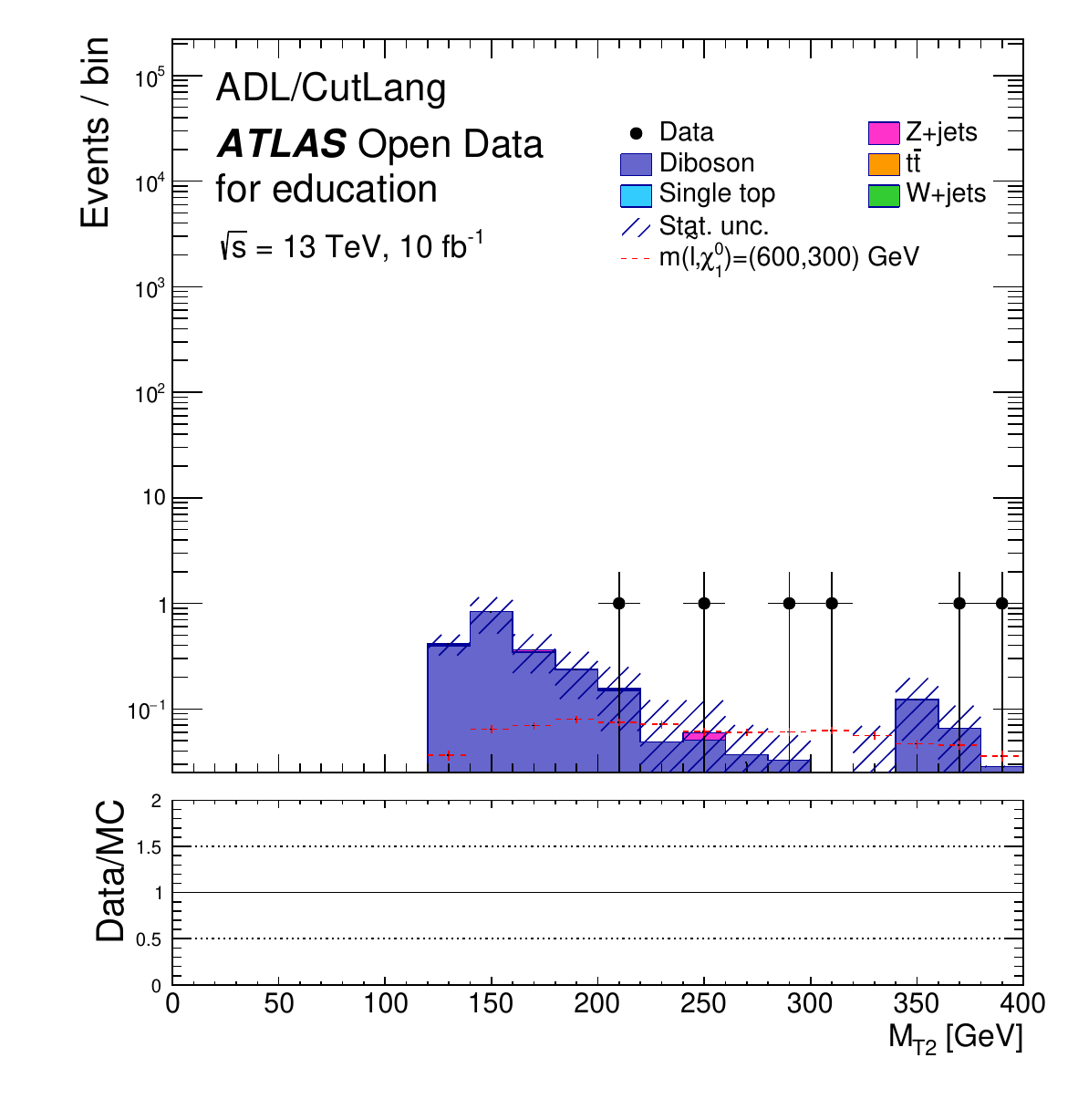}
\includegraphics[width=0.40\textwidth]{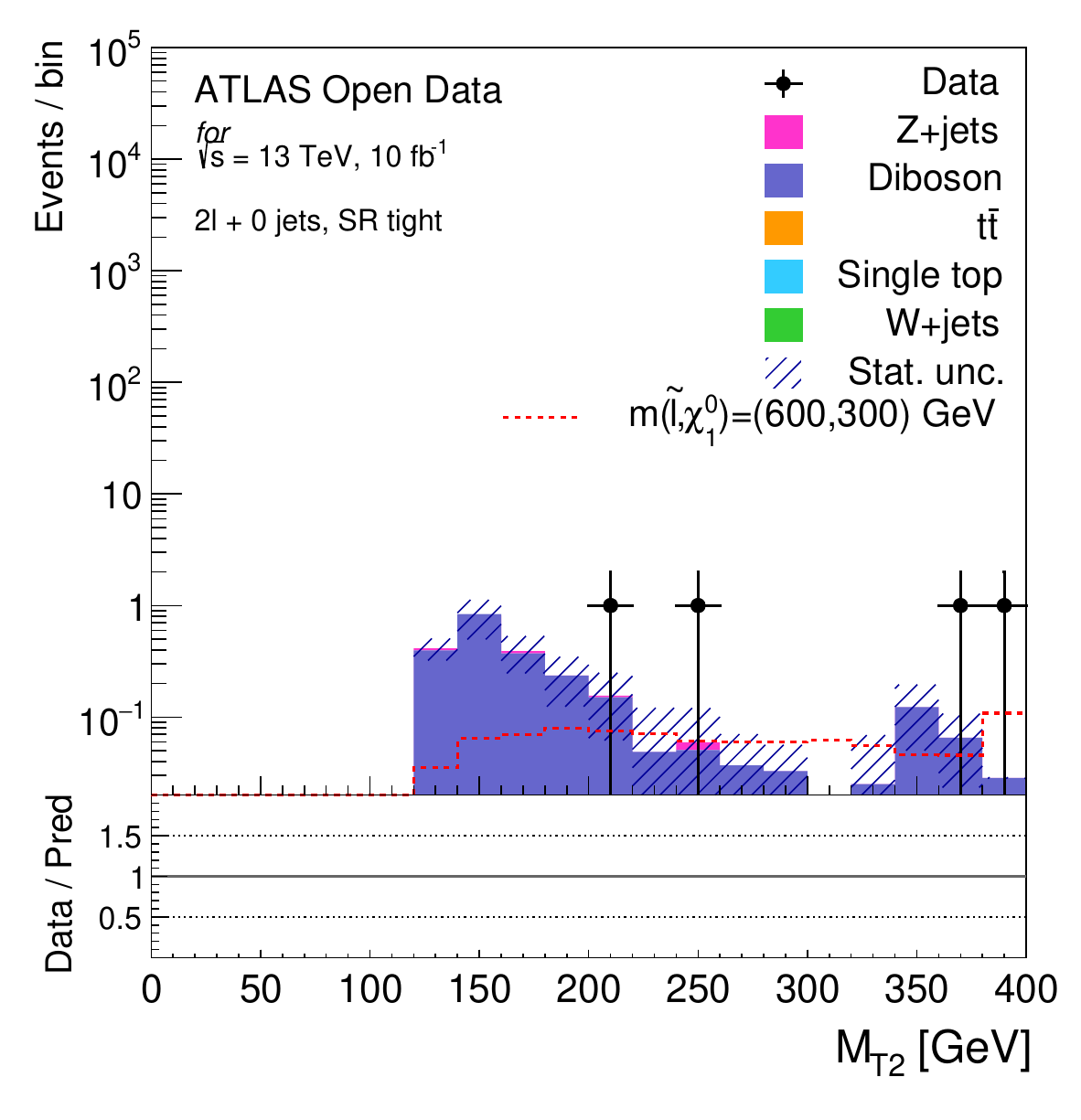}
\caption{Comparison of the $m_{T2}$ distributions in the loose (top) and the tight (bottom) signal regions obtained with the ADL/CutLang (left) and Open Data (right) frameworks in the $\tilde{\ell}^+\tilde{\ell}^- \to \ell^+\ell^-\tilde{\chi}^{0}_{1}\tilde{\chi}^{0}_{1}$ channel. The red dashed distribution represents the signal events normalized to the total background yield for visibility purposes, and the lower pad in each plot represents the Data/MC ratio. The last bin in each plot includes the overflow.} 

\label{fig:CL_vs_ODSUSY_masses}
\end{figure}

Figure \ref{fig:CL_vs_ODSUSY_masses} shows the $M_{T2}$ distributions from ADL/CutLang and the Open Data frameworks for \texttt{SR2loose} and \texttt{SR2tight} regions. Table \ref{tab:cutflow_SUSY_loose} presents the event cutflow for \texttt{SR2loose} region, while Tables \ref{tab:yields_SUSYyields_loose} and \ref{tab:yields_SUSYyields_tight} show the event yields for \texttt{SR2loose} and \texttt{SR2tight} signal regions, respectively. The event yields in these tables are obtained using the integral $M_{T2}$ distributions in both signal regions.

The $M_{T2}$ distributions show consistent behavior between the two frameworks in terms of overall shape and normalization. The event yields in the loose signal region are in good agreement between ADL/CutLang and the Open Data framework, within uncertainties. In the tight signal region, larger relative differences are observed, which can be attributed to limited statistics and the effect of the more stringent selection criteria.

\begin{table}[H]
\begin{center}
\caption{Cutflow of event counts for each sample in the $\tilde{\ell}^+\tilde{\ell}^- \to \ell^+\ell^-\tilde{\chi}^{0}_{1}\tilde{\chi}^{0}_{1}$ channel. The last two rows represent the selections defining the loose signal region.} 
\resizebox{\textwidth}{!}{
\begin{tabular}{|l|c|c|c|c|c|c|}
\hline
Selection & Data & $Z + jets$ & Diboson & $t\bar{t}$ & Single top & $W+jets$ \\ \hline 
ALL & $12205790.00 \pm 3493.68$ & $45960521.00 \pm 6779.42$ & $14254335.00 \pm 3775.49$ & $2910539.00 \pm 1706.03$ & $423766.00 \pm 650.97$ & $1189469.00 \pm 1090.63$ \\
evtweight totalWeight & $12205790.00 \pm 3493.68$ & $10389747.15 \pm 6696.54$ & $50065.46 \pm 41.34$ & $247639.11 \pm 156.12$ & $27733.76 \pm 47.23$ & $503094.49 \pm 2845.93$ \\
trigE == 1 $||$ trigM == 1 & $12205790.00 \pm 3493.68$ & $10389747.15 \pm 6696.54$ & $50065.46 \pm 41.34$ & $247639.11 \pm 156.12$ & $27733.76 \pm 47.23$ & $503094.49 \pm 2845.93$ \\
Size(signalLepts) == 2 & $10470030.00 \pm 3235.74$ & $9893650.86 \pm 6510.04$ & $28257.18 \pm 26.93$ & $89674.64 \pm 93.61$ & $8762.83 \pm 26.17$ & $7755.73 \pm 375.20$ \\
Pt(leadLept) $>$ 25 & $10469368.00 \pm 3235.64$ & $9893540.38 \pm 6510.01$ & $28213.11 \pm 26.91$ & $89392.09 \pm 93.47$ & $8740.31 \pm 26.14$ & $7746.49 \pm 375.00$ \\
Pt(subleadLept) $>$ 20 & $9686835.00 \pm 3112.37$ & $9355458.37 \pm 6329.21$ & $24378.93 \pm 25.01$ & $73795.42 \pm 85.03$ & $7128.52 \pm 23.60$ & $1674.37 \pm 149.63$ \\
Size(JetsA) == 0 & $9641645.00 \pm 3105.10$ & $9333405.32 \pm 6327.50$ & $23919.14 \pm 24.85$ & $68864.02 \pm 82.13$ & $6808.12 \pm 23.06$ & $1658.08 \pm 149.54$ \\
Size(JetsB) == 0 & $9327853.00 \pm 3054.15$ & $9158111.85 \pm 6307.22$ & $22207.03 \pm 24.34$ & $9498.85 \pm 30.51$ & $1960.83 \pm 12.39$ & $1552.02 \pm 148.65$ \\
Size(JetsC) == 0 & $8753415.00 \pm 2958.62$ & $8609545.62 \pm 6064.04$ & $15224.02 \pm 21.83$ & $5298.29 \pm 22.81$ & $1359.87 \pm 10.31$ & $1237.21 \pm 146.96$ \\
Size(JetsD) == 0 & $8753415.00 \pm 2958.62$ & $8609545.62 \pm 6064.04$ & $15224.02 \pm 21.83$ & $5298.29 \pm 22.81$ & $1359.87 \pm 10.31$ & $1237.21 \pm 146.96$ \\
pdgID(leadLept) + pdgID(subleadLept) $==$ 0 & $8686878.00 \pm 2947.35$ & $8542976.31 \pm 6041.41$ & $14670.37 \pm 21.52$ & $4820.64 \pm 21.75$ & $1263.46 \pm 9.93$ & $924.26 \pm 118.36$ \\
MT2 $>$ 100 & $461.00 \pm 21.47$ & $151.38 \pm 15.73$ & $188.50 \pm 3.51$ & $20.91 \pm 1.43$ & $2.65 \pm 0.44$ & $46.36 \pm 21.08$ \\
mLL $>$ 111 & $60.00 \pm 7.75$ & $5.31 \pm 1.15$ & $18.21 \pm 1.56$ & $7.67 \pm 0.84$ & $0.95 \pm 0.26$ & $33.74 \pm 17.99$ \\

\hline
\end{tabular}
}
\label{tab:cutflow_SUSY_loose}
\end{center}
\end{table}

\begin{table}[H]
\begin{center}
\caption{Event yields after applying the loose signal region selections.}
\begin{tabular}{|l|c|c|}
\hline
Sample & ADL/CutLang & Opendata Framework \\ \hline \hline
Data & $60.00 \pm 7.75$ & $58.00$ \\
$Z+jets$ & $5.31 \pm 1.15$   & $6.00$  \\
Diboson & $18.21 \pm 1.56$ & $18.12$ \\
$t\bar{t}$& $7.67 \pm 0.84$ & $7.68$ \\
Single top & $0.95 \pm 0.26$ & $0.86$ \\
$W + jets$ & $33.74 \pm 18.00$ & $26.01$ \\
\hline
\end{tabular}
\label{tab:yields_SUSYyields_loose}
\end{center}
\end{table}

\begin{table}[!h]
\begin{center}
\caption{Event yields after applying the tight signal region selections.}
\begin{tabular}{|l|c|c|}
\hline
Sample & ADL/CutLang & Opendata Framework \\ \hline \hline
Data & $6.00 \pm 2.45$ & $4.00$ \\
$Z+jets$ & $0.04 \pm 0.03$   & $0.02$  \\
Diboson & $2.39 \pm 0.41$ & $2.44$ \\
$t\bar{t}$& $0.00 \pm 0.00$ & $0.00$ \\
Single top & $0.00 \pm 0.00$ & $0.00$ \\
$W + jets$ & $7.75 \pm 7.75$ & $0.00$ \\
\hline
\end{tabular}
\label{tab:yields_SUSYyields_tight}
\end{center}
\end{table}

\subsection{Beyond standard model 
\texorpdfstring{$Z'$}{Z'} search in the 
\texorpdfstring{$Z' \rightarrow t\bar{t}$}{Z' -> tt} final state} \label{section_Zprime}
One of many hypothetical particles predicted by beyond the Standard Model (BSM) theories is $Z'$ boson. This analysis provides a representative example of the advantages of ADL/CutLang object definition structure, as it requires multiple angular criteria on both small-$R$ jets and large-$R$ jets. The baseline analysis objects are based on the standard object definitions described in Section \ref{section_objects} with a minor modification: $Z'$ analysis requires one lepton with $p_T > 30$ GeV, instead of 25 GeV. For the selection of b-tagged jets, the MV2c10 algorithm at the 70\% working point is used \cite{ATL-PHYS-PUB-2016-012}. 

Since the reconstruction of leptonically and hadronically decaying top-quark candidates depends on specific angular requirements between small-$R$ and large-$R$ jets, these criteria are explicitly defined within the object blocks to ensure consistent event selection. The ADL syntax describing this analysis is shown below:  

\begin{courier} \begin{lstlisting}
object goodDRJets 
  take goodJets
  select dR(goodJets,goodLepts[0]) < 2

object goodBJets
  take goodJets
  select bTag(goodJets) == 1

object goodFJets 
  take FJET
  select AbsEta(FJET) < 2
  select m(FJET) > 50
  select Pt(FJET) > 250

object goodDRBJets 
  take goodBJets
  select dR(goodBJets, goodDRJets[0]) < 0.01

object topLRJets 
  take goodFJets
  select m(goodFJets) > 100
  select Pt(goodFJets) > 300 && Pt(goodFJets) < 1500
  select tau32(goodFJets) < 0.75
  select dR(goodFJets, goodDRJets[0]) > 1.5
  select dPhi(goodFJets, goodLepts[0]) > 1
  select dR(goodFJets, anyof(goodBJets)) < 1 OR Size(goodDRBJets) >= 1

define MTW : sqrt(2*Pt(goodLepts[0])*MET*(1 - cos(dPhi(goodLepts[0],METLV[0]))))

define toplep : goodDRJet[0] goodLepts[0]
define tophad : topLRJets[0]
define ttbar : toplep tophad

\end{lstlisting} \end{courier}

\texttt{goodDRJets} represent the small-$R$ jets used in the reconstruction of the leptonically decaying top-quark candidate. These jets are required to be in close angular proximity to the lepton in the event. The \texttt{goodBJets} are derived from the baseline \texttt{goodJets} collection, with a $b$-tagging requirement applied via the \texttt{bTag()} function. This function is one of the predefined keywords in ADL/CutLang and is equivalent to applying \texttt{jet\_MV2c10(goodJets) > 0.8244273}.

The baseline large-$R$ jet objects are defined as \texttt{goodFJets} and are used to study the kinematic distributions of the leading large-$R$ jet at an intermediate stage of the analysis. The top-tagged large-$R$ jets are defined as \texttt{topLRJets}. In addition to basic mass and transverse momentum requirements, these jets must satisfy the $N$-subjettiness ratio $\tau_{32} < 0.75$, implemented via the \texttt{tau32()} function in ADL.

The multiplicity of \texttt{topLRJets} depends on several angular selections between analysis objects, including the $\Delta R$ between the large-$R$ jets and the leading small-$R$ jet, as well as the $\Delta \phi$ between the large-$R$ jet and the lepton. Furthermore, a separation requirement between large-$R$ jets and $b$-tagged jets is imposed. Since multiple combinations can arise, all possible $\Delta R$ pairings are evaluated using the \texttt{anyof} keyword within the object block.

Additionally, a $\Delta R$ requirement between $b$-tagged jets and the leading small-$R$ jet is enforced through a dedicated object, \texttt{goodDRBJets}. This condition is incorporated into the \texttt{topLRJets} definition via the multiplicity of \texttt{goodDRBJets}.

The event selection in this analysis is relatively straightforward and relies primarily on threshold requirements on $E_T^{\mathrm{miss}}$ and the combined quantity $E_T^{\mathrm{miss}} + M_T^W$. The $M_T^W$, is defined using the standard expression, consistent with the definition used in the other analyses. The remaining selections are based on object multiplicities, while the combinatorial aspects of the reconstruction are handled within the object definitions. These combinatorial selections are efficiently implemented using CutLang. The ADL syntax corresponding to the event selection is provided below. The leptonic top-quark candidate is defined as the vector sum of the leading small-$R$ jet and the leading lepton, without including the neutrino momentum for simplicity. The \texttt{topLRJets} object represents the hadronically decaying top-quark candidate in the event. The $t\bar{t}$ system is then constructed as the vector sum of the leptonic and hadronic top-quark candidates.

\begin{courier} \begin{lstlisting}
region Zprime
  select ALL
  weight evtWeight totalWeight
  select Size(FJET) >= 1
  select MET > 20
  select trigE == 1 OR trigM == 1
  select Size(goodLepts) == 1
  select (MET + MTW) > 60 
  select Size(goodBJets) >= 1
  select Size(goodDRJets) >= 1
  select Size(topLRJets) == 1
  select bTag(goodDRJets[0]) == 1

\end{lstlisting} \end{courier}

\begin{figure}[!ht]
\centering
\includegraphics[width=0.40\textwidth]{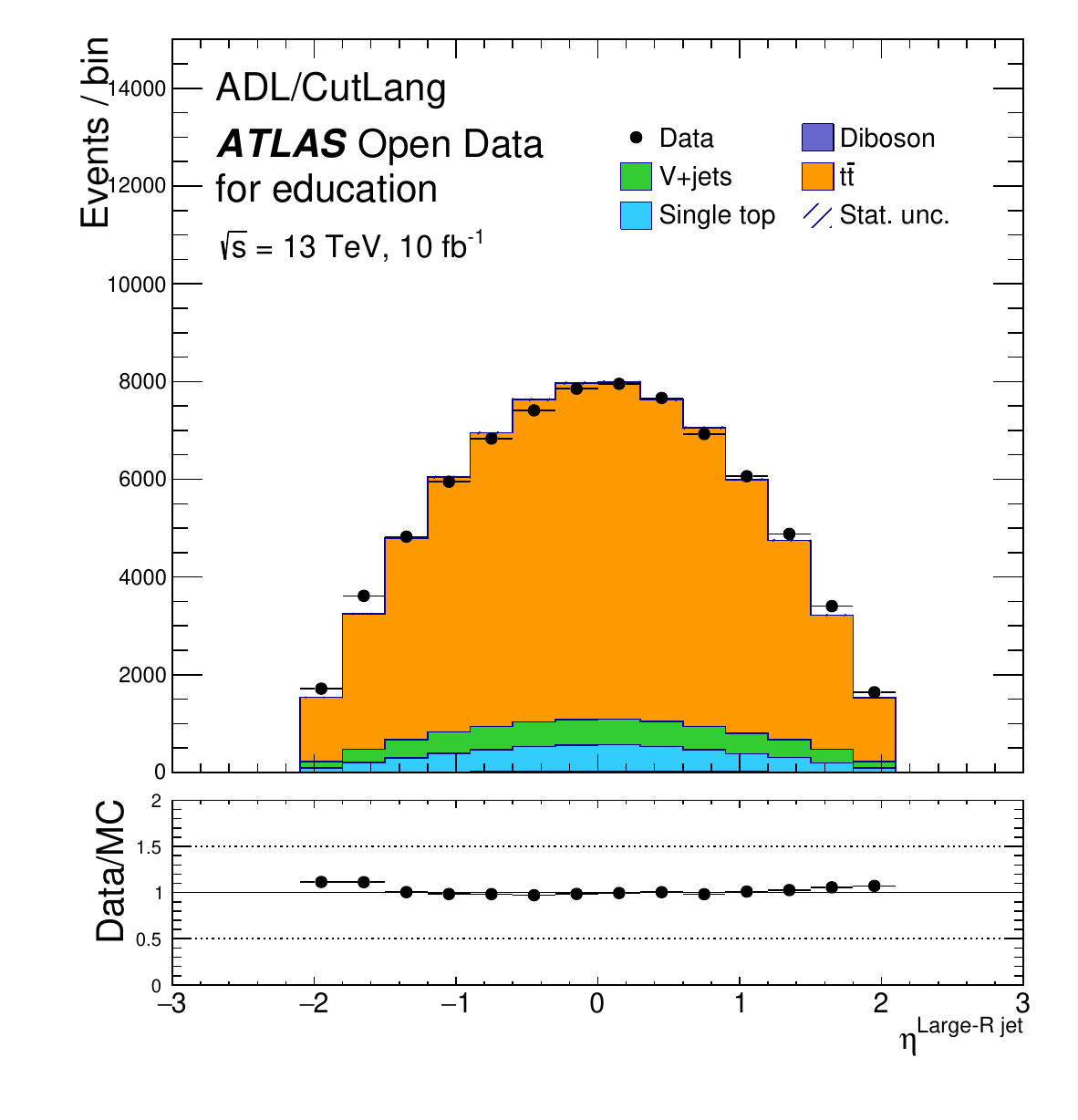}
\includegraphics[width=0.40\textwidth]{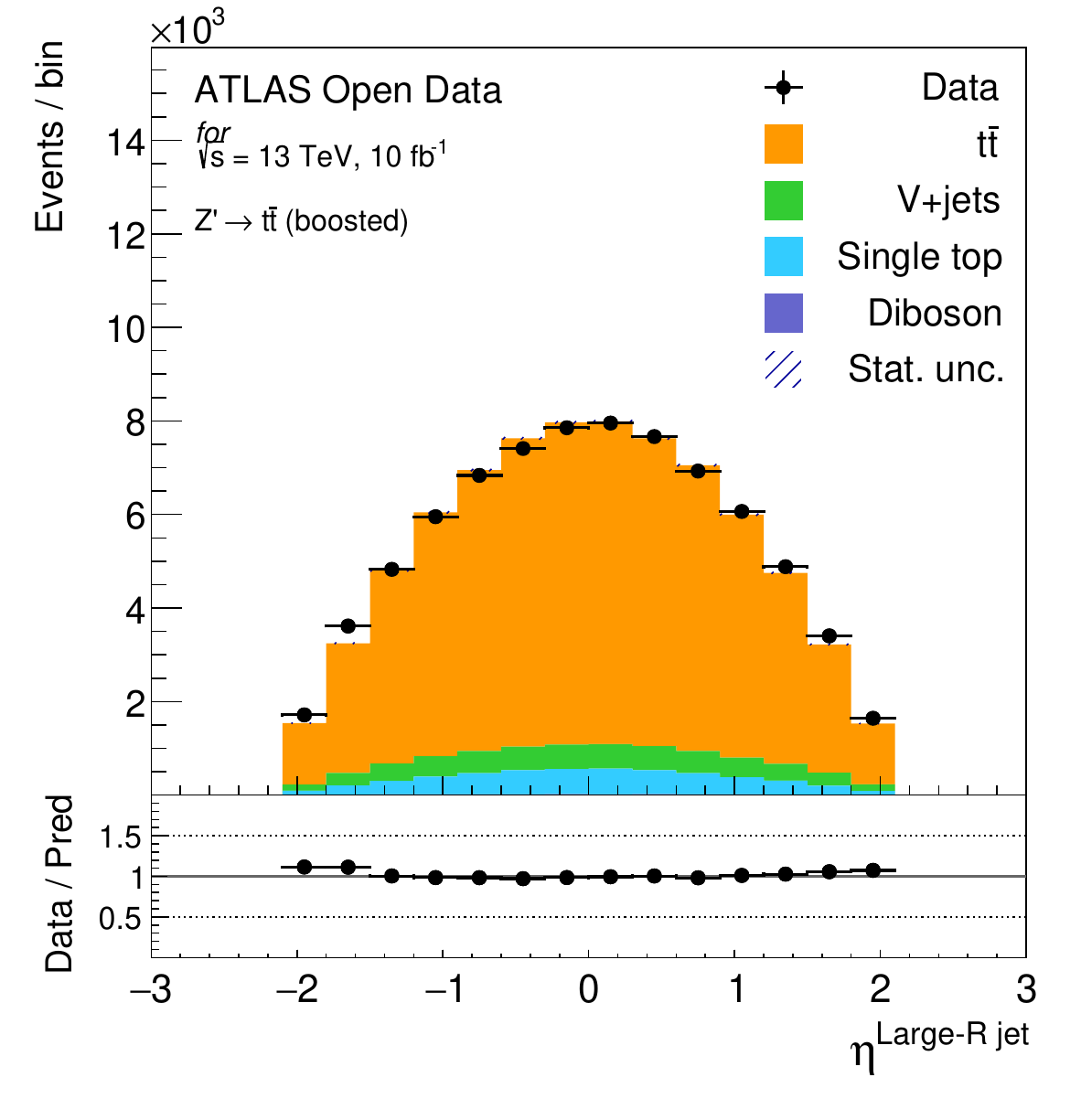}
\includegraphics[width=0.40\textwidth]{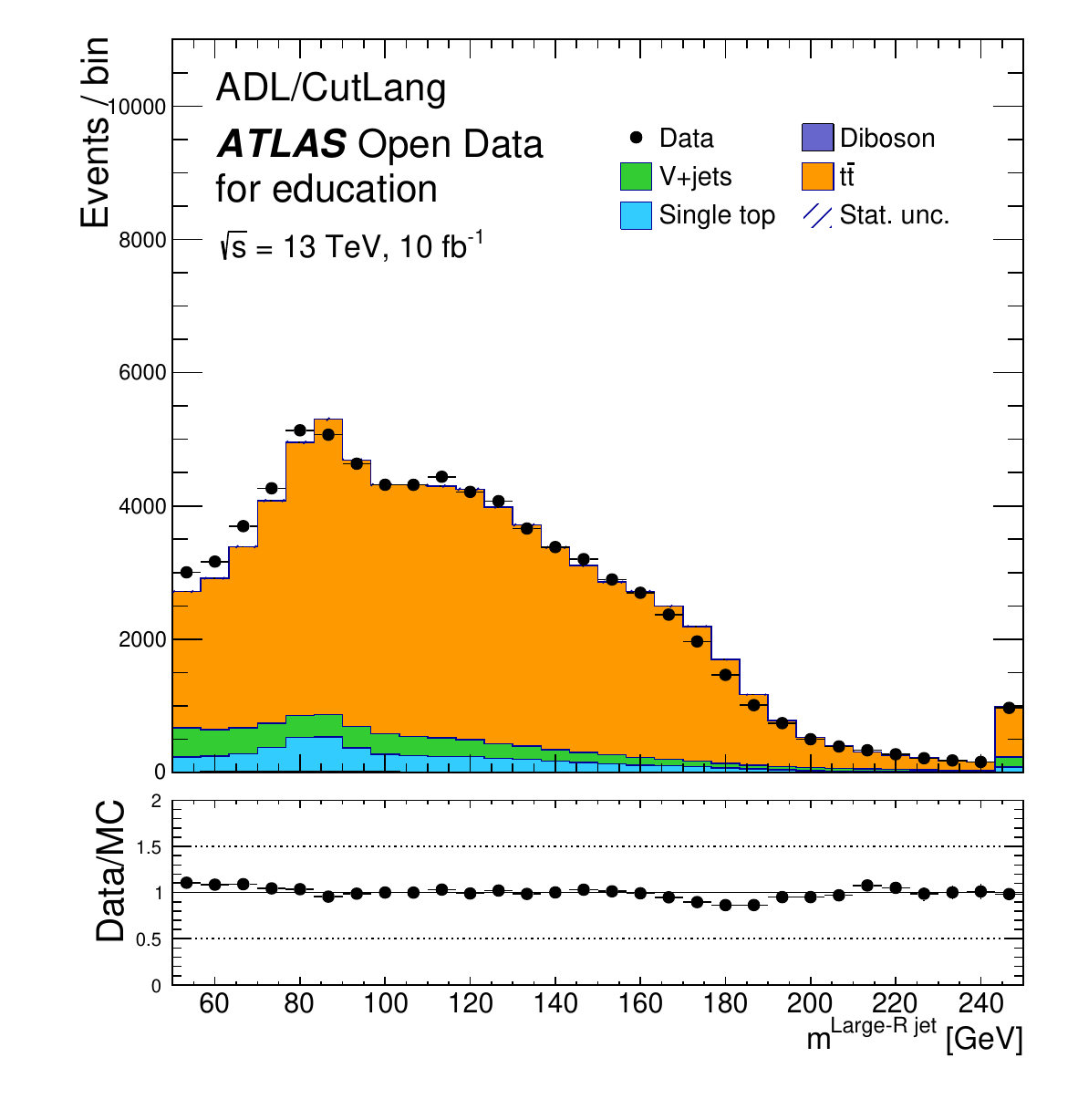}
\includegraphics[width=0.40\textwidth]{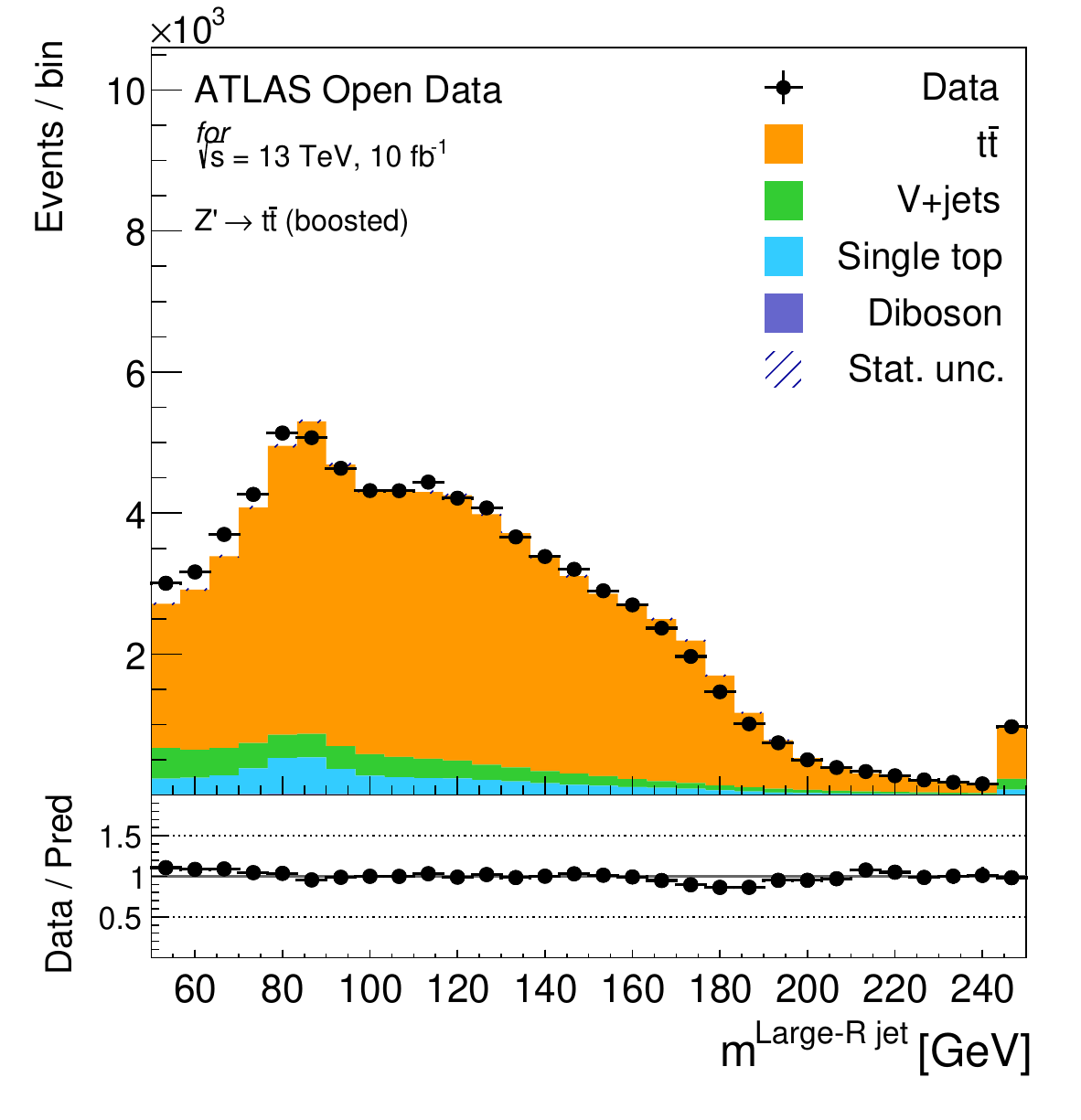}
\caption{Comparison of the $\eta$ (top) and mass (bottom) distributions of the leading large-$R$ jet after Size(goodDRjets) $\geq$ 1 selection obtained from the ADL/CutLang (left) and Open Data (right) frameworks in the $Z' \rightarrow t\bar{t} \rightarrow W^+b,W^-\bar{b} \rightarrow q\bar{q}'b, \ell^- \bar{\nu} \bar{b}$ channel. The lower pad in each plot represents the Data/MC ratio. The last bin in each plot includes the overflow.}
\label{fig:CL_Zboosted_kinematics_1}
\end{figure}

\begin{figure}[!ht]
\centering
\includegraphics[width=0.40\textwidth]{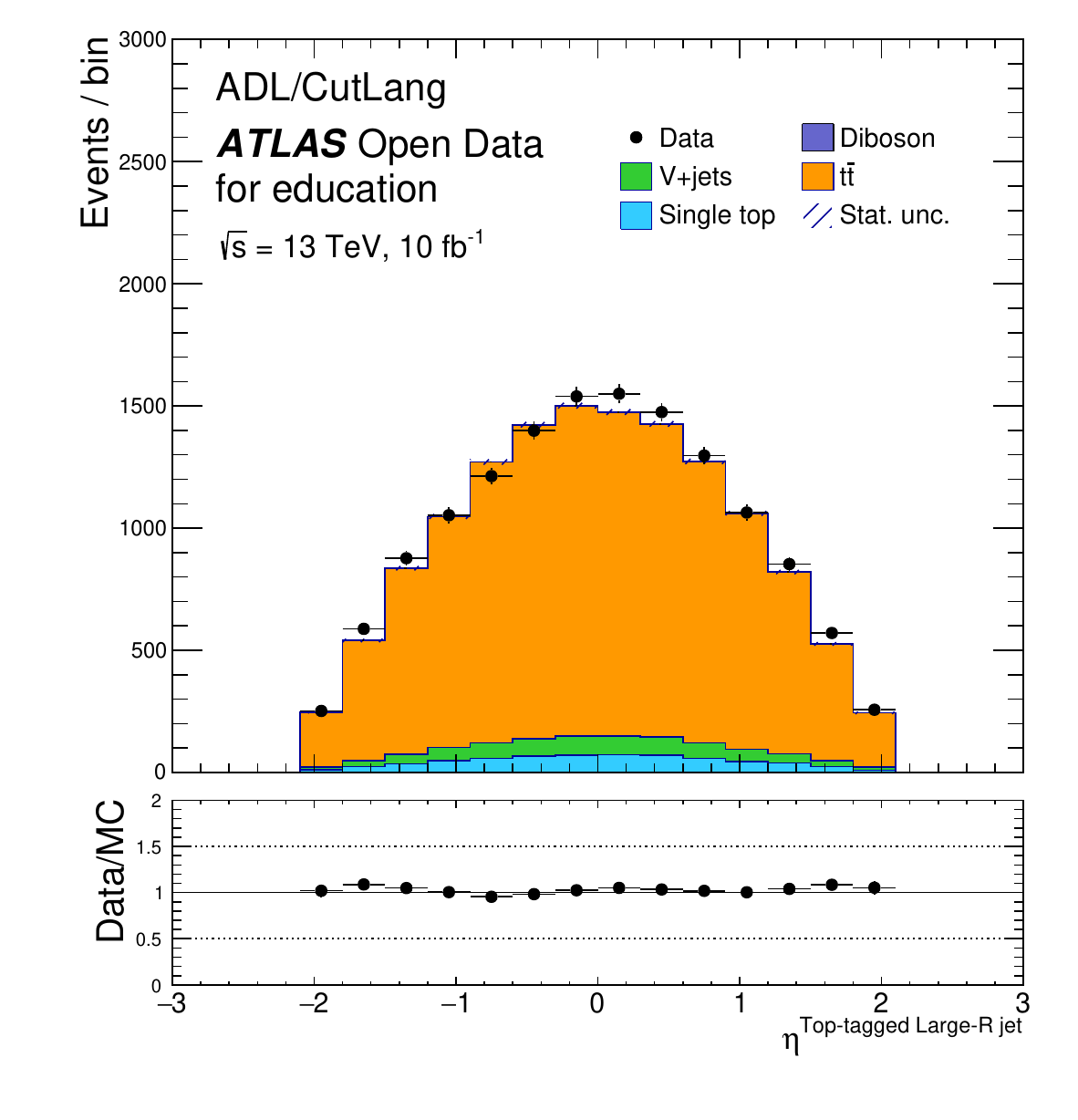}
\includegraphics[width=0.40\textwidth]{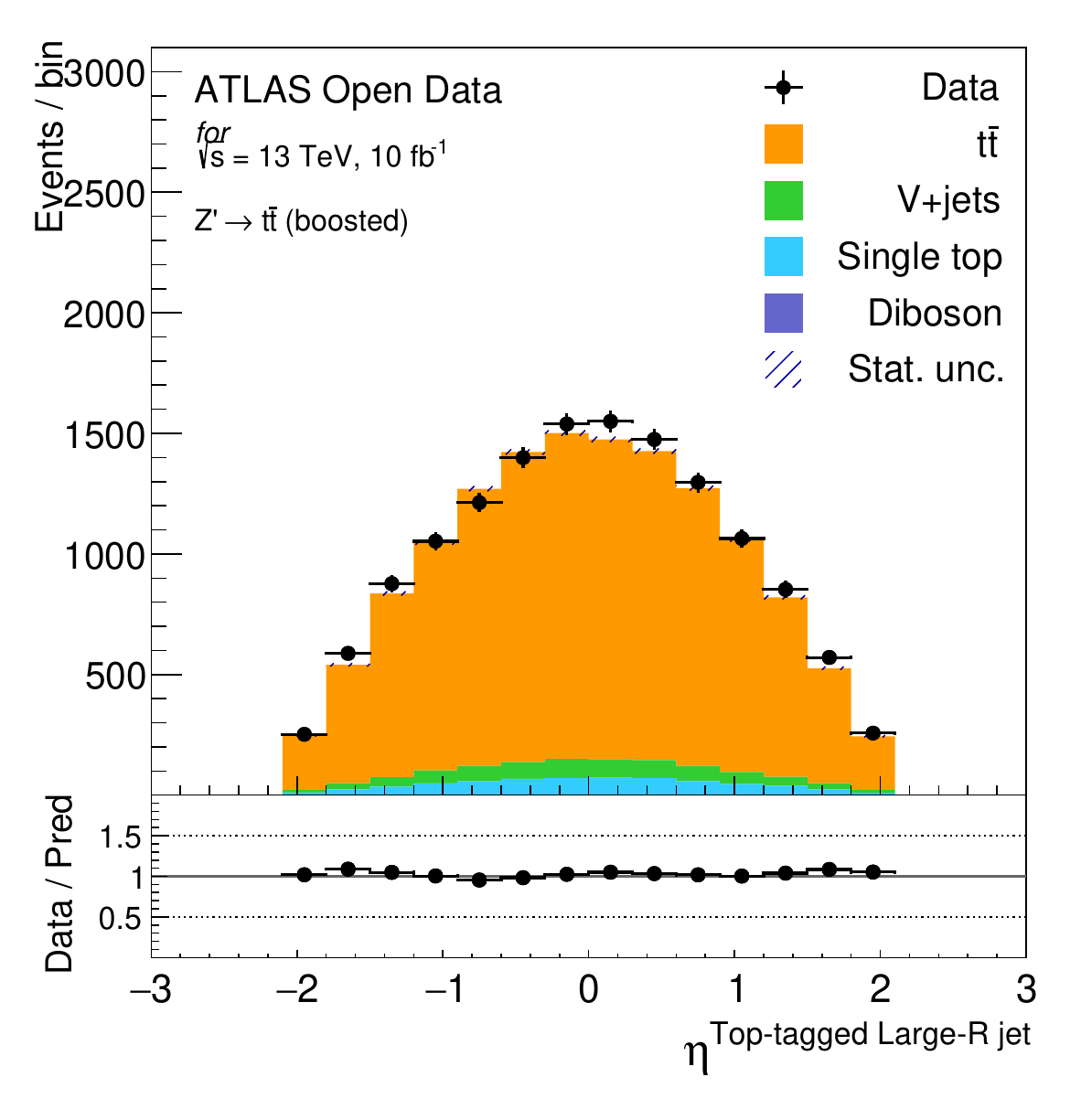}
\includegraphics[width=0.40\textwidth]{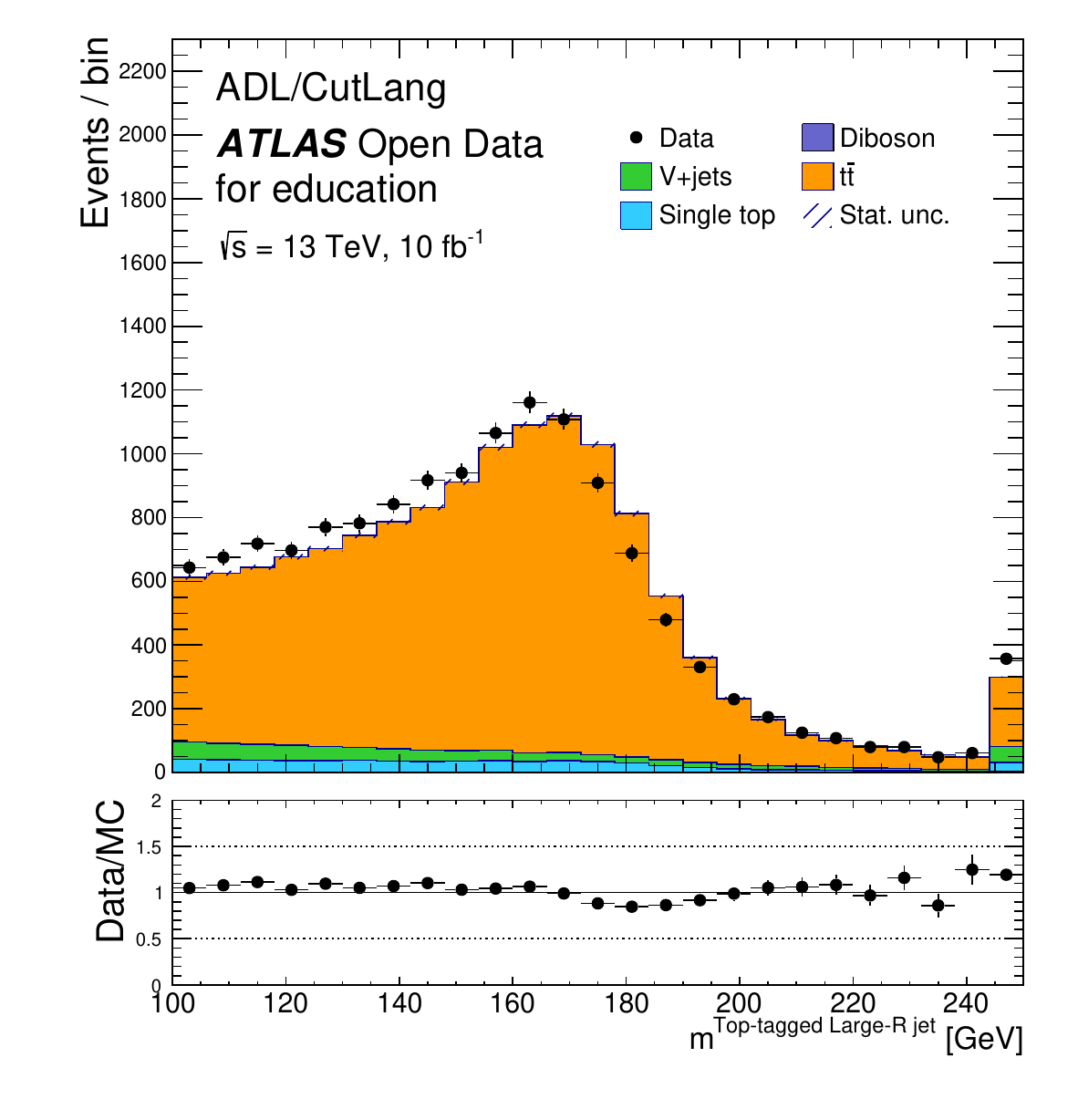}
\includegraphics[width=0.40\textwidth]{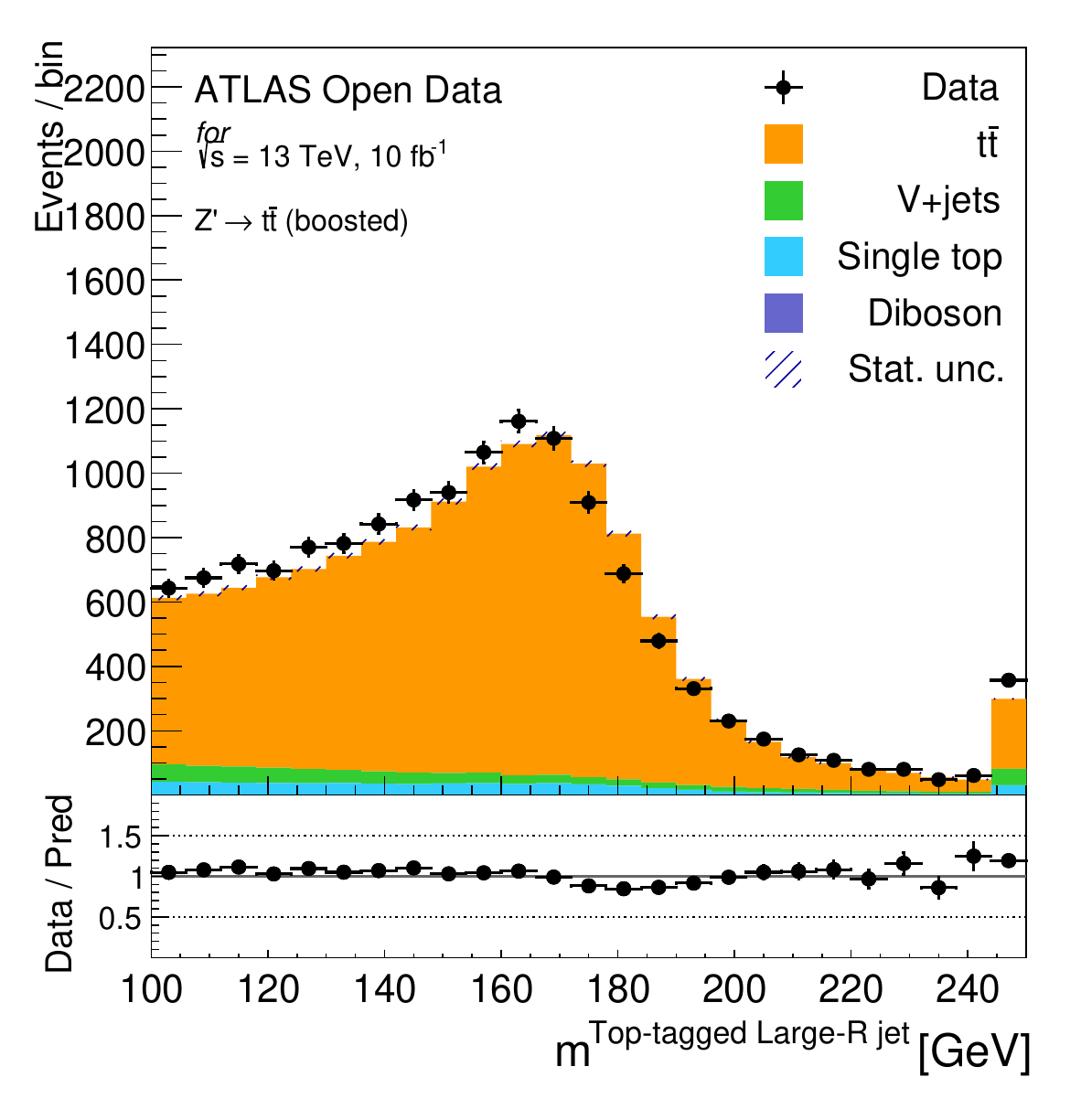}
\caption{Comparison of the $\eta$ (top) and mass (bottom) distributions of the top-tagged large-$R$ jet after Size(topLRjets) $==$ 1 selection obtained with the ADL/CutLang (left) and Open Data (right) frameworks in the $Z' \rightarrow t\bar{t} \rightarrow W^+b,W^-\bar{b} \rightarrow q\bar{q}'b, \ell^- \bar{\nu} \bar{b}$ channel. The lower pad in each plot represents the Data/MC ratio. The last bin in each plot includes the overflow.}
\label{fig:CL_Zboosted_kinematics_2}
\end{figure}

\begin{figure}[!ht]
\centering
\includegraphics[width=0.40\textwidth]{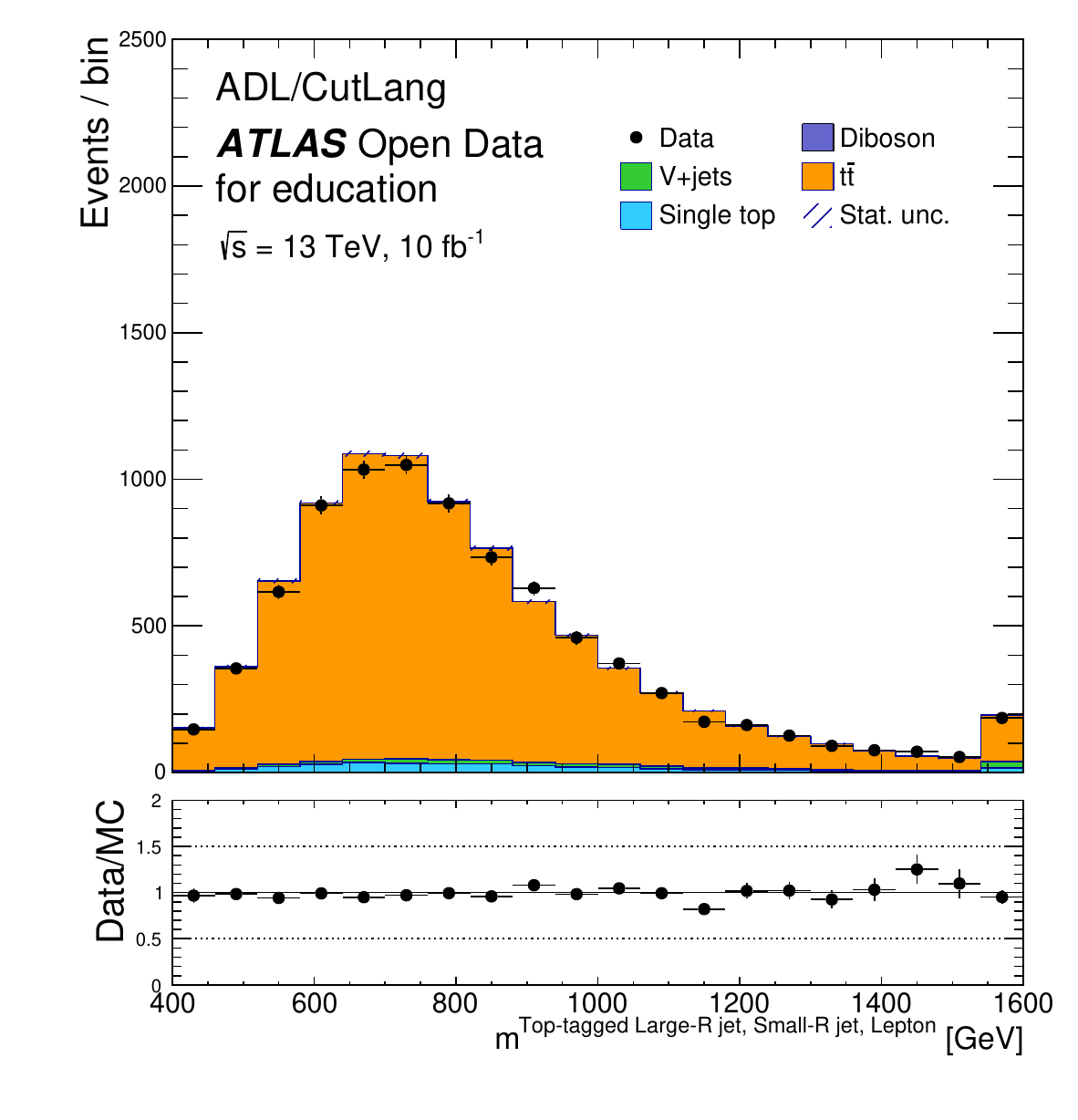}
\includegraphics[width=0.40\textwidth]{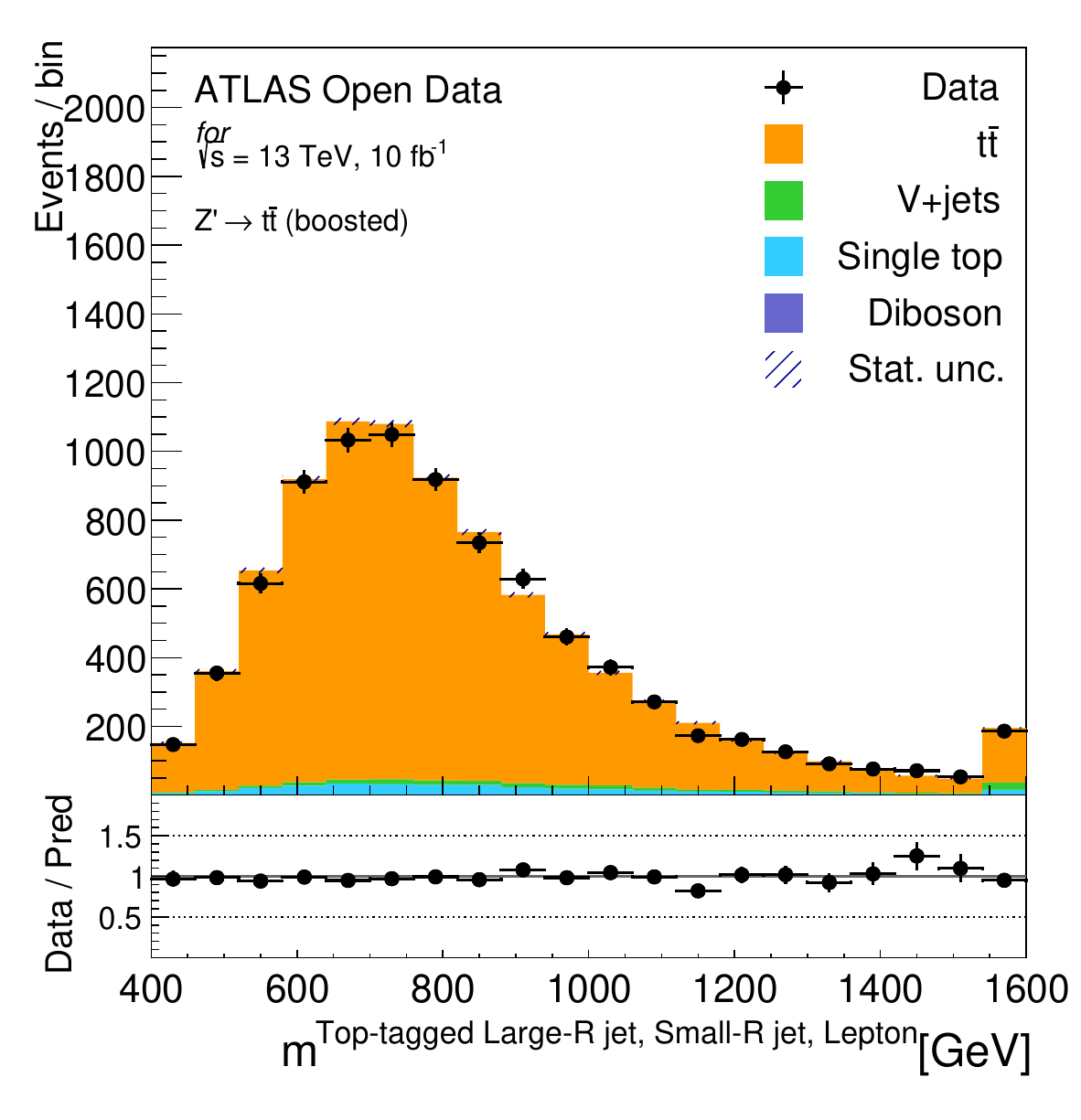}
\caption{The mass of $t\bar{t}$ system (without the neutrino 4-momenta) distribution after bTag(goodDRJets[0]) $==$ 1 selection, obtained with the ADL/Cutlang (left) and Open Data (right) frameworks in the $Z' \rightarrow t\bar{t} \rightarrow W^+b,W^-\bar{b} \rightarrow q\bar{q}'b, \ell^- \bar{\nu} \bar{b}$ channel. The lower pad in each plot represents the Data/MC ratio. The last bin in each plot includes the overflow.}
\label{fig:CL_Zboosted_ttbarm}
\end{figure}

The $\eta$ and mass distributions of the leading large-$R$ jet after the \texttt{Size(goodDRJets) $\geq$ 1} selection are shown in Figure \ref{fig:CL_Zboosted_kinematics_1}. The corresponding distributions for the top-tagged large-$R$ jet, after requiring \texttt{Size(topLRJets) == 1}, are presented in Figure \ref{fig:CL_Zboosted_kinematics_2}. Figure \ref{fig:CL_Zboosted_ttbarm} shows the approximate mass of the $t\bar{t}$ system from both frameworks after the \texttt{bTag(goodDRJets[0]) == 1} selection. Additionally, Tables \ref{tab:yields_Zboosted1}, \ref{tab:yields_Zboosted2}, and \ref{tab:yields_Zboosted3} present a comparison of the event yields between the ADL/CutLang and Open Data frameworks. For a fair comparison of the weighted event yields at each selection stage, the integrals of the corresponding mass distributions shown in each figure are used.

These results demonstrate that ADL/CutLang is in excellent agreement with the Open Data framework across both data and simulated samples, in terms of distribution shapes, event normalization, data-to-MC event ratio, and weighted event yields. This also shows that ADL/CutLang is capable of describing and executing analyses involving complex angular combinations between multiple objects.

\begin{table}[!ht]
\begin{center}
\caption{Cutflow of event counts for each sample in the $Z' \rightarrow t\bar{t} \rightarrow W^+b,W^-\bar{b} \rightarrow q\bar{q}'b, \ell^- \bar{\nu} \bar{b}$ channel.}
\resizebox{0.95\textwidth}{!}{
\begin{tabular}{|l|c|c|c|c|c|}
\hline
Selection & Data & Single Top & V + jets & Diboson & $t\bar{t}$ \\ \hline 
ALL & $466034.00 \pm 682.67$ & $219990.00 \pm 469.03$ & $9679378.00 \pm 3111.17$ & $1238452.00 \pm 1112.86$ & $1242954.00 \pm 1114.88$ \\
evtWeight totalWeight & $466034.00 \pm 682.67$ & $13682.10 \pm 34.28$ & $217747.51 \pm 585.75$ & $5421.90 \pm 14.74$ & $108616.54 \pm 104.83$ \\
Size(FJET) $\geq$ 1 & $466034.00 \pm 682.67$ & $13682.10 \pm 34.28$ & $217747.51 \pm 585.75$ & $5421.90 \pm 14.74$ & $108616.54 \pm 104.83$ \\
MET $>$ 20 & $411666.00 \pm 641.61$ & $13054.45 \pm 33.48$ & $197885.14 \pm 577.57$ & $5018.81 \pm 14.22$ & $103468.10 \pm 102.29$ \\
trigE $==$ 1 OR trigM $==$ 1 & $411666.00 \pm 641.61$ & $13054.45 \pm 33.48$ & $197885.14 \pm 577.57$ & $5018.81 \pm 14.22$ & $103468.10 \pm 102.29$ \\
Size(goodLep) $==$ 1 & $330990.00 \pm 575.32$ & $11623.63 \pm 31.64$ & $177543.99 \pm 559.54$ & $3825.01 \pm 13.34$ & $90546.87 \pm 95.73$ \\
Pt(goodLepts) descend & $330990.00 \pm 575.32$ & $11623.63 \pm 31.64$ & $177543.99 \pm 559.54$ & $3825.01 \pm 13.34$ & $90546.87 \pm 95.73$ \\
(MET + MTW) $>$ 60 & $312771.00 \pm 559.26$ & $11254.96 \pm 31.12$ & $170311.06 \pm 557.97$ & $3698.93 \pm 13.10$ & $87371.33 \pm 94.03$ \\
Size(goodBJets) $\geq$ 1 & $98688.00 \pm 314.15$ & $8312.69 \pm 26.83$ & $13251.27 \pm 41.54$ & $325.87 \pm 4.25$ & $72147.27 \pm 85.62$ \\
Size(goodDRJets) $\geq$ 1 & $85910.00 \pm 293.10$ & $6790.74 \pm 24.33$ & $8882.01 \pm 27.46$ & $247.43 \pm 3.53$ & $67398.50 \pm 82.74$ \\
Size(topLRJets) == 1 & $13988.00 \pm 118.27$ & $610.86 \pm 7.07$ & $682.69 \pm 5.94$ & $22.34 \pm 1.00$ & $13600.48 \pm 37.09$ \\
bTag(goodDRJets[0]) $==$ 1& $8475.00 \pm 92.06$ & $313.65 \pm 5.12$ & $175.90 \pm 2.76$ & $6.97 \pm 0.57$ & $8936.55 \pm 30.08$ \\
\hline
\end{tabular}
}
\label{tab:cutflow_Zboosted}
\end{center}
\end{table}

\begin{table}[!ht]
\begin{center}
\caption{Event yields after Size(goodDRjets) $\geq$ 1 selection for each sample in the $Z' \rightarrow t\bar{t} \rightarrow W^+b,W^-\bar{b} \rightarrow q\bar{q}'b, \ell^- \bar{\nu} \bar{b}$ channel.}
\begin{tabular}{|l|c|c|}
\hline
Sample & ADL/CutLang & Opendata Framework \\ \hline \hline
Data & $76742.00 \pm 277.02$ & $76743.00$ \\
Single Top & $4919.83 \pm 20.45$ & $4919.76$ \\
V + jets & $5412.84 \pm 19.34$ & $5412.81$ \\
Diboson & $192.49 \pm 3.12$ & $192.49$ \\
$t\bar{t}$ & $65827.75 \pm 81.85 $ & $65827.81$ \\ \hline
\end{tabular}
\label{tab:yields_Zboosted1}
\end{center}
\end{table}

\begin{table}[!ht]
\begin{center}
\caption{Event yields after Size(topLRjets) $==$ 1 selection for each sample in the $Z' \rightarrow t\bar{t} \rightarrow W^+b,W^-\bar{b} \rightarrow q\bar{q}'b, \ell^- \bar{\nu} \bar{b}$ channel.}
\begin{tabular}{|l|c|c|}
\hline
Sample & ADL/CutLang & Opendata Framework \\ \hline \hline
Data & $13988.00 \pm 118.27$ & $13988.00$ \\
Single Top & $610.86 \pm 6.88$ & $610.78$ \\
V + jets & $682.69 \pm 5.89$ & $682.68$ \\
Diboson & $22.34 \pm 0.95$ & $22.34$ \\
$t\bar{t}$ & $12376.43 \pm 33.39$ & $12376.43$ \\ \hline
\end{tabular}
\label{tab:yields_Zboosted2}
\end{center}
\end{table}

\begin{table}[!ht]
\begin{center}
\caption{Event yields after bTag(goodDRJets[0]) $==$ 1 selection for each sample in the $Z' \rightarrow t\bar{t} \rightarrow W^+b,W^-\bar{b} \rightarrow q\bar{q}'b, \ell^- \bar{\nu} \bar{b}$ channel.}
\begin{tabular}{|l|c|c|}
\hline
Sample & ADL/CutLang & Opendata Framework \\ \hline \hline
Data & $8433.00 \pm 91.83$ & $8433.00$ \\
Single Top & $311.71 \pm 4.99$ & $311.71$ \\
V + jets & $174.60 \pm 2.70$ & $174.60$ \\
Diboson & $6.95 \pm 0.56$ & $6.95$ \\
$t\bar{t}$ & $8093.67 \pm 27.10$ & $8093.66$ \\ \hline
\end{tabular}
\label{tab:yields_Zboosted3}
\end{center}
\end{table}

\clearpage

\section{Conclusions}
In this study, nine different analyses were performed using ADL/CutLang on proton-proton collision data provided by ATLAS Open Data, corresponding to an integrated luminosity of 10 fb$^{-1}$ at 13 TeV center-of-mass energy. The results were compared with those  obtained using the ATLAS Open Data C++ Framework. The corresponding ADL files for each analysis have been shared on GitHub.
Overall, consistent results were achieved, validating the analyses implemented with ADL/CutLang.

The CutLang runtime interpreter, which enables event processing, execution of analysis-specific functions, and histogramming, provided results largely matching those obtained with the ATLAS Open Data Framework. 
Table \ref{tab:all_yields_Data/SM} provides an overview of the results from both ADL/CutLang and ATLAS Open Data frameworks. Overall, Data/MC ratio from most analyses show fully consistent results between the two frameworks. However, some minor discrepancies were observed in certain analyses, especially in those which require a complicated $\chi^2$ minimization. 
Such complicated algorithms might be difficult to implement and test using C++, especially in educationally oriented studies such as this one.

This scenario in particular and this study in general highlight the advantages of decoupling the analysis algorithm from its software implementation. The use of a previously tested and debugged computational framework ensures consistent and reliable results.
This paper also demonstrates the flexibility and expressiveness of the ADL and its runtime interpreter, CutLang: a range of SM and BSM analyses  have been successfully implemented without the need for new keywords or functions. This study further demonstrates that ADL/CutLang is a practical system for performing open data analyses, particularly in an educational context.

\begin{table}[!h]
\begin{center}
\caption{Comparison of the Data/MC ratio obtained using the ADL/CutLang and Open Data frameworks.}
\resizebox{0.5\textwidth}{!}{
\begin{tabular}{|l|cc|cc|}
\hline
& \multicolumn{2}{|c|}{Data/Monte Carlo} \\ \hline 
Analysis & ADL/CutLang & Open Data Framework\\
\hline
(\hyperref[section_Zanalysis]{4.1}) Zboson  & 0.977 & 0.977 \\
(\hyperref[section_Wanalysis]{4.2}) Wboson  & 1.013 & 1.013 \\
(\hyperref[section_Stopanalysis]{4.3}) SingleTop  & 1.014 & 1.014 \\
(\hyperref[section_ttbaranalysis]{4.4}) TTbar  & 1.047 & 1.047 \\
(\hyperref[section_WZ]{4.5}) WZ  & 1.053 & 1.053 \\
(\hyperref[section_HZZ]{4.6}) HZZ  & 0.997 & 0.993 \\
(\hyperref[section_ZZ]{4.7}) ZZ  & 1.091 & 1.079 \\
(\hyperref[section_susy]{4.8}) SUSY  & 0.911 & 0.989 \\
(\hyperref[section_Zprime]{4.9}) ZPrime  & 0.982 & 0.982 \\
\hline
\end{tabular}
}
\label{tab:all_yields_Data/SM}
\end{center}
\end{table}

\section{Acknowledgement}
This work is being supported by the CHIST-ERA OpenMAPP project under grant 223N165 from The Scientific and Technological Research Council of Türkiye (TUBITAK). The authors acknowledge the efforts of the ATLAS Collaboration to record or simulate, reconstruct, and distribute the Open Data used in this paper, and to develop and support the software with which it was analyzed. Finally, the authors would like to thank Sezen Sekmen, V. Erkcan Özcan and Gökhan Ünel for their valuable contributions.

\bibliographystyle{ieeetr}
\bibliography{references}

@misc{ATLASCppFramework13TeV,
  title        = "{ATLAS Outreach C++ Framework: 13 TeV}",
  url          = "https://github.com/atlas-outreach-data-tools/atlas-outreach-cpp-framework-13tev",
  note         = "Accessed: 2025-02-04",
  howpublished = {\url{https://github.com/atlas-outreach-data-tools/atlas-outreach-cpp-framework-13tev}},
}

@misc{ADL_CERN_WEB,
  author       = {{H. B. Prosper} and {S. Sekmen} and {G. Unel}},
  title        = "{Analysis Description Language (ADL) Web Portal}",
  url          = "https://adl.web.cern.ch/index.html",
  note         = "Accessed: 2025-02-04"
}

@article{Unel_2018,
   title={CutLang: A Particle Physics Analysis Description Language and Runtime Interpreter},
   volume={233},
   ISSN={0010-4655},
   url={http://dx.doi.org/10.1016/j.cpc.2018.06.023},
   DOI={10.1016/j.cpc.2018.06.023},
   journal={Computer Physics Communications},
   publisher={Elsevier BV},
   author={Ünel, Gökhan and Sekmen, Sezen},
   year={2018},
   month=Dec, pages={215–236} }

@article{Unel_2021,
   title={CutLang v2: Advances in a Runtime-Interpreted Analysis Description Language for HEP Data},
   volume={4},
   ISSN={2624-909X},
   url={http://dx.doi.org/10.3389/fdata.2021.659986},
   DOI={10.3389/fdata.2021.659986},
   journal={Frontiers in Big Data},
   publisher={Frontiers Media SA},
   author={Unel, G. and Sekmen, S. and Toon, A. M. and Gokturk, B. and Orgen, B. and Paul, A. and Ravel, N. and Setpal, J.},
   year={2021},
   month=jun
}

@article{Adiguzel_2021,
doi = {10.1088/1361-6404/abdf67},
url = {https://dx.doi.org/10.1088/1361-6404/abdf67},
year = {2021},
month = {feb},
publisher = {IOP Publishing},
volume = {42},
number = {3},
pages = {035802},
author = {Adiguzel, A and Cakir, O and Kaya, U and Ozcan, V E and Ozturk, S and Sekmen, S and Turk Cakir, I and Unel, G},
title = {CutLang as an analysis description language for introducing students to analyses in particle physics},
journal = {European Journal of Physics}
}

@article{BRUN199781,
title = {ROOT — An object oriented data analysis framework},
journal = {Nuclear Instruments and Methods in Physics Research Section A: Accelerators, Spectrometers, Detectors and Associated Equipment},
volume = {389},
number = {1},
pages = {81-86},
year = {1997},
note = {New Computing Techniques in Physics Research V},
issn = {0168-9002},
doi = {https://doi.org/10.1016/S0168-9002(97)00048-X},
url = {https://www.sciencedirect.com/science/article/pii/S016890029700048X},
author = {Rene Brun and Fons Rademakers}
}

@techreport{ATL-OREACH-PUB-2020-001,
      author        = "{ATLAS Collaboration}",
      collaboration = "ATLAS",
      title         = "{Review of the 13 TeV ATLAS Open Data release}",
      institution   = "CERN",
      reportNumber  = "ATL-OREACH-PUB-2020-001",
      address       = "Geneva",
      year          = "2020",
      url           = "https://cds.cern.ch/record/2707171",
      note          = "All figures including auxiliary figures are available at https://atlas.web.cern.ch/Atlas/GROUPS/PHYSICS/PUBNOTES/ATL-OREACH-PUB-2020-001",
}

@misc{cutlang_source,
  title        = "{CutLang GitHub repository}",
  url          = "https://github.com/unelg/cutlang",
  note         = "Accessed: 2026-04-25",
  howpublished = {\url{https://github.com/unelg/cutlang}},
}

@misc{Opendata_web,
  title        = "{ATLAS Open Data and Tools for Education}",
  url          = "https://opendata.atlas.cern",
  note         = "Accessed: 2026-03-02",
  howpublished = {\url{https://opendata.atlas.cern}},
}

@misc{cms_opendata_guide,
  title        = "{CMS Open Data Guide}",
  url          = "https://cms-opendata-guide.web.cern.ch",
  note         = "Accessed: 2026-04-25",
  howpublished = {\url{https://cms-opendata-guide.web.cern.ch}},
}

@article{atlas_singletop8tev,
    author = "Aaboud, Morad and others",
    collaboration = "ATLAS",
    title = "{Probing the W tb vertex structure in t-channel single-top-quark production and decay in pp collisions at $ \sqrt{s}=8 $ TeV with the ATLAS detector}",
    eprint = "1702.08309",
    archivePrefix = "arXiv",
    primaryClass = "hep-ex",
    reportNumber = "CERN-EP-2017-011",
    doi = "10.1007/JHEP04(2017)124",
    journal = "JHEP",
    volume = "04",
    pages = "124",
    year = "2017"
}

@article{atlas_singletop13tev,
    author = "Aaboud, Morad and others",
    collaboration = "ATLAS",
    title = "{Measurement of the inclusive cross-sections of single top-quark and top-antiquark $t$-channel production in $pp$ collisions at $\sqrt{s}$ = 13 TeV with the ATLAS detector}",
    eprint = "1609.03920",
    archivePrefix = "arXiv",
    primaryClass = "hep-ex",
    reportNumber = "CERN-EP-2016-197",
    doi = "10.1007/JHEP04(2017)086",
    journal = "JHEP",
    volume = "04",
    pages = "086",
    year = "2017"
}

@techreport{ATL-PHYS-PUB-2016-012,
      author        = "{ATLAS Collaboration}",
      collaboration = "ATLAS",
      title         = "{Optimisation of the ATLAS $b$-tagging performance for the
                       2016 LHC Run}",
      institution   = "CERN",
      reportNumber  = "ATL-PHYS-PUB-2016-012",
      address       = "Geneva",
      year          = "2016",
      url           = "https://cds.cern.ch/record/2160731",
      note          = "All figures including auxiliary figures are available at   https://atlas.web.cern.ch/Atlas/GROUPS/PHYSICS/PUBNOTES/ATL-PHYS-PUB-2016-012",
}

@misc{ADL4LHC_ATLAS13TeV,
  author       = "{{K. Karaca} and {K. Sahan} and {A. Sansar}}",
  title        = "{ADL Implementations for 13 TeV ATLAS Open Data}",
  year         = "2025",
  url          = "https://github.com/ADL4HEP/ADL4LHCOpenData/tree/main/ATLAS-OTRC-2019-01",
  note         = "Accessed: 2025-02-04",
  howpublished = {\url{https://github.com/ADL4HEP/ADL4LHCOpenData/tree/main/ATLAS-OTRC-2019-01}},
}

\clearpage

\appendix
\section{List of ATLAS Open Data samples}\label{appendix_opendata_samples}
A list of Standard Model simulation samples in the ATLAS 13 TeV 2020 Open dataset is provided in Table \ref{tab:opendata_samples}. These samples include different skimming options such as one lepton, two leptons, three leptons, four leptons, two photons, and one large-$R$ jet with one lepton in the final states.

\begin{table}[!ht]
\begin{center}
\caption{A list of Standard Model (SM) simulation samples in the ATLAS 13 TeV Open Data used in this study. Taken from Ref. \cite{Opendata_web}.}
\begin{tabular}{|c|c|}
\hline
Sample & Dataset ID \\
\hline
$W+\mathrm{jets}$ & $361100$--$361105$ \\
$Z+\mathrm{jets}$ & $361106$--$361108$ \\
Diboson & $363356$, $363358$--$363360$, $363489$--$363493$ \\
$Z+\mathrm{jets}$, filtered & $364100$--$364141$ \\
$W+\mathrm{jets}$, filtered & $364156$--$364197$ \\
$t\bar{t}$ & $410000$ \\
Single top & $410011$--$410014$, $410025$--$410026$ \\
\hline
$H\rightarrow\gamma\gamma$ & $341081$, $343981$, $345041$, $345318$--$345319$ \\
$H\rightarrow WW$ & $345323$--$345324$ \\
SUSY signal & $392985$ \\
\hline
\end{tabular}
\label{tab:opendata_samples}
\end{center}
\end{table}

\section{Predefined functions in CutLang used in the Open Data Analyses}\label{cutlang_funcs}

A list of keywords corresponding to predefined functions in CutLang used in this study is provided in Table~\ref{tab:ADL_functions}.

\begin{table}[!ht]
\caption{Predefined functions in CutLang and their ADL syntax \cite{ADL_CERN_WEB}.} 
\label{tab:ADL_functions}
\centering
\begin{tabular}{|c|c|}
\hline
Syntax & Meaning \\
\hline 
\multicolumn{2}{|c|}{\textit{Object attributes}} \\
\hline
\texttt{m()} & Mass of \\
\texttt{q()} & Charge of\\
\texttt{Phi()} & Phi of \\
\texttt{Eta()} & Eta of \\
\texttt{AbsEta()} & Absolute value of Eta of \\
\texttt{Rap()} & Rapidity of\\
\texttt{Pt()} & $p_T$ of \\
\texttt{Pz()} & $p_Z$ of \\
\texttt{E()} & Energy of \\
\texttt{P()} & Momentum of \\
\texttt{pdgID()} & PDGID of a particle \\
\texttt{bTag()} & is the jet b-tagged? \\
\hline
\multicolumn{2}{|c|}{\textit{Computing new quantities}} \\
\hline
\texttt{dR()} & Angular distance between \\
\texttt{dPhi()} & Phi difference between \\
\texttt{dEta()} & Eta difference between \\
MET & Missing transverse energy in the event \\
fHT() & Sum of the transverse momenta of objects \\
\hline
\end{tabular}
\end{table}

\end{document}